\documentclass[aps,12pt,prb,english,showpacs]{revtex4}
\usepackage[dvips]{graphicx}
\usepackage{babel} 
\usepackage{color}
\usepackage{ifthen}  
\usepackage{longtable}
\usepackage{natbib}
\usepackage{dcolumn}
\usepackage{amssymb,amsmath}
\usepackage{axodraw4j}
\usepackage{pstricks}
\def\Tr{\text{Tr}}
\def\la{\langle}
\def\ra{\rangle}
\def\be{\begin{equation}}
\def\ee{\end{equation}}
\def\bea{\begin{eqnarray}}
\def\eea{\end{eqnarray}}
\def\bean{\begin{eqnarray*}}
\def\eean{\end{eqnarray*}}
\def\hpsid{{\hat \psi} ^{\dag}}
\def\hpsi{\hat \psi}
\def\psid{\psi ^{\dag}}
\def\x{{\bf x}}
\def\r{{\bf r}}

\def\y{{\bf y}}
\def\z{{\bf z}}

\def\ve{\varepsilon}
\def\vt{\vartheta}
\def\3di{{\cdot}}
\def\4di{{\scriptstyle \circ}}

\def\PH{{\mathcal G}}    
\def\p{{\mathbf p}}
\def\q{{\mathbf q}}
\begin{document}
\renewcommand{\baselinestretch}{0.95}

\draft
\title{Derivation of the Density Functional \\via Effective Action
  }


\author{Yi-Kuo Yu}
\affiliation{National Center for Biotechnology Information, National Library of Medicine\\
National Institutes of Health, Bethesda, MD 20894, USA}
\begin{abstract}
A rigorous derivation of the density functional in the Hohenberg-Kohn theory is presented. 
With no assumption regarding the magnitude of the electric coupling constant $e^2$ (or correlation), 
this work provides a firm basis for first-principles calculations. 
Using the auxiliary field method, in which $e^2$ need not
be small, we show that the bosonic loop expansion of the exchange-correlation functional can be reorganized
so as to be expressed entirely in terms of the Kohn-Sham single-particle orbitals and energies. 
 The excitations of the many-particle system can be obtained within the same formalism.  
 We also explicitly demonstrate at zero-temperature the single-particle limit,  
  the weak-coupling limit of the energy functional, and its application to homogeneous electron gas. 
\end{abstract}

\maketitle

\renewcommand{\baselinestretch}{0.75}
\section{Introduction} 
At low energy scale, interactions among electrons 
largely determine the structure, phases, and stability of matter. 
 Although this fact is well known, pragmatic first-priniciples/quantum-mechanical 
 calculations to determine various properties of many-electron systems are often hindered by
 two factors. First, in most condensed matter systems, the typical interaction energy between electrons
  (the electric coupling constant $e^2$ divided by average electron-electron separation) is often larger 
  than the typical kinetic/Fermi energy of electrons. The results is that a perturbative expansion using
   $e^2$ as the expansion parameter may not be fruitful. This is particularly true for  
    strongly correlated systems. Second, there is an exponential increase in the number of 
   degrees of freedom as the number of electrons involved
   increases. When the number of electrons becomes large, according to Kohn,\cite{Kohn_99} calculations 
 based on constructing many-electron wave functions soon lose accuracy and
  will be stopped by an ``exponential wall".
  It is thus imperative to have a method that goes beyond the conventional perturbative scheme
   using $e^2$ as the expansion parameter and whose computational complexity does not
   grow exponentially with the number of electrons involved.

In 1964, Hohenberg and Kohn~\cite{HK_64} proved a theorem stating that there exists 
a {\it unique} description of the ground state  of a many-body system in terms of 
the expectation value of the particle-density operator. This theorem started the 
 development of the density functional theory (DFT), which offers a possibility 
 of finding the ground state energy $E_g$ by minimizing the energy functional $E_\upsilon$
that depends on the charge density $n$ only:
\be
E_g=\min\limits_n E_\upsilon \left[ n\right],  \label{enfunc}
\ee
with $\upsilon$ representing the external one-particle potential of the system. The electronic
 density $n_g$, which minimizes the energy functional $E_\upsilon[n]$, is the ground state electronic density. 
Hohenberg and Kohn showed that the energy functional $E_\upsilon[n]$ can be decomposed into
\be \label{first.func} 
E_\upsilon \left[ n \right] = \int d\r \, \upsilon (\r) \, n(\r) + {\mathcal F}\left[ n \right]  \; ,
\ee
with ${\mathcal F} \left[ n \right]$ being a universal functional independent of the 
external potential $\upsilon$. Mermin\cite{Mermin_65} extended this theorem to  
 finite temperature with $E_\upsilon$ in (\ref{first.func}) replaced by the grand potential,
 and ${\mathcal F}[n]$ replaced by a different
   universal functional. The electron density $n_T$, minimizing the grand potential functional, 
   corresponds to the electron density at thermal equilibrium.  

To make practical use of the DFT, however, a recipe to compute the energy functional is needed. 
Kohn and Sham\cite{KS_65} proposed a decomposition scheme, aiming to 
 express the energy functional $E_\upsilon[n]$ via an auxiliary, {\it noninteracting} 
 system that yields a particle density identical to that of the physical ground state. 
 For a typical nonrelativistic many-fermion system, 
 described by the Hamiltonian  
\bea
\hat H &=&\int d{\bf x}\hpsid ({\x})\left( -\frac 1{2m}%
\nabla ^2+\upsilon _{\rm ion}({\x}) -\mu \right) \hpsi ({\x})  \nonumber \\
&&\ +\frac{e^2}2\int \int \frac{\hpsid({\x})\hpsid({\y})\hpsi ({\y})\hpsi ({\x})}{|{\x}-{\y}|%
}d{\x}d{\y},  \label{Ho}
\eea
 the Kohn-Sham decomposition takes the form
\bea
E_\upsilon \left[ n\right] &=&T_{0}\left[ n\right] +\int \upsilon _{\rm ion}(\x )\, n\left( \x\right) d{\x}
-\mu N_e  \nonumber \\
&& + \frac{e^{2}}{2}\int \int \frac{n({\x})n({\y})}{|{\x}-{\y}|}d\x d\y 
+ E_{xc}\left[ n\right] ,  \label{KSfunctional}
\eea
 where the chemical potential $\mu$ is introduced to ensure $\int\! n(\x) \, d\x = N_e$, with $N_e$ being the 
 number of electrons.
Here $T_{0}\left[ n\right] $ is the kinetic energy of an auxiliary system of
 noninteracting fermions that yields the electron density $n\left( {\bf x}\right) $, 
and the density functional $E_{xc}\left[ n\right] $ is the so-called exchange-correlation energy functional.
Given $E_{xc}\left[ n\right]$ and provided that it is differentiable, one may 
minimize the functional $\left( \ref{KSfunctional}\right)$ to arrive at 
the familiar Kohn-Sham single-particle equations\cite{HKS_90} 
\bea
&& \left( -\frac{\nabla^2}{2m} + \upsilon _{\rm ion}(\x) + \int \frac{n(\y)}{|\x-\y|} d\y + 
 \frac{\delta E_{xc}[n]}{\delta n(\x)}  - \mu \right) \psi_i(\x) = \epsilon_i \psi_i(\x) \label{KS_sptcle} \\
&& \hspace{0.5in} n(\x) = \sum_{i=1}^{N_e} \psi_i^*(\x)\psi_i(\x) \; . \label{KS_dsc} 
\eea
  All of the many-particle complexity 
  is now completely hidden in the exchange-correlation energy functional. 

Although $T_0[n] + E_{xc}[n]$ is universal,\cite{HK_64} there exists no
 simple means thus far to obtain it. As a consequence, various {\it ad hoc} exchange-correlation density functionals 
 have been suggested/needed to yield acceptable results in different settings~\cite{AG_04,SFR_07,TDWetal_09,KL_09} 
 when employing the Kohn-Sham scheme.   
 Limitations of these {\it approximate} functionals have been discussed.\cite{KK_08,CM-SY_08} 
 Some of the failures while using {\it ad hoc} density functionals can be attributed to 
 misuse of the density-functional theory.  For example, 
 it is sometimes neglected that the electron density $n(\x)$ achievable 
  via introduction of a source potential 
  must obey $\int n(\x) d\x = N_e$ and 
  does not cover the functional space $\{ n(\x) \ge 0 \}$, a problem also known 
  as the $\upsilon$-representability.~\cite{Kohn_83,Lieb_82,Levy_82}  
 As pointed out in reference~\onlinecite{VF_95},    
 neglecting these constraints may lead to conclusions~\cite{Janak_78} that are
  not always valid.

 The main objective of the DFT is to 
 describe a many-body system in terms of the expectation value of the particle density operator.
 In fact, the use of the expectation value of a suitable operator to describe a many-body system
 via Legendre transformation, first introduced 
  into quantum field theory by Jona-Lasinio,\cite{Jona-Lasinio_64} is known as the effective action formalism.
    As the temperature approaches zero, the effective potential
 becomes the ground state energy. This connection suggests that effective action formalism 
  can be used to achieve the general goal of the DFT: describing a many-particle system
 in terms of the expectation value of the density operator. A number of publications~\cite{FKSY_94,FKYetal_95,VF_97_arxiv,PS_02} 
 showed that at zero temperature the effective action plus $\mu N_e$ is the ground state energy, 
 linking effective action to the DFT.   
 
 Existing methods of expressing DFT via effective action formalism 
  can be classified roughly into two categories: either (a) by introducing
  an auxiliary field or (b) by using a perturbative scheme assuming the electric coupling
   as a small parameter. The former category includes a method developed by Fukuda {\it et al.}~\cite{FKSY_94}  
  and that developed by Polonyi and Sailer.~\cite{PS_02} The latter scheme is used by Valiev and Fernando\cite{VF_97_arxiv}
 and by Fukuda {\it et al.}~\cite{FKYetal_95}.

The strengths of the auxiliary field approach often come from the simplicity 
 of the effective action  expression and from the fact that in principle each term already includes infinitely many
Feynman diagrams.\cite{Jackiw_74} However, as pointed out by Fukuda {\it et al.},\cite{FKSY_94}    
 the auxiliary field approach seems to lack a direct connection to the Kohn-Sham scheme. 
  Valiev and Fernando~\cite{VF_96} introduced an auxiliary field to compute
   the exchange-correlation energy. However, the source term they introduced is coupled to the
    auxiliary field instead of the electron density operator. Furthermore, 
   as pointed out by the authors themselves,~\cite{VF_97_arxiv}
 an  artificial decomposition of the auxiliary field into a sum of the Hartree potential, 
  the exchange correlation potential, and the remaining fluctuations is needed to write down the exchange-correlation energy.

There are two advantages when one uses electric coupling in the perturbative (diagrammatic) expansion 
without introducing an auxiliary field. First, under the effective action formalism of this type, 
    a direct connection to the Kohn-Sham scheme can be made.~\cite{VF_97_arxiv,FKYetal_95} 
  Second, there exist other $e^2$ expansion-based developments that can be used to 
  obtain $T_0[n] + E_{xc}[n]$, the universal density functional (UDF). 
  For example, with increasingly complex incorporation of KS
    orbitals and energies at each order of $e^2$, 
    G\"orling and Levy~\cite{GL_94} wrote $E_g$ as a perturbation series. 
 Employing the Luttinger-Ward~\cite{LW_60} method that uses $e^2$ as the 
  perturbative expansion parameter to calculate electron self-energy, 
  Sham and Schl\"uter expressed $E_{xc}$ as a rather convoluted implicit functional of
  electron density.\cite{SS_83,Sham_85} As shown by Tokatly and Pankratov,\cite{TP_01} these methods mentioned above 
  can be expressed diagrammatically. 
  However, the problem associated with $e^2$ based expansion remains. 
  As described in reference~\onlinecite{Negele_Orland_book}, the expansion 
  using $e^2$ is only good when 
 $e^2$ is very small. Whether one can treat $e^2$ as a small parameter or not depends on the 
  kinetic/Fermi energy of the electrons and the strength as well as the magnitude of variation of 
  the single-particle potential involved.  As a matter of fact,
   the success~\cite{DvLvB_06} in employing the GW approximation~\cite{Hedin_65} indicates 
 that not treating $e^2$ as small may lead to results closer to experimental outcomes.  

 In principle, the problem associated with assuming $e^2$ small can be tamed by summing an 
 infinite subset of Feynman diagrams. However,  
 as pointed out by Hedin,~\cite{Hedin_65} it is nontrivial to devise a  systematic 
 resummation scheme where each new term is free from divergence even if $e^2 \gg 1$ and for which  
 the sum of the new terms, each containing an infinitely many Feynman diagrams,  
 accounts  completely and non-redundantly for all conventional $e^2$ expansion diagrams. 
In this paper, without assuming $e^2$ small we develop an auxiliary field method 
that makes a direct connection to the Kohn-Sham scheme and at the same time provides equivalently
 a systematic resummation scheme. 
 
 A commonly used approximation for
 the density functional is the so-called 
 local density approximation (LDA) in which the exchange-correlation energy is 
  approximated by a linear functional $E_{xc} [n] \approx \int d\r \, n(\r) \; e_{xc}(n(\r))$, where 
 $e_{xc}(n)$ is a function of the local density (not a functional of the density profile). 
  See reference~\onlinecite{Kohn_99} for a nontechnical review. 
  This approximation ignores the nonlocal effect of the density profile, i.e., it assumes that $\delta E_{xc}[n]/\delta n(\r)$ 
   only depends on the value of $n$ at $\r$ but not on the density $n(\r' \ne \r)$ at 
   locations other than $\r$. To complement the LDA by incorporating nonlocal
   density dependence, Polonyi and Sailer~\cite{PS_02} proposed 
  the $l$-local approximation for the density functional, based on an idea 
   very similar to the cluster expansion in statistical physics. 
   Using this method to obtain explicit expressions for the approximate functional 
   with $l \ge 3$, however, becomes increasingly challenging due to the necessity of 
   going through the coupling constant integration     
   as required by the Hellmann-Feynman theorem.\cite{Hellmann_37,Feynman_39}  

Another route to developing density functionals is via the so-called
 optimal effective potential (OEP) methods.~\cite{SH_53,TS_76,PGG_96}
 These methods typically start by introducing {\it a priori} an approximate, explicitly~\cite{KP_03}
  orbital-dependent functional. (The approximate functional can be either Hartree-Fock 
  or a more elaborated form.) The procedure then continues with  
   a minimization of the functional via varying single-particle 
   KS orbitals and associated energies. Recently, the definition of OEP methods has been 
 generalized~\cite{Casida_95b,vBDvLS_05} to include 
  functionals dependent on either Green's function, the self-energy, or the KS potential. 
 Since our effective action based functional is based on self-consistently obtaining 
 the KS potential, it falls exactly in the latter category. The 
  generalized definition of OEP methods is probably becoming the standard definition now.   
 A general characteristic of OEP methods is that the functional arguments 
 --be they KS orbitals/energies, Green's functions, self energies, or KS potentials-- 
 are obtained via self-consistent procedure. Therefore, even with correction terms derived from 
  pertrubative expansion, the self-consistency condition for OEP methods
  distinguishes them from regular pertrubative methods. The important matter here is whether
   an OEP functional can be systematically improved and possibly be asymptotically exact 
  or not.    
 
   Based on effective action formalism, the OEP functional proposed here is asymptotically exact
    and can be shown to give rise to the desired UDF. Containing all the $l$-local interaction
     vertices, our method can provide equivalent approximate functionals for $l\ge 3$ 
     without going through the Hellmann-Feynman theorem.   
  Since we  focus on describing the proposed approach in a manner as self-contained as possible, 
 we have included  a non-negligible amount of standard materials available in 
  existing literature/textbooks while keeping only a small portion of the existing literature 
  that we deem closely related to the present manuscript.  
  Readers interested in gaining a broad background are  
 referred to references~\onlinecite{KK_08} and~\onlinecite{BdGdCetal_01,Furnstahl_07,JG_89,BM_07} 
 for reviews on the extensive body of literature in the DFT and related many-body approaches. 
 Expert readers should note that new developments are mainly provided in sections IIIB-D. 
  Although section IV and end of section V also contain some 
  useful developments, they typically rederive/re-express  known results 
   within our framework and/or provide contrast with existing methods. 
  Section VI contains some insight of problems shared in post Hartree corrections.

This paper is otherwise organized as follows. We first establish the notation in section~\ref{sec:notation}, 
followed by the development of the general formalism in section~\ref{sec:formulation}. 
The purpose of subsection~\ref{subsec:recipe} is to provide a computational recipe 
  and to give some perspectives on computational complexity: no novelty is claimed here.  
 In section~\ref{sec:case_studies}, we discuss a number of case studies: the emergence 
 of the universal functional ${\mathcal F}[n]$ in Eq.~(\ref{first.func}) at arbitrary temperature,
  the behavior of  the effective potential and the single electron limit at zero temperature, 
  the screening effect, as well as the case of homogeneous electron gas.  
  In section~\ref{sec:excitation}, we then discuss the excitations of the system, and 
  make comparisons with existing studies along this direction. 
   An alternative formalism to obtain the effective action is then discussed in section~\ref{sec:saddle}.  
   We conclude with the discussion and future directions section, in which we also provide some more relations/comparisons
    to other methods as well as some technical remarks.

\section{Notation}
\label{sec:notation}
Let us first define useful notation to lighten the exposition of the
 mathematical formulas.

We define a three dimensional integral contraction by a dot
\[
 a \3di b \equiv \int d\x \; a(\x)\, b(\x)
\]  
where $a$ and $b$ may be single or composite fields. That is,
 with $m\ge 1$ and $n\ge 1$, $a(\x)$ and $b(\x)$ may represent
\bea
a(\x) & = & a_1(\x) \ldots a_m(\x) \; ,\nonumber \\
{\rm and~~~}b(\x) & = & b_1(\x) \ldots b_n(\x) \; . \nonumber
\eea

Similarly, with a kernel $M$, we may define 
\[
a\3di b \3di M = a \3di M \3di b = M \3di a \3di b
 = \iint d\x d\y \; M(\x,\y) a(\x)\, b(\y) \; .
\]
Note that in all expressions $a$ is in front of $b$, and that is important when there
 are Grassmann variables involved. 
Evidently, one may generalize this notation to include higher order kernels. That is, one may have
\[
a \3di b \3di c \3di M = a \3di b \3di M \3di c =  a \3di M \3di b \3di c = M \3di a  \3di b \3di c 
 = \iiint d\x d\y d\z \; M(\x,\y,\z) \, a(\x)\, b(\y)\, c(\z)\; .
\]

We define the four dimensional integral contraction by an
 open circle 
\[
 a\4di b \equiv \int dx \; a(x)\, b(x) \; ,
\]  
with 
\bea
x &=& (\tau, \x) \; , \nonumber \\ 
0 &\le &\tau \le \beta  \; ,\nonumber \\ 
{\rm and~~~} \int dx &=& \int_0^\beta d\tau \int d\x \nonumber \; .
\eea
Again, $a$ and $b$ may be single or composite fields. That is,
 with $m\ge 1$ and $n\ge 1$, $a(x)$ and $b(x)$ may represent
\bea
a(x) & = & a_1(x) \ldots a_m(x) \; , \nonumber \\
{\rm and~~~}b(x) & = & b_1(x) \ldots b_n(x) \; .\nonumber
\eea

Similarly, we may define 
\[
a\4di b \4di M = a \4di M \4di b = M \4di a \4di b
 = \iint dx dy \; M(x,y) a(x)\, b(y) \; , 
\]
and 
\[
M \4di a  \4di b \4di c 
 = \iiint dx dy dz \; M(x,y,z) \, a(x)\, b(y)\, c(z)\; .
\]

\section{Relevant formulation} \label{sec:formulation}
Consider the following generic fermionic Hamiltonian with $s$ denoting the spins
\begin{eqnarray}
\hat H [\hpsid, \hpsi]&=& \sum_s \int d{\bf x}\; \hpsid_s(\x) \left( -\frac{\nabla ^2}{2m}
+\upsilon _{\rm ion}({\x, s})-\mu_s \right) \hpsi_s (\x)  \nonumber \\
&&\ +\frac{1}2 \sum_{s,s'}\iint {\hpsid_s (\x)\hpsid_{s'}(\y) U(\x-\y) \hpsi_{s'} (\y)\hpsi_s (\x)}
 d\x d\y,  \label{H.0} \\ 
 & =& \sum_s \int d{\bf x}\; \hpsid_s(\x) \left( -\frac{\nabla ^2}{2m}
+\upsilon _{\rm ion}({\x,s}) - \frac{U({\mathbf 0})}{2} -\mu_s \right) \hpsi_s (\x)  \nonumber \\
&&\ +\frac{1}2\iint {\left( \sum_s \hpsid_s (\x)\hpsi_s (\x)\right) U(\x-\y)\left( \sum_s \hpsid_s(\y)\hpsi_s (\y) \right)} \nonumber \\
& = & \sum_s \int d{\bf x}\; \hpsid_s(\x) \left( -\frac{\nabla ^2}{2m}
+\upsilon _{\rm ion}({\x,s}) - \frac{U({\mathbf 0})}{2} -\mu_s \right) \hpsi_s (\x) \nonumber \\
&& \  + \frac{1}{2} (\sum_s\hpsid_s \hpsi_s) \3di U \3di  (\sum_s \hpsid_s \hpsi_s) \; .
 \label{H.1}
\end{eqnarray}
From this point on, we absorb $-\frac{U({\mathbf 0})}{2}$ into $\upsilon _{\rm ion}(\x, s)$. 

To lighten the notation, we will first ignore the spin degree of freedom 
but will comment on its effect when clarifications are needed. 
Let $\beta$ be the inverse temperature. 
The partition function $Z \equiv \Tr [e^{-\beta \hat H} ]$ contains all the information 
 one needs. To probe the system in terms of the particle density, one often introduces a classical source term $J(\x)$
 coupled to $\hpsid(\x) \hpsi(\x)$. The partition function now becomes a functional of the source $J$, and we 
 write
\[
Z [J] \Rightarrow e^{-\beta W[J]} = \Tr \left[ e^{-\beta \left[\hat H + J \3di (\hpsid \hpsi) \right]} \right] \; .
\]

It is easy to show that 
\be \label{J_n}
\frac{\delta W[J]}{\delta J(\x)} = \frac{\Tr \left[ \hpsid(\x) \hpsi(\x) e^{-\beta \left[\hat H + J \3di (\hpsid \hpsi) \right]} \right]}{Z[J]} = \la {\hat n}(\x) \ra_J \equiv n_J(\x) 
\ee
 Eq.~(\ref{J_n}) expresses $n$ in terms of $J$, or more generally
 expresses $n_s$ (charge density of spin $s$) in terms of sources $J_{s'}$ of all spins.  
 Given $\upsilon_{\rm ion}(\x)$ and $U(\x-\y)$ (for Coulomb interaction $U(\x-\y) = \frac{e^2}{|\x-\y|}$), 
  each time-independent configuration of $\{ J(\x) \}$ 
 generates a time-independent charge density distribution $\{ n(\x) \}$. However, it is not guaranteed that every configuration
 of $\{ n(\x) \}$ is reachable by varying $J$.  
 The functional variation on $\{ n(\x) \}$ is thus limited to the subset of
 $\{ n(\x) \}$ reachable by considering various $\{ J(\x) \}$. In this stationary case, the
  effective action $\Gamma[n]$ is defined as the Legendre transformation of $W[J]$, 
 \[
\Gamma [n_J] = W[J] - J \3di n_J \; , 
 \]  
where  the subscript $J$ indicates that the domain of $\Gamma[n]$
is the set of density profiles reachable by varying $J$, or the so-called $\upsilon$-representable~\cite{Kohn_83,Lieb_82,Levy_82} densities.

We now show the equivalence between the effective action and 
 the energy functional $E_\upsilon[n]$ in (\ref{enfunc}).  
Since 
\[
e^{-\beta W[J]} =  \Tr \left[ e^{-\beta \left[\hat H + J \3di (\hpsid \hpsi) \right]} \right] \;, 
\]
 at zero temperature limit  $W[J]$ is simply the ground state energy corresponding
 to the Hamiltonian $\hat H_J \equiv \hat H + J \3di (\hpsid \hpsi)$, 
 \bea
\hat H_J &=&\int d{\bf x}\hpsid ({\x})\left[ -\frac 1{2m}%
\nabla ^2+ \left(\upsilon _{\rm ion}({\x}) + J(\x) \right) -\mu \right] \hpsi ({\x})  \nonumber \\
&&\ +\frac{e^2}2\int \int \frac{\hpsid({\x})\hpsid({\y})\hpsi ({\y})\hpsi ({\x})}{|{\x}-{\y}|%
}d{\x}d{\y},  \label{H_J}
\eea
 while the electron density $n_J(\x)$ is obtained by integrating all but one spatial 
 variable of the ground state wave function corresponding to $\hat H_J$.
 Evidently, when $J=0$, $\Gamma[n]|_{n=n_g} = W[J]|_{J=0} = E_g$ where $E_g$ stands for  
 the ground state energy corresponding to $\hat H$ and $n_g$ represents the 
 electron density at the physical ($J=0$) ground state. When $J \ne 0$ the 
 corresponding electronic density $n[J\ne 0]$ is different from $n_g$,  
 and $\Gamma[n_J]$ represents the expectation value of the original 
 Hamiltonian $\hat H$, calculated using the ground state wave function corresponding to 
 a different Hamiltonian, namely, $\hat H + J \3di (\hpsid \hpsi)$.
 Since $n_J \ne n_g$, $\Gamma[n]|_{n=n[J]} > \Gamma [n]|_{n=n_g}$ by the definition of the ground state. 
 This means that $\Gamma[n]$ reaches its minimum at $n_g$, and various other electron density 
 profiles $n[J]$ producible by introducing different $J$ form the domain of argument for $\Gamma[n]$. 
Thus $\Gamma[n]$ has all the properties attributed to the energy functional $E_\upsilon[n]$ in
(\ref{enfunc}). Since the theorem of Hohenberg and Kohn states that this functional
is unique, it must be equal to $E_\upsilon[n]$. 
 As we will show in section~\ref{sec:zero.T}, $\Gamma[n]$ can be decomposed exactly in the 
 same manner as in (\ref{KSfunctional}).

 Within allowable configurations of $\{ n(\x) \}$, if one is able to invert the relation
  (\ref{J_n})  to obtain, say, $J[n]$, then the explicit construction of the effective action becomes possible. 
 In principle, this can be done via an inversion method.\cite{FKYetal_95}
 Using this scheme, Valiev and Fernando~\cite{VF_97_arxiv} proposed a perturbative expansion in terms of $e^2$ to 
  express the exchange-correlation functional as a sum of an infinite number of Feynman diagrams
 and the diagrams' derivatives with respect to the Kohn-Sham potential.

 We approach this problem from two different routes, both involving the 
 introduction of an auxiliary field.\cite{FKSY_94,PS_02} As will be described in
  secton~\ref{sec:saddle_point}, the second route does not 
 have an exact correspondence to the Kohn-Sham decomposition, but has the advantage
 that the correction terms may be obtained without further functional derivatives. 
  The first route, as will be described later in this section, 
 gives a recipe equivalent to the Kohn-Sham decomposition, together with a way
   to calculate the exchange-correlation functional in a self-consistent manner. 

 This auxiliary field approach was pursued in an earlier publication,~\cite{FKSY_94} but
 there it was concluded that it seems infeasible to make a direct connection 
 to the Kohn-Sham scheme. Using the inversion method,~\cite{FKYetal_95} we show explicitly how 
 the connection to the Kohn-Sham scheme can be made. One advantage of the
 auxiliary field method is that each Feynman diagram here corresponds to the sum of 
 infinitely many Feynman diagrams in standard perturbative field theory calculations~\cite{Jackiw_74} such as
 used in reference~\onlinecite{VF_97_arxiv}. Subsections~\ref{subsec:Path_Integral} through \ref{subsec:FDPH0} 
 detail the proposed approach. Subsection~\ref{subsec:recipe} lays out the computational procedure 
  to give some perspectives on computational complexity.  

\subsection{Path Integral}
\label{subsec:Path_Integral}
To accommodate a time-dependent probe and to deal with excitations, we  
 express $Z$ as a path integral over Grassmann fields and we have 
\begin{equation}
e^{-\beta W[J]} \equiv Z[J]  =\int D\psid D\psi \; \exp \left\{ -S\left[ \psid,\psi \right]  - J\4di (\psid \psi) \right\} \; ,  \label{genfunc}
\end{equation}
with 
\be
S \left[ \psi ^{\dagger },\psi \right] =
 \psid \4di G_0^{-1} \4di \psi +\frac{1}2 \left(\psid \psi \right) \4di\, U \4di \left(\psid \psi \right)\; ,   \label{action}
\ee
where $\psi ^{(\dagger) }$ denote Grassmann fields with $\psi ^{(\dagger)}(\beta,\x) = -\psi ^{(\dagger)}(0,\x)$, and
\bea
 G_0^{-1}(x,x') &\equiv&  \la x | G_0^{-1} | x' \ra \nonumber \\ 
 &= &  \left( {\partial \tau }-\frac{\nabla ^2}{2m}+\upsilon _{\rm ion}({\bf x}) - \mu \right) 
 \la x | x' \ra 
 = \left( {\partial \tau }-\frac{\nabla ^2}{2m}+\upsilon _{\rm ion}({\bf x}) - \mu \right)
\delta (x-x') \; , \nonumber
\eea
and 
\bea
U(x,x') &=& U(x-x')=\delta \left( \tau -\tau^{\prime}\right) U(\x-\x'), \nonumber \\ 
\delta(x-x') &=& \delta(\tau-\tau') \delta (\x - \x') \nonumber \; .
\eea
For a time-independent source, 
$J(\tau,\x) = J(\x)$ (i.e., $J \4di (\psid \psi) = \int dx J(\x) \psid(x) \psi(x)$). 
For time-dependent probes, $J(x)$ becomes $\tau$-dependent, and 
  $J \4di (\psid \psi) = \int dx J(x) \psid(x) \psi(x)$. 
It is straightforward to verify that 
\be \label{W2n}
\frac{\delta (\beta W\left[ J \right])}{\delta J(x)} 
 = \la \hpsid(x)\hpsi(x) \ra_J = \la \hat n(x) \ra_J \equiv n_J(x)\; .
\ee
This quantity is important for later development. 

The quartic fermionic interaction in (\ref{action}) can be disentangled via introducing an auxiliary real field $\phi$
 with $D \phi \equiv \prod_x  \frac{d\phi(x)}{\sqrt{2\pi}}$. Note that 
\be \label{St.id} 
1 = \sqrt{\det U} \int D\phi \;  e^{-\frac{1}{2}\phi \4di \, U \4di \phi } =  \sqrt{\det U}\int D\phi  \; 
e^{-\frac{1}{2}(\phi+Y)\4di \, U \4di (\phi+Y)} \; ,
\ee  
for an arbitrary field $Y$, provided that $Y(x)$ and $Y(x')$ always commute. Since $U(x,x')$ is diagonal in $\tau$,
 it suffices that they commute at equal Euclidean times. Let us set $Y(x) = i\psid(x)\psi(x)$, which satisfies
  the equal time commutation requirement, and then multiply (\ref{genfunc}) by (\ref{St.id}) to obtain 
\be
Z\left[J\right] =\int D\phi D\psid D\psi \;
 \exp \left\{ -{S}\left[ \phi,\psid,\psi \right]  \right\} \; ,   \label{genfunc.1}
\ee
where
\be \label{action.1}
{S}\left[ \phi,\psid,\psi \right] = -\frac{1}2 \Tr \ln (U) 
+ \frac{1}2 \phi \4di \, U \4di \phi + \psid \4di \, G^{-1}_{\!(\phi-iu^{-1}\4di J)}\,  \4di \psi
\; ,
\ee
with 
\be
G^{-1}_{\!(\phi-iU^{-1}\4di J)}(x,x') 
 =  \left( \partial_\tau -\frac{\nabla ^2}{2m}
+\upsilon _{\rm ion}({\x}) -\mu + i (U \4di \phi)_x + J(x)  \right)\delta(x-x') \; . \label{GJ}
\ee   
If we make a change of variable $\phi' \equiv \phi - i U^{-1} \4di J$ and then rename $\phi'$ by $\phi$, 
 we may rewrite (\ref{genfunc.1}-\ref{GJ})
 as 
\bea
Z\left[J\right] &=& e^{\frac{1}{2} J\4di U^{-1} \4di J} \; 
\int D\phi D\psid D\psi \; \exp \left\{ -{S}_J\left[ \phi,\psid,\psi \right]  \right\}  \; , \label{genfunc.2} \\
{S}_J\left[ \phi,\psid,\psi \right] &=& -\frac{1}2 \Tr \ln (U) + \frac{1}2 \phi \4di \, U  \4di \phi + 
 i \phi \,\4di J  +  \psid \4di {G}_\phi^{-1} \4di \psi \; ,
\label{action.2} \\
\la x \vert G_\phi^{-1} \vert x' \ra &\equiv & {G}_\phi^{-1}(x,x') 
 =   \left( \partial_\tau -\frac{\nabla ^2}{2m}
+\upsilon _{\rm ion}({\x}) -\mu  + i (U \4di \phi)_x  \right)\delta(x-x') \, . \label{Green.general} 
\eea

Integrating over the Grassmann fields in (\ref{genfunc.2}), 
we obtain an effective theory in terms of $\phi$
\be
 e^{-\beta W[J]} = Z\left[J\right] = 
 e^{\frac{1}2 J \4di \, U^{-1} \4di J}
\int D\phi  \; \exp \left\{ -I\left[ \phi \right] - iJ\4di \phi  \right\}   \label{genfunc.3}
\ee
where 
\be \label{I.phi}
I[\phi] = -\frac{1}2 \Tr \ln (U) + \frac{1}2 \phi \4di \, U \4di \phi  - \Tr \ln (G_{\phi}^{-1}) \; .
\ee
Let us introduce a new notation $W_\phi[J]$ via 
\be \label{z1}
e^{-\beta W_\phi[J]} \equiv \int D\phi  \; \exp \left\{ -I\left[ \phi \right] - iJ\4di \phi  \right\} \; .
\ee
We describe later how to evaluate (\ref{z1}) using well-developed
 functional integral techniques. 
Evidently, we have 
\be
\beta W\left[ J \right]  = \beta W_\phi\left[J \right] - \frac{1}{2} J \4di \, U^{-1} \4di J \; .
\ee

The expectation value, $n_J(x)$, of the density operator in the presence of a source term $J$ 
 is given by 
\be \label{n2phiJ4d}
n_J(x) =  \frac{\delta (\beta W[J])} {\delta J(x)} = \frac{\delta (\beta W_\phi[J])}{\delta J(x)} - ({U}^{-1}\4di J)_x \equiv i\varphi(x)  - (U^{-1}\4di J)_x \; ,
\ee
 where 
\be
(U^{-1}\4di J)_x \equiv \int dy\,  U^{-1}(x,y) J (y) \; , \nonumber
\ee
and the expectation value of the auxiliary field is defined by
\be \label{phiJ4.def}
i \varphi(x) \equiv \la i\phi(x) \ra_J = {\delta (\beta W_\phi) \over \delta J(x)} = 
 \frac{\int D\phi  \left(i\phi(x)\right) e^{-I[\phi] -iJ\4di \phi}}{\int D\phi  
 e^{-I[\phi] -iJ\4di \phi}}\; ,
\ee
providing a relation between $J$ and $i\varphi$.

Eq.~(\ref{n2phiJ4d}) tells us that at 
the physical limit  $J\to 0$, $i\varphi $ is the same as $n$.
Since $n$ is a real number, this implies that
 the expectation value of $\phi(\x)$ is an imaginary number, which also implies that when viewed in the complex 
 plane of $\phi(\x)$,  the saddle point of the integrand is located where the $\phi(\x)$s are imaginary numbers. 

Now let us write down the effective action. At finite temperature, 
the effective action is defined as the Legendre transform of
 $\beta W[J]$:
\be \label{effective_action}
\Gamma [ n ] \equiv \beta W\left[ J \right] - {\delta (\beta W[J]) \over \delta J} \4di J = 
\beta W_\phi[J] - \frac{1}{2} J \4di \, U^{-1} \4di J - n\4di J \; .
\ee
Note that the functional derivative of $\Gamma[n]$ with respect to $n$ reads
\be\label{Gamma2J}
\frac{\delta \Gamma[n]}{\delta n} = \left[ \frac{\delta (\beta W[J])}{\delta J} -n \right] \4di \frac{\delta J}{\delta n}
 - J = -J  \; ,
\ee
 because ${\delta (\beta W[J])}/{\delta J} = n$ by Eq.~(\ref{W2n}).
The effective action formalism requires one to express the probe $J$ in terms of
  an expectation value of some sort, such as the electron density $n$, classical field
 $i\varphi$, or some equivalent quantity. 
Below, we will first calculate $\beta W_\phi[J]$, and
 then use the system's electron density as the variable and
 make an {\it explicit} connection to the Kohn-Sham decomposition.

\subsection{Evaluation of $e^{-\beta  W_\phi [J]}$ via one-particle irreducible diagrams}
 As shown by Jackiw,\cite{Jackiw_74} it is possible to express $W_\phi [J]$ as a diagrammatic
 expansion containing only one-particle irreducible diagrams. 
 The main idea is to shift the field $\phi$ by $\varphi$, $\phi \to \varphi + \phi$, 
  and note that $J$ is a functional of $\varphi$ via (\ref{phiJ4.def}). We then 
rewrite 
\[
e^{-\beta W_\phi[J]} \equiv e^{-I[\varphi] - iJ \4di\, \varphi}
 \int D\phi  \; e^{ -I\left[ \phi + \varphi \right] + I\left[ \varphi \right] - iJ\4di \phi } 
 \equiv e^{-I[\varphi] - iJ \4di \, \varphi} Z_{*}[J] \; 
\]
where 
\be 
-\ln Z_{*}[J] \equiv \beta W_{*} [J] = -\ln \left[ \int D\phi  \; e^{ -I\left[ \phi + \varphi \right] 
+ I\left[ \varphi \right] - iJ\4di \phi } 
 \right] \label{W*} \; , 
\ee
leading to 
\be \label{Wphi.in.*}
\beta W_\phi[J] = I[\varphi] + J \4di (i\varphi ) + \beta W_{*}[\varphi ] \; .
\ee

Note that 
\[
i\varphi(x) = {\delta (\beta W_\phi[J]) \over \delta J(x)}
 =  i\varphi (x)+ \int  dy \left[ {\delta I \over \delta (i\varphi (y))} + { \delta (\beta W_{*}) \over  \delta (i\varphi (y))}  + J (y) 
\right] {\delta (i\varphi (y)) \over \delta J(x)}  \; ,
\]
 leading to 
\be \label{J.eq}
{\delta I \over \delta (i\varphi (y))} + { \delta (\beta W_{*}) \over 
 \delta (i\varphi (y))}  = -  J (y)   \; .
\ee
Using an implicit method and replacing $-J$ in (\ref{W*}) by the left-hand side (LHS)
 of (\ref{J.eq}), Jackiw~\cite{Jackiw_74} showed that $\beta W_{*}[\varphi]$ 
 is the sum of all connected one-particle-irreducible (1PI) 
 vacuum graphs governed by the action
\[ 
 -I[\phi+\varphi] + I[\varphi] +  \phi \, \4di {\delta I[\varphi]\over \delta \varphi} \; .
 \] 

To evaluate the expression above, we first rewrite (\ref{Green.general}) as
\be \label{Green.general.0}
G_{\phi+\varphi}^{-1}(x,x') = G_\varphi^{-1}(x,x') + i \delta(x-x')
 b(x)
 \equiv G^{-1}_\varphi(x,x') +  V(x,x') \; ,
\ee
with 
\bea 
b &=& U\, \4di \, \phi \; ,\label{b.def} \\
{\rm and~~~~}G_{\varphi}^{-1}(x,x') &=& \left[ \partial_\tau -\frac{\nabla^2}{2m} + \upsilon _{\rm ion}({\x}) -\mu
 + U \4di (i\varphi) \right] \delta (x-x') \; . \label{Green_phi}
\eea
We may then write down
\[
G_{\phi+\varphi}^{-1} = G^{-1}_\varphi \left[ \, {\mathbf I} + G_\varphi \4di {\mathbf V} \right] \; ,
\]
and 
\be \label{Green.general.expand}
\ln \left( G_{\phi+\varphi}^{-1} \right) = 
\ln \left( G^{-1}_\varphi \right) + \sum_{k=1}^\infty \frac{(-1)^{k-1}}{k}
 \left[ G_\varphi \4di  {\mathbf V} \right]^k \; .
\ee
Note also that 
\[
\left[ G_\varphi \4di {\mathbf V} \right]_{x,z} 
 = \int\!\! dy \, G_\varphi (x,y) V(y,z)
 =  \int dy G_\varphi (x,y) \delta(y-z) \left( i b(y) \right)
 = G_\varphi (x,z) \left(i b(z)\right)  \; .
\]

Consequently, 
\bea 
\Tr \ln \left( G_{\phi+\varphi}^{-1} \right) & = &  \Tr \ln \left( G^{-1}_\varphi \right)
 + \int \! dx_1 \,G_\varphi(x_1,x_1) (i b(x_1))  \nonumber \\
&& \hspace*{-50pt}
 - \frac{1}{2} \int \! dx_1 dx_2  \, G_\varphi(x_1,x_2)  G_\varphi(x_2,x_1) 
 (i b(x_1))(i b(x_2))  \nonumber \\
&& \hspace*{-50pt} + \sum_{k=3}^\infty \frac{(-1)^{k-1}}{k} \int \! dx_1 \ldots dx_k   
 G_\varphi(x_k,x_1)\ldots G_\varphi(x_{k-1},x_k)
  (i b(x_1))\ldots(i b(x_k)) \; .
\label{Green.general.trace}
\eea
Therefore, 
\be \label{Jackiw_main}
-I[\phi+\varphi] + I[\varphi] + \phi \, \4di {\delta I[\varphi]\over \delta \varphi}
= - \frac{1}{2}b\, \4di \tilde {\mathcal D}^{-1} \4di\,  b  + \sum_{k=3}^\infty
  I^{(k)}[\varphi] \4di \,b_1 \ldots \4di\, b_k 
\ee 
with 
\be \label{tsD.def} 
\tilde {\mathcal D}^{-1} = U^{-1} - D \; ,
\ee
\be   \label{D.def}
D(x,y) = G_{\varphi}(x,y)\, G_{\varphi}(y,x) \; ,
\ee
and 
\be 
 I^{(k)}[\varphi]  \4di\, b_1 \ldots \4di\, b_k
\equiv   \frac{(-1)^{k-1}}{k} \int \! dx_1 \ldots dx_k   
G_{\varphi}(x_k,x_1) \ldots G_{\varphi}(x_{k-1},x_k) \left[
  (ib(x_1))\ldots(ib(x_k)) \right] \label{I_kernel} \; .
\ee 
As a side note, the quantity $D(x,y)$ defined in (\ref{D.def}) is also called the polarization
 associated with the Green's function $G_{\varphi}$, since it can be shown, by using a derivation identical  
 to that leading to (\ref{dndJ.tdep}),  that  $D(x,y) = - \delta G_\varphi(x,x)/\delta J(y)$
 represents the reaction rate of density (given by $n(x) \equiv -G_\varphi(x,x)$) 
 due to the influence of the potential. 
According to Jackiw's results,\cite{Jackiw_74} Eq.~(\ref{Jackiw_main}) means that $\beta W_*[\varphi]$ is given by
\be \label{W*.eval}
\beta W_* [\varphi ] = \Tr \ln (U) + \frac{1}{2} \Tr \ln \left( \tilde {\mathcal D}^{-1} \right) 
 - \sum_{n=1}^\infty \frac{1}{n!} \la \left[ \sum_{k=3}^\infty
 I^{(k)}[\varphi] \4di \, b_1 \ldots \4di\, b_k  \right]^n \ra_{\rm 1PI,~conn.} 
\ee 
where the $\Tr \ln (U)$ term comes from the Jacobian of changing the variable from $\phi$ to $b$
 in (\ref{W*}),  
the angular bracket indicates the following average 
\be \label{ave.def} 
\la \hat O \ra \equiv \frac{ \int D[b] \, \hat O \, \exp\left( -\frac{1}{2} b \4di \tilde {\mathcal D}^{-1} \4di b \right) }
{\int D[b]\, \exp \left(-\frac{1}{2} b \4di \tilde {\mathcal D}^{-1} \4di b\right) } \; ,
\ee
and the subscript ``${\rm 1PI,~conn.}$" means to include only connected, one-particle-irreducible diagrams.
 In our context, a one-particle-irreducible diagrams refers to a diagram that cannot be separated into 
  two by cutting a propagator line representing $\tilde {\mathcal D}$.

Substituting (\ref{I.phi}) and (\ref{W*.eval}) into (\ref{Wphi.in.*}), we obtain
\bea
\beta W_\phi [J] &=&  \frac{1}{2} \varphi \4di \, U \4di \varphi
- \Tr \ln \left( G_{\varphi}^{-1} \right) + i J \4di \varphi \nonumber \\
&& + \frac{1}{2} \Tr \ln \left( \tilde {\mathcal D}^{-1} \4di \, U \right) 
 - \sum_{n=1}^\infty \frac{1}{n!} \la \left[ \sum_{k=3}^\infty
 I^{(k)}[\varphi] \4di \,b_1 \ldots \4di\, b_k  \right]^n \ra_{\rm 1PI,~conn.} \; . \label{W.fukuda}
\eea

Note that Fukuda {\it et al.}~\cite{FKSY_94}  obtained an expression similar to (\ref{W.fukuda}) 
and used it to derive an effective action as a functional of $i\varphi$, which coincides with $n_J$
 only at $J=0$. 
 
We wish to keep $n_J$ as the functional variable.  Let us first  
note that $\beta W[J] = \beta W_\phi[J] - {1\over 2} J \4di\, U^{-1} \4di J $. By employing 
  the identity (\ref{n2phiJ4d})
\[
n_J = i\varphi - U^{-1} \4di J \;, 
\]
we can now write $\beta W[J]$ as 
\bea
\beta W [ J]  &=& 
- \Tr \ln \left( G_{\varphi}^{-1} \right)   - \frac{1}{2} n_J \4di \, U \4di n_J 
+ \frac{1}{2} \Tr \ln \left( \tilde{\mathcal D}^{-1} \4di \, U \right) \nonumber \\
&& \hspace*{15pt}  - \sum_{n=1}^\infty \frac{1}{n!} \la \left[ \sum_{k=3}^\infty
 I^{(k)}[\varphi] \4di \,b_1 \ldots \4di\, b_k  \right]^n \ra_{\rm 1PI,~conn.} \;,
 \label{W.varphi.2}
\eea
and 
\bea
\Gamma[n]  &=& 
- \Tr \ln \left( G_{\varphi}^{-1} \right)   - \left[ J +  \frac{1}{2} n_J \4di \, U \right] \4di n_J 
+ \frac{1}{2} \Tr \ln \left( \tilde{\mathcal D}^{-1} \4di \, U \right) \nonumber \\
&& \hspace*{15pt}  - \sum_{n=1}^\infty \frac{1}{n!} \la \left[ \sum_{k=3}^\infty
 I^{(k)}[\varphi] \4di \,b_1 \ldots \4di\, b_k  \right]^n \ra_{\rm 1PI,~conn.} \;.
 \label{G1.1}
\eea
For later convenience we introduce a parameter $\lambda$ to denote the order.
Specifically, we write 
\bea
\beta W [ J]  &=& 
- \Tr \ln \left( G_{\varphi}^{-1} \right)  -  \frac{1}{2} n_J \4di \, U \4di n_J   
+ \frac{\lambda}{2}  \Tr \ln \left( \tilde {\cal D}^{-1} \4di \, U \right)  \nonumber \\
&& \hspace*{15pt} - \lambda \sum_{n=1}^\infty \frac{1}{n!} \la \left[ {1\over \lambda} \sum_{k=3}^\infty
 \lambda^{k\over 2} I^{(k)}[\varphi] \4di \,b_1 \ldots \4di\, b_k  \right]^n \ra_{\rm 1PI,~conn.} \;.
 \label{W.varphi.3} \\
 &\equiv& \beta \tilde W_0[J] + \beta \sum_{l=1}^\infty \lambda^l\, W_l[J] \label{W.loop}
\eea
The bookkeeping parameter $\lambda$ will be set to $1$ in the end. The exponent associated with the
parameter $\lambda$ plays the role of the number of loops introduced. For example, the term 
$\frac{1}{2} \Tr \ln \left( \tilde {\cal D}^{-1} \4di \, U\right)$
 consists of one-loop contributions. 
 The last part of (\ref{W.varphi.3}) contains diagrams of two loops or higher. 
The explicit appearance of the $n_J\4di \, U \4di n_J$ term in (\ref{W.varphi.2}) and (\ref{W.varphi.3}) 
suggests the possibility of a connection to the Kohn-Sham decomposition of the DFT.

\subsection{Inversion Method by Loop Order} 
\label{sec:inversion}
We now come to the point of departure from typical treatments of auxiliary field approach
 and are ready to make a direct connection to the Kohn-Sham scheme.   
Let us first define a new free fermion propagator
\be \label{G_KS}
\PH_0^{-1}(x,x') = \left[ \partial_\tau -\frac{\nabla^2}{2m} + \upsilon _{\rm ion}({\x}) -\mu + J_0(x) \right] \delta (x-x') \; ,
 \ee
where $J_0$ is chosen such that the free fermion system has the same particle density as the physical system
 (where Coulomb interactions exist) with source potential $J$
\be \label{KS_n_free}
-\PH_0(x,x) = n_J(x) \; .
\ee
The existence of $J_0$ is scrutinized below. From the perspective of the Kohn-Sham decomposition, Eq.~(\ref{KS_n_free})
 corresponds to Eq.~(\ref{KS_dsc}). In the presence of a source term $J$, Eq.~(\ref{H_J}) shows that it is
  equivalent to making $\upsilon_{\rm ion} \to \upsilon_{\rm ion} + J $. Comparing with Eq.~(\ref{KS_sptcle}), we see immediately
  if one were to choose 
\be   \label{KS_J0}
 J_0 = J + U\3di n_J + \left. \frac{\delta E_{xc}[n]}{\delta n}\right|_{n=n_J} \; ,
\ee
 the requirement (\ref{KS_n_free}) can be fulfilled. We will therefore call the corresponding free
  fermion system the Kohn-Sham system. The presence of $J_0$, of course, depends crucially
  on the differentiability of $E_{xc}[n]$, whose existence (not differentiability) was proven.\cite{HK_64} 
 
To bring out the Kohn-Sham quantities (orbitals and energies) in our loop expansion,  
let us first use a variant of (\ref{n2phiJ4d})
\[
u \4di (i\varphi) = J + U\4di n_J
\]
and replace $U \4di (i\varphi)$ by $J + U\4di n_J$ in the expression of propagator $G_{\varphi}$. 
Specifically, we write (\ref{Green_phi}) as 
\be  
G_{\varphi}^{-1}(x,x') 
= \left[ \partial_\tau -\frac{\nabla^2}{2m} + \upsilon _{\rm ion}({\x}) -\mu + J(x)
 + U \4di n_J \right] \delta (x-x') \label{Green_phi_n} \; . 
\ee

The critical step is to decompose the source $J$ in a particular way,
\be \label{source.decomp}
J \equiv (J_0 - U\4di n_J) + J' \equiv \tilde J_0 + J' \; . 
\ee
Therefore, from Eqs.~(\ref{n2phiJ4d}), (\ref{Green_phi}) and (\ref{source.decomp}) we have 
\be 
G_\varphi^{-1}(x,x') = \PH_0^{-1}(x,x')+J'(x)\, \delta(x-x') \; ,
\ee
and the Kohn-Sham propagator $\PH_0$ appears as we expected. 
 As will be described in section~\ref{sec:KS_systematics}, $\PH_0(x,x')$ can be expressed 
  in terms of the Kohn-Sham quantities. Therefore, the idea is to expand 
  $G_\varphi$ around $\PH_0$ provided that $J'$ can also be expressed via 
   Kohn-Sham quantities. Comparing (\ref{source.decomp}) with (\ref{KS_J0}), we find that
 formally speaking $J'[n] = -\frac{\delta E_{xc}[n]}{\delta n}$.    

This source decomposition is introduced here for the first time in the auxiliary field approach. 
(A similar decomposition has been used~\cite{FKYetal_95,VF_97_arxiv} 
  in the perturbative expansion in powers of $e^2$.) 
   
The idea of inversion is to obtain $J[n]$, that is, to find the corresponding $J$ for each 
configuration of electron density $n$ in the domain of $\Gamma[n]$. One then substitutes
 $J[n]$ into $\Gamma[n] = \beta W[J[n]] - J[n]\4di n$ to express $\Gamma[n]$ using the 
  density profile $n$ as the natural variable.     
For each given density profile $n$ within the domain of $\Gamma$, Eq.~(\ref{KS_n_free}) 
 determines the corresponding $\tilde J_0$. The collection of such relations forms $\tilde J_0[n]$. 
  Similarly, if for every given $n$ one can find the corresponding $J'$, one obtains
   $J'[n]$ and the goal of inversion is achieved. 
 When evaluating $\Gamma[n]$, one employs one density configuration at a time.
  That said, when we expand $W[J[n]] = W[\tilde J_0[n] + J'[n]]$ in powers of $J'$  
 within the expression $\Gamma[n] = \beta W[J] - J\4di n$,  
we will keep $n_J$ fixed, instead of treating it as a functional of $\tilde J_0$. 

We now examine the loop expansion of $W[J]$ carefully. Eq.~(\ref{W.varphi.3}) tells us that
\be
\beta W[J] = \beta (\tilde W_0 - W_0) + \beta \sum_{l=0}^\infty \lambda^l \, W_l[J+ U\4di n_J]  \; ,
\ee    
where $\beta (\tilde W_0 - W_0) = -\frac{1}{2} n_J\4di \, U \4di n_J$ and 
$\beta W_0[J+ U\4di n_J] = -\Tr \ln (G_\varphi^{-1})$. Note that for any $W_{l}$ term, 
its $J$ dependence is through the propagator $G_\varphi$, which always has 
$J + U\4di n_J$ as the natural variable. Our reasoning earlier indicates that when 
 we expand $W_l[J+ U\4di n_J] = W_l[\tilde J_0 + U\4di n_J  + J'] = W_l[J_0 + J']$ 
 in powers of $J'$  within the expression $\Gamma[n] = \beta W[J] - J\4di n$,   
we may write the expansion in the following way (and forget about needing to keep $n_J$ fixed) 
\be \label{W_l.exp}
W_l[J_0+J'] = W_l[J_0] + \frac{\delta W_l[J_0]}{\delta J_0} \4di J' 
+ \frac{1}{2!} J' \4di \frac{\delta^2 W[J_0]}{\delta J_0 \, \delta J_0} \4di J'
+ \ldots  \;\; .
\ee 
 
With (\ref{W_l.exp}), we may express $\beta W[J]$ as a double series
\be \label{W.db} 
 \beta W[J] = \beta (\tilde W_{00}-W_{00}) + \beta \sum_{i,k} W_{ik}{J'}^{k} \lambda^i\; ,
\ee
where each $W_{ik}$ involves the $k$'th derivative of $W_i$. In particular, $\tilde W_{00}$ is given
by (with $n_J \to n$ hereafter) 
\be \label{W.00}
\beta \tilde W_{00} = \beta W_{00} - \frac{1}{2} n \4di \, U \4di n  = 
 -\Tr \ln (\PH_0^{-1}) - \frac{1}{2} n \4di \, U \4di n \; ,
\ee
and  $W_{01}$ is given by 
\be \label{W0.dJ0}
\beta W_{01}[J_0] =  -\Tr \left( \frac{\delta \ln (\PH_0^{-1})}{\delta J_0(x)} \right)
 = - \!\!\int\!\! dz dy \,\PH_0(z,y) \delta (y-x) \delta (y-z) = -\PH_0(x,x)   =  n  \; ,
\ee
in view of (\ref{KS_n_free}).

Instead of looking at the expansion of $W_l$ in powers of $J'$ 
 within the effective action expression,  we now take a moment to
 look at the $\tilde W_{00}$ term in the double series expansion of $\beta W[J]$. 
Consider the functional derivative of $\beta \tilde W_{00}$ with respect to 
 the source term $\tilde J_0$. Here, one is asking the response of $\beta \tilde W_{00}$
  with respect to change in $\tilde J_0$. Evidently, when $\tilde J_0$ changes, its corresponding
  density $n$ has to vary as well.  
Using the chain rule of differentiation and (\ref{W0.dJ0}), we obtain 
\be \label{W_00.dtJ0}
\frac{\delta (\beta \tilde W_{00})} {\delta \tilde J_0} 
 =  \frac{\delta (\beta W_{00})}{\delta J_0}\4di \frac{\delta J_0}{\delta \tilde J_0}
    - n_ \4di \, U \4di \frac{\delta n}{\delta \tilde J_0}
     = n \4di ({\mathbf I}+  U \4di \frac{\delta n}{\delta \tilde J_0})
      - n_ \4di \, U \4di \frac{\delta n}{\delta \tilde J_0} = n \; .
\ee  
 This suggests that we define 
\be\label{tGamma.0.def}
\tilde \Gamma_0[n] = \beta \tilde W_{00}[\tilde J_0] - \tilde J_0 \4di n 
 = -\Tr \ln (\PH_0^{-1}) - \frac{1}{2} n \4di \, U \4di n - \tilde J_0 \4di n\; ,
\ee
 the Legendre transformation of $\beta W_{00}[\tilde J_0]$,   
leading to 
\be \label{tJ0.def}
\frac{\delta \tilde \Gamma_0[n]}{\delta n} = -\tilde J_0 \; .
\ee

Comparing (\ref{tJ0.def}) with (\ref{Gamma2J}), 
 we find \vspace*{-2pt}
 \be \label{Gamma.int.der}
\frac{\delta (\Gamma[n]- \tilde \Gamma_0[n])}{\delta n} = -J'  \;.
\vspace*{-2pt} \phantom{12}
 \ee
The idea now is to develop a series for $\Gamma[n]$ led by $\tilde \Gamma_0[n]$. 
 Subtracting (\ref{tGamma.0.def}) from $\Gamma[n] = \beta W[J] - J \4di n$, 
 we have \vspace*{-1pt}
\be \label{Gamma.xc.def}
\Gamma[n] - \tilde \Gamma_0[n] = \beta W[J] - \beta \tilde W_{00}[\tilde J_0] - J' \4di n \; ,
\vspace*{-1pt} \phantom{12}
\ee
in which the last two terms on the right-hand side (RHS) exactly cancel the terms in $\tilde W_{00}$ and
$W_{01}$ contributing to $\beta W[J]$. So the series for $\Gamma - \tilde \Gamma_0$ is just 
(\ref{W.db}) with those two terms removed. Next we convert the double sum in (\ref{W.db}) 
 into a single sum by expanding $J'$ as a series in $\lambda$. We write \vspace*{-2pt} 
 \be \label{Jp.def}
 J'[n] = \sum_{l=1}^\infty J_l [n] \lambda^l \; ,
 \vspace*{-2pt} \phantom{12}
 \ee
where the precise expressions for $J_1, J_2,\ldots$ are as yet undetermined since (\ref{Jp.def})
is not a loop expansion. We substitute (\ref{Jp.def}) formally into (\ref{Gamma.xc.def}) and (\ref{W.db}) 
to obtain a series \vspace*{-4pt} \phantom{12} 
\be \label{Gamma.loop}
\Gamma[n]-\tilde \Gamma_0[n] = \sum_{l=1}^\infty \Gamma_l[n] \; \lambda^l \; , 
\vspace*{-2pt} \phantom{12}
\ee
in which each $\Gamma_l$ is defined explicitly in terms of the $J_k$, $\beta W_{k \le l}[J_0]$,
and their derivatives.  
Because $W_{01}$ is missing from (\ref{Gamma.xc.def}), any occurrence of $J_k$ is accompanied by at
least one other factor $J_{k'}$ or else by an occurrence of some $W_{i>0}$, and hence by
a power of $\lambda$ higher than the $k$'th. In other words, the expression for $\Gamma_{l \ge 1}$ involves only $J_k$ with $ k < l$. We finally remove the indeterminacy in (\ref{Jp.def}) by imposing 
 (\ref{Gamma.int.der}) order by order in $\lambda$, leading to (for $l \ge 1$) 
 \be \label{J_l} 
 \frac{\delta \Gamma_l[n]}{\delta n} = - J_l \; .
 \ee

Since $\Gamma_{l\ge 1}$ involves only $J_{k < l}$,  all
the $J_l$ and $\Gamma_l$ can be found explicitly by applying (\ref{Gamma.loop}) and (\ref{J_l}) alternately. Evidently, it is the source decomposition (\ref{source.decomp})
 that allows us to obtain exact correspondence to the Kohn-Sham scheme. 
 Below we will provide an explicit formula for 
$\Gamma_l[n]$ in terms of $W_l[J_0]$ and their functional derivatives.

To obtain an explicit expression for $\Gamma_l[n]$, 
let us substitute the LHS of (\ref{Gamma.xc.def}) by the RHS of (\ref{Gamma.loop}) 
  and apply (\ref{W.db}) as well as (\ref{Jp.def}) to the RHS of (\ref{Gamma.xc.def}). Then, by equating
   the coefficients associated with $\lambda^l$ on both sides of (\ref{Gamma.xc.def}), 
 we obtain (for $l \ge 1$) 
\bea
\Gamma _l\left[ n \right] &=& \beta W_l\left[ J_0\right]  
 +\sum_{k=1}^{l-1}\frac{\delta (\beta W_{l-k}\left[ J_0\right]) }{\delta J_0}\4di J_k
 \nonumber  \\
&& +\sum_{m=2}^l\frac 1{m!}\sum_{k_1,\ldots ,k_m\geq 1}^{k_1+\ldots +k_m\leq l}
\frac{\delta ^m(\beta W_{l-\left( k_1+\ldots +k_m\right) }\left[ J_0\right] )}{%
\delta J_0  \ldots \delta J_0 }\4di J_{k_1} \4di \cdots \4di J_{k_m} \label{Gamma_order_l.new}\; .
\eea

For $l=1$, we see that 
\be \label{Gn.1}
\Gamma_1[n] = \beta W_1[J_0] = \frac{1}{2}  
\Tr \ln \left( \tilde {\cal D}_{\!\! J \to \tilde J_0}^{-1} \4di \, U \right)  \equiv 
\Tr \ln \left( \tilde {\cal D}_{0}^{-1} \4di \, U \right) \; . 
\ee
We observe that 
\[
\tilde {\cal D}_{0}^{-1} \equiv \tilde {\cal D}_{\!\! J \to \tilde J_0}^{-1} = U^{-1} - D_{\!\! J \to \tilde J_0}  
= U^{-1} - D_0 
\]

We then have 
\[
J_1 = -\frac{\delta \Gamma_1[n]}{\delta n} = - \frac{\delta J_0}{\delta n}  \4di 
\frac{\delta (\beta W_1[J_0])}{\delta J_0}
 \; .
\]
Note that $-\delta J_0/\delta n$ can be written as
\be \label{idc.exp} 
-\frac{\delta J_0(x)}{\delta n(y)} = -\left( \frac{\delta n(y)}{\delta J_0(x)} \right)^{-1} 
= -\left( \frac{\delta^2 (\beta W_0[J_0])}{\delta J_0(x) \delta J_0(y)} \right)^{-1}
 = \frac{\delta^2 \Gamma_0[n]}{\delta n(x) \delta n(y)} 
\ee
where 
\be \label{Gamma.0.def}
\Gamma_0[n] = \beta W_0[J_0] - J_0 \4di n = \tilde \Gamma_0[n] - \frac{1}{2} n\4di \, U \4di n  \; ,
\ee
and the inverse is in the functional matrix sense.
Using (\ref{KS_n_free}), we can evaluate $\delta n(x)/\delta J_0(y)$ by
\bea 
\frac{\delta n(x)}{\delta J_0(y)}& =& - \frac{\delta \PH_0(x,x)}{\delta J_0(y)}
= \int dz dz' \PH_0(x,z) \frac{\delta \PH_0^{-1}(z,z')}{\delta J_0(y)} \PH_0(z',x) \nonumber \\
&=& \PH_0(x,y) \PH_0(y,x) = D_{J \to \tilde J_0}(x,y) \equiv D_0(x,y) \label{dndJ.tdep} \; .
\eea
 Therefore,
\[
J_1 =  - D_0^{-1} \; \4di \frac{\delta (\beta W_1[J_0])}{\delta J_0}\; .
\]
Since $-\PH_0(x,x) = n(x)$ represents the electron density of the KS system,
 we will call $D_0(x,y)$ the polarization associated with the KS system.
 Note that if one were to approximate the effective action $\Gamma [n]$ by 
  $\Gamma [n] = \tilde \Gamma_0 [n] + \Gamma_1[n]$, the displayed equation 
   above is the OEP equation for GW-OEP~\cite{HvB_07}, while eq.~(\ref{Gn.1}) 
   is the corresponding exchange-correlation functional. 

Once $J_1$ is known, one can find $\Gamma _2\left[ n\right] $
\bea
\Gamma _2\left[ n \right] &=& \beta W_2\left[ J_0\right] +\frac{\delta \left( \beta W_1\left[
J_0\right] \right)}{\delta J_0 }\4di J_1
 +\frac {1}{2} J_1 \4di \frac{\delta ^2\left( \beta W_0\left[ J_0\right] \right) }{\delta
J_0 \delta J_0 } \4di J_1  \nonumber \\
&=& \beta W_2\left[ J_0\right] - \frac{1}{2} \frac{\delta \left( \beta W_1\left[
J_0\right] \right)}{\delta J_0 }\4di D_0^{-1} \4di \frac{\delta \left( \beta W_1\left[
J_0\right] \right)}{\delta J_0 } \; ,\label{G2.n}
\eea
and $J_2$ now can be computed via $ -\delta \Gamma_2[n]/\delta n$
\be \label{J2}
J_2 = D_0^{-1} \4di \frac{\delta (\beta W_2 [J_0])}{\delta J_0} + D_0^{-1} \4di 
 \frac{\delta^2 (\beta W_1[J_0])}{\delta J_0 \delta J_0} \4di J_1 
 + \frac 12 D_0^{-1} \4di \frac{ \delta^3 \left( \beta W_0[J_0] \right)}{
 \delta J_0 \delta J_0 \delta J_0 } \4di J_1 \4di J_1 \; .
\ee
The explicit expression of $J_2$ leads to $\Gamma_3[n]$ and so on. It should now be clear how the strategy goes.
 We are assuming that the functionals $\{ W_l[J_0] \}$ and their derivatives with respect to $J_0$ are known.
 This information can indeed be obtained by standard, albeit tedious, many-body perturbation method. 
 One then uses Eq.~(\ref{Gamma_order_l.new}) to express $\Gamma_l$ in terms of the known functionals and $J_k$ with $k \le l-1$.
 Once $\Gamma_l$ is obtained, one can then use Eq.~(\ref{J_l}) to obtain $J_l$, which then facilitates the calculation of $\Gamma_{l+1}$
 via (\ref{Gamma_order_l.new}) and so on.

We now take a moment to organize the terms of effective action $\Gamma[n]$. 
 From (\ref{Gamma.loop}) and (\ref{Gamma.0.def}), we know that 
\be \label{Gamma.id}
\Gamma[n] = \tilde \Gamma_0[n] + \sum_{l=1}^\infty \Gamma_l [n] = \Gamma_0[n] 
+ \frac{1}{2} n \4di \, U \4di n + \sum_{l=1}^\infty \Gamma_l [n] \; .
\ee 
The first term on the RHS of (\ref{Gamma.id}) indeed corresponds to the effective action of the free KS system, the second 
 term is exactly the Hartree energy.  It is thus natural for us to define the last part, sum of $\Gamma_l[n]$, as
\be \label{Gamma.xc.def.1}
\Gamma_{xc}[n] \equiv \sum_{l=1}^\infty \Gamma_l[n] \; . 
\ee

At the physical condition (i.e., when the source is absent), we should have $\delta \Gamma/\delta n = - J = 0$.  
 Knowing that 
\[
\frac {\delta  \tilde \Gamma_0[n]}{\delta n} = - \tilde J_0  = -J_0 + U\4di n\; ,
\]
we conclude that at the physical condition, 
\be \label{Gn.equil}
\frac{\delta \Gamma_{xc}[n] }{\delta n} = \tilde J_0 = J_0 - U\4di n \; .
\ee
Note that the potential $J_0$ 
  together with the original non-interacting part of the Hamiltonian leads to the exact particle 
 density of the interacting system. This implies that $J_0$ contains both the Hartree term and the exchange-correlation
 potential. Since $J_0 = \tilde J_0 + U\4di n_T$, with $U\4di n_T$ being the Hartree term, 
 the term $\tilde J_0$ plays the role of exchange-correlation potential, as evidenced by (\ref{Gn.equil}). 
  The quantum mechanical effects are completely contained in $\Gamma_{xc}[n]$, defined in (\ref{Gamma.xc.def.1}).

Note that a different density profile other than that corresponding to the physical ground state 
can be induced by introducing a nonzero $J$. Let us again call the corresponding density
 profile $n_J$. When $J\ne 0$, we learn from (\ref{H_J}) that it is equivalent to replacing $\upsilon$ by $\upsilon + J$.
  Eq.~(\ref{KS_sptcle}) then tells us that $J_0$ must contain  $J$, the Hartree term and the exchange-correlational potential
 as shown in (\ref{KS_J0}). 
 In this case, one writes $J_0 = J + (\tilde J_0 - J) +  U \4di n_J$.  
 With $U\4di n_J$ being the Hartree term, $(\tilde J_0 -J)$ must represent 
 the exchange-correlation potential corresponding to the configuration $n_J$, and 
\be \label{ex.general.1} 
\frac{\delta \left( \Gamma_{xc}[n] \right)}{\delta n} = \tilde J_0  - J \;. 
\ee  
This then leads to 
\[
\frac{\delta \Gamma[n]}{\delta n} = -J \; ,
\]
what we expected when $J \ne 0$.

In terms of real computation, since the $n$ dependence is through $J_0$, we may rewrite Eq.~(\ref{Gn.equil})
 as 
\[
 \frac{\delta J_0}{\delta n} \4di \frac{\delta \left(  \Gamma_{xc}\left [J_0[n] \right] \right)}{\delta J_0} 
 = \tilde J_0 =  J_0 - U\4di n
\]
or 
\be \label{OEP}
0 = D_0\, \4di \, \tilde J_0   -
 \frac{\delta \left( \Gamma_{xc} \left[ J_0[n] \right] \right)}{\delta J_0} 
 = D_0\, \4di \, \left(  J_0  - U\4di n \right)  -
 \frac{\delta \left( \Gamma_{xc} \left[ J_0[n] \right] \right)}{\delta J_0} \; ,
\ee
and the condition $\delta \Gamma [n]/\delta n = 0$ is turned into $\delta \Gamma [J_0[n]]/\delta J_0 = 0$
Since the effective action is a strictly convex function of the electron density $n$, there exists no local
 minima. One can therefore solve $\delta \Gamma [J_0[n]]/\delta n= 0$ by steepest descent. Effectively, we may define the direction of steepest descent $\kappa(x)$ by 
 \be \label{steepest_descent}
\kappa (x) \equiv  - \frac{\delta \Gamma \left[ J_0[n] \right]}{\delta J_0(x)} = 
D_0\, \4di \, \left( J_0 - U\4di n \right) -
 \frac{\delta \left( \Gamma_{xc} \left[ J_0[n] \right] \right)}{\delta J_0}
 \ee 
 and then update $J_0(x)$ by $J_0(x) \to J_0(x) + \varsigma \kappa(x)$, with $\varsigma > 0$ being the step size, till convergence is reached,
  that is, when $\kappa(x) \to 0$.  
Note that  eq.~(78) is the standard OEP-equation for consistency. The iterative procedure 
described after (78) is largely identical to the Kuemmel-Perdew~\cite{KP_03} procedure used for 
solving the exchange-only OEP equation.  

\subsection{Diagrammatic Expansion of the Density Functional}
We now examine how one 
 computes the effective action via diagrams. 
 From Eqs.~(\ref{Gamma.loop}),  (\ref{Gn.1}) and (\ref{Gamma.0.def}),  we have
\bea
\Gamma [n] &=& \tilde \Gamma_0 [n] + \sum_{i=1}^\infty \Gamma_i [n]
 = \Gamma_0[n] + \frac{1}{2} n \4di \, U \4di n + \sum_{i=1}^\infty \Gamma_i [n] \nonumber \\
& = &  - \Tr \ln \left( \PH_0^{-1} \right)  - J_0 \4di n 
 + \frac{1}{2} n \4di \, U \4di n  + \frac{1}{2} \Tr \ln \left( \tilde {\cal D}_0^{-1} \4di \, U \right) 
  + \sum_{i=2}^\infty \Gamma_i[n] \label{Gn.1PI} 
\;,
\eea
where 
\bea
\tilde {\cal D}_0^{-1} &=& U^{-1} -  D_0 \label{tDI} \\
D_0 (x,y)  &=&  \PH_0(x,y) \PH_0(y,x) \label{D0}  \; .
\eea
Evidently, we need diagrammatic symbols for $U$, $\PH_0$ and $\tilde {\mathcal D}_0$.
To get to higher-order terms $\Gamma_{i\ge 2}$ of the effective action, we see from Eqs.~(\ref{dndJ.tdep}-\ref{J2}) 
 and the text afterwards that it is necessary to incorporate into the diagrams 
 the inverse density correlator $D_0^{-1}$ and to evaluate functional 
 derivatives with respect to $n$ (or $J_0$). We define the symbols for each line type below
\[
\begin{picture}(255,40)(0,0)
\Line[dash,dashsize=2.5](10,30)(60,30) \Vertex(10,30){1.0} \Vertex(60,30){1.0}
\Line[double](150,30)(200,30) \Vertex(150,30){1.0} \Vertex(200,30){1.0}
\Line[arrow,arrowlength=3.0,arrowwidth=0.8,arrowinset=0.2](10,10)(60,10) 
\Vertex(10,10){1.0} \Vertex(60,10){1.0}
\Photon[](150,10)(200,10){2}{9} \Vertex(150.5,10){1.8} \Vertex(199.5,10){1.8} 
\Text(80,30)[lc]{$U(x,x')$}
\Text(-2,28)[lb]{$x'$} \Text(72,28)[rb]{$x$}
\Text(220,30)[lc]{$D_0^{-1}(x,x')$}
\Text(138,28)[lb]{$x'$} \Text(212,28)[rb]{$x$}
\Text(80,10)[lc]{${\mathcal G}_0(x,x')$}
\Text(-2,8)[lb]{$x'$} \Text(72,8)[rb]{$x$}
\Text(138,8)[lb]{$x'$} \Text(212,8)[rb]{$x$}
\Text(222,10)[lc]{$\tilde {\mathcal D}_0 (x,x')\hspace*{10pt}$ .}
\end{picture} 
\]
The smaller dots associated with the $U$, $D_0^{-1}$, and $\PH_0$ propagators are introduced to 
guide the eyes regarding the starting and ending points of these propagators. 
The $\tilde {\mathcal D}_0$ propagator comes from contracting two $b$ fields, see (\ref{b.def}),
 and thus the bigger dots associated with $\tilde {\mathcal D}_0$ denote the coordinates/locations 
 of those $b$ fields. We will also use the bigger dots to indicate 
  the space-time coordinate of a point of interest. 

 To evaluate functional derivatives of  $W_l[J_0[n]]$ with respect to $n$ (or $J_0$), we note    
 from Eqs.~ (\ref{n2phiJ4d}), (\ref{Green_phi}), (\ref{I_kernel}) and (\ref{W.varphi.3}) 
  that the $ J_0$ dependence comes from $\PH_0(x,y)$ and the functional derivatives 
   associated with the formalism using Eqs.~(\ref{Gamma_order_l.new}-\ref{steepest_descent}) 
 necessarily require evaluations of $\delta \PH_0(x,x')/\delta J_0(y)$. 
 Using a derivation similar to that in (\ref{dndJ.tdep}), we find that
\be \label{dgdJ.tdep} 
\frac{\delta \PH_0(x,x')}{\delta  J_0 (y)} =  - \PH_0(x,y) \, \PH_0(y,x')   \; .
\ee
The propagators $D_0^{-1}$ and $\tilde {\mathcal D}_0$ also contain $\PH_0$ and thus may be differentiated
 with respect to $J_0$.  Note that $\tilde {\mathcal D}_0 = ({\mathbf I}- U\4di D_0)^{-1}\4di \, U$ 
  and $D_0(x,y) = \PH_0(x,y)\PH_0(y,x)$. Employing the identity
\[
\frac{\delta M^{-1}}{\delta J_0} = -M^{-1} \4di \frac{\delta M}{\delta J_0}\4di M^{-1} \;,
\]   
we let $M = ({\mathbf I}- U\4di D_0)$ for the case of $\tilde {\mathcal D}_0$
 and $M = D_0 $ for the case of $D_0^{-1}$ to obtain 
\bea
\frac{\delta \tilde {\mathcal D}_0}{\delta J_0} &=&  \tilde {\mathcal D}_0 \4di \frac{\delta D_0}
{\delta J_0}\4di \tilde {\mathcal D}_0 \; , \label{dtsDdJ.tdep} \\
{\rm and~~~~}\frac{\delta D_0^{-1}}{\delta J_0} &=& - D_0^{-1} \4di \frac{\delta D_0}{\delta J_0 } \4di D_0^{-1} 
 \; .\label{dDdJ.tdep}
\eea 

Eq.~(\ref{dgdJ.tdep}-\ref{dDdJ.tdep}) may be expressed diagrammatically as follows
\bea 
\frac{\delta \PH_0(x,x') }{\delta J_0(y)}
 = \frac{\delta }{\delta J_0(y)} \; 
\begin{picture}(10,40)(0,-3)
\Line[arrow,arrowlength=3.0,arrowwidth=0.8,arrowinset=0.2](5,-30)(5,30)
\Vertex(5,-30){1.0}
\Vertex(5,30){1.0}
\Text(5,-34)[tc]{$x'$}
\Text(5,34)[bc]{$x$}
\end{picture}
&=& -\; \begin{picture}(10,40)(0,-3)
\Line[arrow,arrowlength=3.0,arrowwidth=0.8,arrowinset=0.2](5,-30)(5,0)
\Line[arrow,arrowlength=3.0,arrowwidth=0.8,arrowinset=0.2](5,0)(5,30)
\Vertex(5,0){1.8}
\Vertex(5,-30){1.0}
\Vertex(5,30){1.0}
\Text(5,-34)[tc]{$x'$}
\Text(5,34)[bc]{$x$}
\Text(12,-4)[bc]{$y$}
\end{picture} \; , \nonumber \\
\frac{\delta \tilde{\mathcal D}_0(x,x') }{\delta J_0(y)} = \frac{\delta }{\delta J_0(y)} \; 
\begin{picture}(10,80)(0,-3)
\Photon[](5,-30)(5,30){2}{10} 
\Vertex(5,-30){1.8}
\Vertex(5,30){1.8}
\Text(5,-34)[tc]{$x'$}
\Text(5,34)[bc]{$x$}
\end{picture}
&=&  - \; \begin{picture}(30,40)(0,-3)
\Photon[](15,-30)(15,-10){2}{3}
\Photon[](15,10)(15,30){2}{3}
\Arc[arrow,arrowpos=0.50,arrowlength=3.0,arrowwidth=0.8,arrowinset=0.2,clock](15,0)(10,-90,-180)
\Arc[arrow,arrowpos=0.50,arrowlength=3.0,arrowwidth=0.8,arrowinset=0.2,clock](15,0)(10,-180,-270)
\Arc[arrow,arrowpos=0.50,arrowlength=3.0,arrowwidth=0.8,arrowinset=0.2,clock](15,0)(10,90,-90)
\Vertex(5,0){1.8}
\Vertex(15,-30){1.8}
\Vertex(15,30){1.8}
\Vertex(15,-10){1.8}
\Vertex(15,10){1.8}
\Text(15,-34)[tc]{$x'$}
\Text(15,34)[bc]{$x$}
\Text(11,-4)[bc]{$y$}
\end{picture}
- \; \begin{picture}(30,40)(0,-3)
\Photon[](15,-30)(15,-10){2}{3}
\Photon[](15,10)(15,30){2}{3}
\Arc[arrow,arrowpos=0.50,arrowlength=3.0,arrowwidth=0.8,arrowinset=0.2,clock,flip](15,0)(10,-90,-180)
\Arc[arrow,arrowpos=0.50,arrowlength=3.0,arrowwidth=0.8,arrowinset=0.2,clock,flip](15,0)(10,-180,-270)
\Arc[arrow,arrowpos=0.50,arrowlength=3.0,arrowwidth=0.8,arrowinset=0.2,clock,flip](15,0)(10,90,-90)
\Vertex(5,0){1.8}
\Vertex(15,-30){1.8}
\Vertex(15,30){1.8}
\Vertex(15,-10){1.8}
\Vertex(15,10){1.8}
\Text(15,-34)[tc]{$x'$}
\Text(15,34)[bc]{$x$}
\Text(11,-4)[bc]{$y$}
\end{picture} \; , \nonumber \\
\frac{\delta D_0^{-1}(x,x') }{\delta J_0(y)} 
= \frac{\delta }{\delta J_0(y)} \; 
\begin{picture}(10,80)(0,-3)
\Line[double](5,-30)(5,30)
\Vertex(5,-29.5){1.0}
\Vertex(5,29.5){1.0}
\Text(5,-34)[tc]{$x'$}
\Text(5,34)[bc]{$x$}
\end{picture}
&=&  + \; \begin{picture}(30,40)(0,-3)
\Line[double](15,-30)(15,-10) \Vertex(15,-29.5){1.0}
\Line[double](15,10)(15,30) \Vertex(15,29.5){1.0}
\Arc[arrow,arrowpos=0.50,arrowlength=3.0,arrowwidth=0.8,arrowinset=0.2,clock](15,0)(10,-90,-180)
\Arc[arrow,arrowpos=0.50,arrowlength=3.0,arrowwidth=0.8,arrowinset=0.2,clock](15,0)(10,-180,-270)
\Arc[arrow,arrowpos=0.50,arrowlength=3.0,arrowwidth=0.8,arrowinset=0.2,clock](15,0)(10,90,-90)
\Vertex(5,0){1.8}
\Vertex(15,10){1.5}\Vertex(15,-10){1.0}
\Text(15,-34)[tc]{$x'$}
\Text(15,34)[bc]{$x$}
\Text(11,-4)[bc]{$y$}
\end{picture}
+ \;  \begin{picture}(30,40)(0,-3)
\Line[double](15,-30)(15,-10) \Vertex(15,-29.5){1.0}
\Line[double](15,10)(15,30) \Vertex(15,29.5){1.0}
\Arc[arrow,arrowpos=0.50,arrowlength=3.0,arrowwidth=0.8,arrowinset=0.2,clock,flip](15,0)(10,-90,-180)
\Arc[arrow,arrowpos=0.50,arrowlength=3.0,arrowwidth=0.8,arrowinset=0.2,clock,flip](15,0)(10,-180,-270)
\Arc[arrow,arrowpos=0.50,arrowlength=3.0,arrowwidth=0.8,arrowinset=0.2,clock,flip](15,0)(10,90,-90)
\Vertex(5,0){1.8}
\Vertex(15,10){1.0}\Vertex(15,-10){1.0}
\Text(15,-34)[tc]{$x'$}
\Text(15,34)[bc]{$x$}
\Text(11,-4)[bc]{$y$}
\end{picture} \; ,\nonumber  \\
&& \nonumber
\eea
\phantom{12}\vspace*{20pt}
where the $\mp$ signs come from 
\[
\frac{\delta}{\delta J_0(y)} D_0(z,z') 
= \frac{\delta}{\delta J_0(y)} \begin{picture}(30,20)(0,-3)
\Arc[arrow,arrowpos=0.50,arrowlength=3.0,arrowwidth=0.8,arrowinset=0.2,clock](15,0)(10,90,-90)
\Arc[arrow,arrowpos=0.50,arrowlength=3.0,arrowwidth=0.8,arrowinset=0.2,clock](15,0)(10,-90,90)
\Vertex(15,-10){1.0}
\Vertex(15,10){1.0}
\Text(15,-14)[tc]{$z'$}
\Text(15,14)[bc]{$z$}
\end{picture}
= - \; \begin{picture}(30,20)(0,-3)
\Arc[arrow,arrowpos=0.50,arrowlength=3.0,arrowwidth=0.8,arrowinset=0.2,clock](15,0)(10,90,-90)
\Arc[arrow,arrowpos=0.50,arrowlength=3.0,arrowwidth=0.8,arrowinset=0.2,clock](15,0)(10,-90,0)
\Arc[arrow,arrowpos=0.50,arrowlength=3.0,arrowwidth=0.8,arrowinset=0.2,clock](15,0)(10,0,90)
\Vertex(5,0){1.8}
\Vertex(15,-10){1.0}
\Vertex(15,10){1.0}
\Text(15,-14)[tc]{$z'$}
\Text(15,14)[bc]{$z$}
\Text(11,-4)[bc]{$y$}
\end{picture}
- \; \begin{picture}(30,20)(0,-3)
\Arc[arrow,arrowpos=0.50,arrowlength=3.0,arrowwidth=0.8,arrowinset=0.2,clock,flip](15,0)(10,90,-90)
\Arc[arrow,arrowpos=0.50,arrowlength=3.0,arrowwidth=0.8,arrowinset=0.2,clock,flip](15,0)(10,-90,0)
\Arc[arrow,arrowpos=0.50,arrowlength=3.0,arrowwidth=0.8,arrowinset=0.2,clock,flip](15,0)(10,0,90)
\Vertex(5,0){1.8}
\Vertex(15,-10){1.0}
\Vertex(15,10){1.0}
\Text(15,-14)[tc]{$z'$}
\Text(15,14)[bc]{$z$}
\Text(11,-4)[bc]{$y$}
\end{picture} \; . 
\]
\phantom{12345}\vspace*{2pt}\\
When combined with the inverse density correlator, the graphs above yield
\bea
\frac{\delta \PH_0(x,x')}{\delta n(z)} &=& - \int dy \, D_0^{-1}(z,y)\,  \PH_0(x,y) \, \PH_0(y,x') \; ,
\nonumber \\
\frac{\delta \tilde {\mathcal D}_0(x,x')}{\delta n(z)} &=& - \int dy dx_1 dx_2 
D_0^{-1}(z,y)\,  \tilde {\mathcal D}_0(x,x_1) \left[ \PH_0(x_2,x_1) \PH_0(x_1,y) \PH_0(y,x_2) + \right. 
\nonumber \\
&& \hspace*{20pt} \left. + \PH_0(x_1,x_2) \PH_0(x_2,y) \PH_0(y,x_1) \right] 
\tilde {\mathcal D}_0(x_2,x') \; , \nonumber \\
\frac{\delta D_0^{-1}(x,x')}{\delta n(z)} &=&  \int dy dx_1 dx_2 
D_0^{-1}(z,y)\,  D_0^{-1}(x,x_1) \left[ \PH_0(x_2,x_1) \PH_0(x_1,y) \PH_0(y,x_2) + \right. 
\nonumber \\
&& \hspace*{20pt} \left. + \PH_0(x_1,x_2) \PH_0(x_2,y) \PH_0(y,x_1) \right] 
 D_0^{-1}(x_2,x') \; , \nonumber 
\eea
which are shown diagrammatically below
\bea 
\frac{\delta \PH_0(x,x') }{\delta n(z)}
 = \frac{\delta }{\delta n(z)} \; 
\begin{picture}(10,40)(0,-3)
\Line[arrow,arrowlength=3.0,arrowwidth=0.8,arrowinset=0.2](5,-30)(5,30)
\Vertex(5,-30){1.0}
\Vertex(5,30){1.0}
\Text(5,-34)[tc]{$x'$}
\Text(5,34)[bc]{$x$}
\end{picture}
&=& -\;   \begin{picture}(30,40)(-20,-3)
\Line[arrow,arrowlength=3.0,arrowwidth=0.8,arrowinset=0.2](5,-30)(5,0)
\Line[arrow,arrowlength=3.0,arrowwidth=0.8,arrowinset=0.2](5,0)(5,30)
\Line[double](-15,0)(5,0)
\Vertex(-15,0){1.8}
\Vertex(5,0){1.0}
\Vertex(5,-30){1.0}
\Vertex(5,30){1.0}
\Text(5,-34)[tc]{$x'$}
\Text(5,34)[bc]{$x$}
\Text(-14,-9)[bc]{$z$}
\end{picture} \; , \nonumber \\
\frac{\delta \tilde{\mathcal D}_0(x,x') }{\delta n(z)} = \frac{\delta }{\delta n(z)} \; 
\begin{picture}(10,80)(0,-3)
\Photon[](5,-30)(5,30){2}{10}
\Vertex(5,-30){1.8}
\Vertex(5,30){1.8}
\Text(5,-34)[tc]{$x'$}
\Text(5,34)[bc]{$x$}
\end{picture}
&=&  - \;  \begin{picture}(50,40)(-20,-3)
\Photon[](15,-30)(15,-10){2}{3}
\Photon[](15,10)(15,30){2}{3}
\Arc[arrow,arrowpos=0.50,arrowlength=3.0,arrowwidth=0.8,arrowinset=0.2,clock](15,0)(10,-90,-180)
\Arc[arrow,arrowpos=0.50,arrowlength=3.0,arrowwidth=0.8,arrowinset=0.2,clock](15,0)(10,-180,-270)
\Arc[arrow,arrowpos=0.50,arrowlength=3.0,arrowwidth=0.8,arrowinset=0.2,clock](15,0)(10,90,-90)
\Line[double](-15,0)(5,0)
\Vertex(5,0){1.0} \Vertex(-15,0){1.8}
\Vertex(15,-30){1.8}
\Vertex(15,30){1.8} 
\Vertex(15,-10){1.8}
\Vertex(15,10){1.8}
\Text(15,-34)[tc]{$x'$}
\Text(15,34)[bc]{$x$}
\Text(-14,-9)[bc]{$z$}
\end{picture}
- \;  \begin{picture}(50,40)(-20,-3)
\Photon[](15,-30)(15,-10){2}{3}
\Photon[](15,10)(15,30){2}{3}
\Arc[arrow,arrowpos=0.50,arrowlength=3.0,arrowwidth=0.8,arrowinset=0.2,clock,flip](15,0)(10,-90,-180)
\Arc[arrow,arrowpos=0.50,arrowlength=3.0,arrowwidth=0.8,arrowinset=0.2,clock,flip](15,0)(10,-180,-270)
\Arc[arrow,arrowpos=0.50,arrowlength=3.0,arrowwidth=0.8,arrowinset=0.2,clock,flip](15,0)(10,90,-90)
\Line[double](-15,0)(5,0)
\Vertex(5,0){1.0} \Vertex(-15,0){1.8}
\Vertex(15,-30){1.8}
\Vertex(15,30){1.8}
\Vertex(15,-10){1.8}
\Vertex(15,10){1.8} 
\Text(15,-34)[tc]{$x'$}
\Text(15,34)[bc]{$x$}
\Text(-14,-9)[bc]{$z$}
\end{picture} \; , \nonumber \\
\frac{\delta D_0^{-1}(x,x') }{\delta n(z)} 
= \frac{\delta }{\delta n(z)} \; 
\begin{picture}(10,80)(0,-3)
\Line[double](5,-30)(5,30)
\Vertex(5,-29.5){1.0}
\Vertex(5,29.5){1.0}
\Text(5,-34)[tc]{$x'$}
\Text(5,34)[bc]{$x$}
\end{picture}
&=&  + \;  \begin{picture}(50,40)(-20,-3)
\Line[double](15,-30)(15,-10) \Vertex(15,-30){1.0}
\Line[double](15,10)(15,30) \Vertex(15,30){1.0}
\Arc[arrow,arrowpos=0.50,arrowlength=3.0,arrowwidth=0.8,arrowinset=0.2,clock](15,0)(10,-90,-180)
\Arc[arrow,arrowpos=0.50,arrowlength=3.0,arrowwidth=0.8,arrowinset=0.2,clock](15,0)(10,-180,-270)
\Arc[arrow,arrowpos=0.50,arrowlength=3.0,arrowwidth=0.8,arrowinset=0.2,clock](15,0)(10,90,-90)
\Line[double](-15,0)(5,0)
\Vertex(5,0){1.0} \Vertex(-15,0){1.8}  
\Vertex(15,10){1.0}\Vertex(15,-10){1.0}
\Text(15,-34)[tc]{$x'$}
\Text(15,34)[bc]{$x$}
\Text(-14,-9)[bc]{$z$}
\end{picture}
+ \; \begin{picture}(50,40)(-20,-3)
\Line[double](15,-30)(15,-10) \Vertex(15,-30){1.0}
\Line[double](15,10)(15,30) \Vertex(15,30){1.0}
\Arc[arrow,arrowpos=0.50,arrowlength=3.0,arrowwidth=0.8,arrowinset=0.2,clock,flip](15,0)(10,-90,-180)
\Arc[arrow,arrowpos=0.50,arrowlength=3.0,arrowwidth=0.8,arrowinset=0.2,clock,flip](15,0)(10,-180,-270)
\Arc[arrow,arrowpos=0.50,arrowlength=3.0,arrowwidth=0.8,arrowinset=0.2,clock,flip](15,0)(10,90,-90)
\Line[double](-15,0)(5,0)
\Vertex(5,0){1.0} \Vertex(-15,0){1.8} 
\Vertex(15,10){1.0}\Vertex(15,-10){1.0}
\Text(15,-34)[tc]{$x'$}
\Text(15,34)[bc]{$x$}
\Text(-14,-9)[bc]{$z$}
\end{picture} \; . \nonumber  \\
&& \nonumber
\eea

The diagrammatic differential rules of $\delta/\delta J_0$ and $\delta/\delta n$ are needed
not only for the calculations of higher-order terms $\Gamma_{i\ge 2}$ of the effective action 
 but also for the calculations of excitations, which we will discuss in section~\ref{sec:excitation}.

Before formally introducing the diagrammatic expansion, let us first set the convention that we will use.
 In general, each Feynman graph will carry with it a {\it symmetry} factor, which is inversely 
  proportional to the number of ways to label this graph without changing the topology of the graph.  
 The convention in quantum field theory usually leaves out the symmetry factor, as it may be deduced from
  the graph. To avoid enumeration of the symmetry factors needed, however, 
  we will explicitly provide the symmetry factors for Feynman diagrams to be investigated 
   later.

Let us now look at the effective action (\ref{Gn.1PI}) term by term. 
The Hartree term $n\4di \, U\4di n/2$ is expressed diagrammatically below
\[
\Gamma_{\rm Hartree} = \frac{1}{2} \; \; 
\begin{picture}(70,28)(0,11)
\Arc[arrow,arrowpos=0.50,arrowlength=3.0,arrowwidth=0.8,arrowinset=0.2](14,14)(14,0,360)
\Arc[arrow,arrowpos=0.00,arrowlength=3.0,arrowwidth=0.8,arrowinset=0.2](56,14)(14,0,360)
\Line[dash,dashsize=1.5](28,14)(42,14)
\Vertex(28,14){1.0}
\Vertex(42,14){1.0}
\end{picture} \; \; .
\]

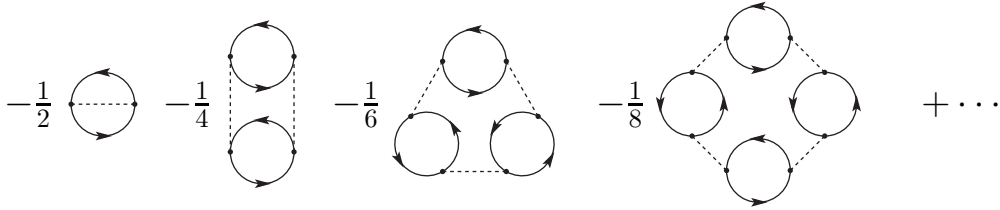
\begin{figure*}[!h]
\begin{center}
\begin{picture}(390,100)(-10,0)
\Text(0,50)[cc]{\large $-\frac{1}{2}$} 
\Arc[arrow,arrowpos=0.50,arrowlength=3.0,arrowwidth=0.8,arrowinset=0.2](28,50)(12,0,180)
\Arc[arrow,arrowpos=0.50,arrowlength=3.0,arrowwidth=0.8,arrowinset=0.2](28,50)(12,180,360)
\Line[dash,dashsize=1.5](16,50)(40,50)
\Vertex(16,50){1.0} \Vertex(40,50){1.0}
\Text(60,50)[cc]{\large $-\frac{1}{4}$} 
\Arc[arrow,arrowpos=0.50,arrowlength=3.0,arrowwidth=0.8,arrowinset=0.2](88,68)(12,0,180)
\Arc[arrow,arrowpos=0.50,arrowlength=3.0,arrowwidth=0.8,arrowinset=0.2](88,68)(12,180,360)
\Arc[arrow,arrowpos=0.50,arrowlength=3.0,arrowwidth=0.8,arrowinset=0.2](88,32)(12,0,180)
\Arc[arrow,arrowpos=0.50,arrowlength=3.0,arrowwidth=0.8,arrowinset=0.2](88,32)(12,180,360)
\Vertex(76,68){1.0}\Vertex(100,68){1.0}
\Vertex(76,32){1.0}\Vertex(100,32){1.0}
\Line[dash,dashsize=1.5](76,68)(76,32)
\Line[dash,dashsize=1.5](100,68)(100,32)
\Text(124,50)[cc]{\large $-\frac{1}{6}$} 
\Arc[arrow,arrowpos=0.50,arrowlength=3.0,arrowwidth=0.8,arrowinset=0.2](150,35)(12,120,300)
\Arc[arrow,arrowpos=0.50,arrowlength=3.0,arrowwidth=0.8,arrowinset=0.2](150,35)(12,300,120)
\Arc[arrow,arrowpos=0.50,arrowlength=3.0,arrowwidth=0.8,arrowinset=0.2](186,35)(12,60,240)
\Arc[arrow,arrowpos=0.50,arrowlength=3.0,arrowwidth=0.8,arrowinset=0.2](186,35)(12,240,60)
\Arc[arrow,arrowpos=0.50,arrowlength=3.0,arrowwidth=0.8,arrowinset=0.2](168,66.177)(12,0,180)
\Arc[arrow,arrowpos=0.50,arrowlength=3.0,arrowwidth=0.8,arrowinset=0.2](168,66.177)(12,180,0)
\Vertex(156,66.177){1.0} \Vertex(180,66.177){1.0}
\Vertex(192,45.3923){1.0}\Vertex(144,45.3923){1.0}
\Vertex(180,24.6077){1.0}
\Vertex(156,24.6077){1.0}
\Line[dash,dashsize=1.5](180,66.177)(192,45.3923)
\Line[dash,dashsize=1.5](156,66.177)(144,45.3923)
\Line[dash,dashsize=1.5](180,24.6077)(156,24.6077)
\Text(224,50)[cc]{\large $-\frac{1}{8}$}
\Arc[arrow,arrowpos=0.50,arrowlength=3.0,arrowwidth=0.8,arrowinset=0.2](250,50)(12,-90,90)
\Arc[arrow,arrowpos=0.50,arrowlength=3.0,arrowwidth=0.8,arrowinset=0.2](250,50)(12,90,-90)
\Arc[arrow,arrowpos=0.50,arrowlength=3.0,arrowwidth=0.8,arrowinset=0.2](300,50)(12,-90,90)
\Arc[arrow,arrowpos=0.50,arrowlength=3.0,arrowwidth=0.8,arrowinset=0.2](300,50)(12,90,-90)
\Arc[arrow,arrowpos=0.50,arrowlength=3.0,arrowwidth=0.8,arrowinset=0.2](275,25)(12,0,180)
\Arc[arrow,arrowpos=0.50,arrowlength=3.0,arrowwidth=0.8,arrowinset=0.2](275,25)(12,180,360)
\Arc[arrow,arrowpos=0.50,arrowlength=3.0,arrowwidth=0.8,arrowinset=0.2](275,75)(12,0,180)
\Arc[arrow,arrowpos=0.50,arrowlength=3.0,arrowwidth=0.8,arrowinset=0.2](275,75)(12,180,360)
\Vertex(250,62){1.0}\Vertex(250,38){1.0} \Vertex(300,62){1.0}\Vertex(300,38){1.0}
\Vertex(263,25){1.0}\Vertex(287,25){1.0} \Vertex(263,75){1.0}\Vertex(287,75){1.0}
\Line[dash,dashsize=1.5](250,62)(263,75)
\Line[dash,dashsize=1.5](250,38)(263,25)
\Line[dash,dashsize=1.5](300,62)(287,75)
\Line[dash,dashsize=1.5](300,38)(287,25)
\Text(354,50)[cc]{\large $+\cdots$}
\end{picture}
\end{center}
\caption[]{Feynman diagrams corresponding to $\frac{1}{2}\Tr \ln (\tilde {\mathcal D}_0^{-1} \4di \, U)$. 
The first term corresponds to $-\frac{1}{2 \cdot 1}\Tr \,(D_0\4di \, U)$, the second term corresponds to 
$-\frac{1}{2\cdot 2}\Tr \,(D_0\4di \, U)^2$,
 the third term corresponds to $-\frac{1}{2 \cdot 3}\Tr \, (D_0\4di \, U)^3$, 
 and the fourth term corresponds to $-\frac{1}{2 \cdot 4}\Tr \, (D_0\4di \, U)^4$. 
 In general, the diagram corresponding to $-\Tr \, (D_0\4di \, U)^n$ contains $n$ bubbles strung by $n$ $U$ propagators with the symmetry  factor $\frac{1}{2n}$.
} \label{fig:polarization}
\end{figure*}

As remarked by Jackiw,~\cite{Jackiw_74} each term in the effective action expansion represents an 
infinite number of Feynman diagrams in regular perturbative field theoretic calculations. 
We use the $\frac{1}{2}\Tr \ln (\tilde {\mathcal{D}_0}^{-1} \4di \, U )$ term as an explicit example  
\[
\frac{1}{2}\Tr \ln (\tilde {\mathcal{D}_0}^{-1} \4di \, U ) =\frac{1}{2} \Tr \ln \left[ {\mathbf I} 
- D_0\4di \, U \right]
 = -\sum_{n=1}^\infty \frac{\Tr \left( D_0 \4di \, U \right)^n}{2n} \;,
\] 
where each term in the summation corresponds to a vacuum diagram, see Fig~\ref{fig:polarization}. 

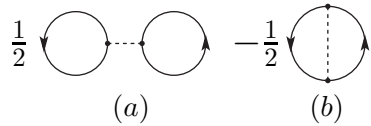
\begin{figure}[!h]
\begin{center}
\begin{picture}(134,40)(0,-6)
\Text(0,20)[lc]{\large $\frac{1}{2}$}
\Arc[arrow,arrowpos=0.50,arrowlength=3.0,arrowwidth=0.8,arrowinset=0.2](25,20)(12,0,360)
\Line[dash,dashsize=1.5](37,20)(50,20)
\Vertex(37,20){1.0} 
\Vertex(50,20){1.0}
\Arc[arrow,arrowpos=0.50,arrowlength=3.0,arrowwidth=0.8,arrowinset=0.2](62,20)(12,-180,180)
\Text(84,20)[lc]{\large $-\frac{1}{2}$} 
\Arc[arrow,arrowpos=0.50,arrowlength=3.0,arrowwidth=0.8,arrowinset=0.2](120,20)(14,-90,90)
\Arc[arrow,arrowpos=0.50,arrowlength=3.0,arrowwidth=0.8,arrowinset=0.2](120,20)(14,90,-90)
\Line[dash,dashsize=1.5](120,6)(120,34)
\Vertex(120,6){1.0} 
\Vertex(120,34){1.0}
\Text(46,-10)[bc]{($a$)}
\Text(120,-10)[bc]{($b$)}
\end{picture}
\end{center}
\caption[]{The Feynman diagrams corresponding to the Hartree term (a) and
 the lowest order exchange term (b), the first graph from Fig.~\ref{fig:polarization}. 
 The numerical factor associated with
  each diagram is shown explicitly. 
} \label{fig:HF}
\end{figure}

If we pull together the lowest order diagrams in $e^2$, we find the combination
\[
\frac{1}{2}n\4di \, U \4di n - \frac{1}{2} \Tr (D_0 \4di \, U)
\] 
that corresponds to the Feynman diagrams in Fig.~\ref{fig:HF}. 

To compute $\Gamma_2$, we need to first calculate $J_1$. Since $\Gamma_1[n] = \beta W_1[J_0]
 = \frac{1}{2} \Tr \ln ({\mathbf I} - D_0 \4di \, U )$ 
\bea
J_1(z) &=& -\frac{\delta \Gamma_1}{\delta n(z)} = 
- \frac{1}{2} \Tr \left[ ({\mathbf I} -D_0 \4di \, U )^{-1} \4di \left(-\frac{\delta D_0}{\delta n(z)} 
\right) \4di \, U \right] \nonumber \\
&=& \frac{1}{2} \Tr \left[ U\4di ({\mathbf I} -D_0 \4di \, U )^{-1} \4di \frac{\delta D_0}{\delta n(z)} 
 \right] = \frac{1}{2} \Tr \left[ \tilde {\mathcal D}_0 \4di \frac{\delta D_0}{\delta n(z)} \right]
  \nonumber \\
&=& -\int dy\, dx\, dx'\, D_0^{-1}(z,y) \tilde {\mathcal D}_0(x,x') \PH_0(x,x')\PH_0(x',y)\PH_0(y,x)  
\; , \label{J1.explicit}
\eea
which can also be expressed diagrammatically as
\[
J_1(z) = -\; 
\begin{picture}(48,20)(0,0)
\Line[double](3,2.5)(20,2.5)
\Arc[arrow,arrowpos=0.50,arrowlength=3.0,arrowwidth=0.8,arrowinset=0.2](34,2.5)(14,-90,90)
\Arc[arrow,arrowpos=0.50,arrowlength=3.0,arrowwidth=0.8,arrowinset=0.2](34,2.5)(14,90,0)
\Arc[arrow,arrowpos=0.50,arrowlength=3.0,arrowwidth=0.8,arrowinset=0.2](34,2.5)(14,0,-90)
\Vertex(3.5,2.5){1.8}
\Vertex(20,2.5){1.0}
\Photon[](34,-11.5)(34,16.5){2}{6}
\Vertex(34,-11.5){1.8}
\Vertex(34,16.5){1.8}
\Text(3.5,-1)[tc]{$z$}
\end{picture} \; \;\; .
\]
From Eq.~(\ref{G2.n}), we know that $\Gamma_2[n] = \beta W_2[J_0] - \frac{1}{2} J_1 \4di D_0^{-1} \4di J_1$.
We note from Eq.~(\ref{W.varphi.3}) and (\ref{W.loop}) that $\beta W_2[J_0]$ corresponds to 
 the $\lambda^2$ diagrams in 
 \[  
 - \lambda \sum_{n=1}^\infty \frac{1}{n!} \la \left[ {1\over \lambda} \sum_{k=3}^\infty
 \lambda^{k\over 2} I^{(k)}[\varphi] \4di \,b_1 \ldots \4di\, b_k  \right]^n \ra_{\rm 1PI,~conn.} \;, 
 \] 
 the  last part of Eq.~(\ref{W.varphi.3}). 
 There are two combinations of $n$ and $k$ that can give rise to $\lambda^2$. 
 The first one is to have $n=1$ and $k=4$, while the second one is to have $n=2$ and $k=3$.
 The first possibility generates two distinct graphs, while the second possibility generates
  three distinct diagrams. For $n=1$ and $k=4$, we have the following diagrams with their symmetry factors specified
\[
\begin{picture}(180,80)(0,0)
\Text(30,72)[bc]{($a$)}
\Arc[arrow,arrowpos=0.50,arrowlength=3.0,arrowwidth=0.8,arrowinset=0.2](30,42)(25,0,90)
\Arc[arrow,arrowpos=0.50,arrowlength=3.0,arrowwidth=0.8,arrowinset=0.2](30,42)(25,90,180)
\Arc[arrow,arrowpos=0.50,arrowlength=3.0,arrowwidth=0.8,arrowinset=0.2](30,42)(25,180,270)
\Arc[arrow,arrowpos=0.50,arrowlength=3.0,arrowwidth=0.8,arrowinset=0.2](30,42)(25,270,360)
\Vertex(30,67){1.8}
\Vertex(30,17){1.8}
\Vertex(5,42){1.8}
\Vertex(55,42){1.8}
\Text(30,10)[tc]{\large $- \frac{(i)^4}{1!} \frac{(-1)^3}{4}$}
\Photon[](30,67)(30,17){2}{7.5}
\Photon[](5,42)(55,42){2}{7.5} 
\Text(130,72)[bc]{($b$)}
\Arc[arrow,arrowpos=0.50,arrowlength=3.0,arrowwidth=0.8,arrowinset=0.2](130,42)(25,0,90)
\Arc[arrow,arrowpos=0.50,arrowlength=3.0,arrowwidth=0.8,arrowinset=0.2](130,42)(25,90,180)
\Arc[arrow,arrowpos=0.50,arrowlength=3.0,arrowwidth=0.8,arrowinset=0.2](130,42)(25,180,270)
\Arc[arrow,arrowpos=0.50,arrowlength=3.0,arrowwidth=0.8,arrowinset=0.2](130,42)(25,270,360)
\Vertex(130,67){1.8}
\Vertex(130,17){1.8}
\Vertex(105,42){1.8}
\Vertex(155,42){1.8}
\Text(130,10)[tc]{{\large $- \frac{(i)^4}{1!} \frac{(-1)^3}{4}$}{\small$2$}} 
\Photon[](130,67)(155,42){2}{5.5}
\Photon[](105,42)(130,17){2}{5.5}  
\end{picture} \; \; \; .
\]
 For $n=2$ and $k=3$, we have the following diagrams with their symmetry factors specified
\bea 
(a) & 
\begin{picture}(100,22.5)(0,-4)
\Arc[arrow,arrowpos=0.50,arrowlength=3.0,arrowwidth=0.8,arrowinset=0.2](20,0)(20,90,-90)
\Arc[arrow,arrowpos=0.50,arrowlength=3.0,arrowwidth=0.8,arrowinset=0.2](20,0)(20,-90,0)
\Arc[arrow,arrowpos=0.50,arrowlength=3.0,arrowwidth=0.8,arrowinset=0.2](20,0)(20,0,90)
\Vertex(20,-20){1.8} \Vertex(20,20){1.8} \Vertex(40,0){1.8}
\Arc[arrow,arrowpos=0.50,arrowlength=3.0,arrowwidth=0.8,arrowinset=0.2,flip](80,0)(20,90,0)
\Arc[arrow,arrowpos=0.50,arrowlength=3.0,arrowwidth=0.8,arrowinset=0.2,flip](80,0)(20,0,-90)
\Arc[arrow,arrowpos=0.50,arrowlength=3.0,arrowwidth=0.8,arrowinset=0.2,flip](80,0)(20,-90,90)
\Vertex(80,-20){1.8} \Vertex(80,20){1.8} \Vertex(60,0){1.8}
\Photon[](20,-20)(20,20){2}{8}
\Photon[](80,-20)(80,20){2}{8}
\Photon[](40,0)(60,0){2}{4}
\end{picture}  
& \hspace*{10pt} -\frac{3(i)^6}{2!} \frac{(-1)^2}{3} \frac{(-1)^2}{3} \; , \nonumber\\
(b) & 
\begin{picture}(100,42.5)(0,-4)
\Arc[arrow,arrowpos=0.50,arrowlength=3.0,arrowwidth=0.8,arrowinset=0.2](20,0)(20,60,-60)
\Arc[arrow,arrowpos=0.50,arrowlength=3.0,arrowwidth=0.8,arrowinset=0.2](20,0)(20,-60,0)
\Arc[arrow,arrowpos=0.50,arrowlength=3.0,arrowwidth=0.8,arrowinset=0.2](20,0)(20,0,60)
\Vertex(30,-17.32){1.8} \Vertex(30,17.32){1.8} \Vertex(40,0){1.8}
\Arc[arrow,arrowpos=0.50,arrowlength=3.0,arrowwidth=0.8,arrowinset=0.2,flip](80,0)(20,60,0)
\Arc[arrow,arrowpos=0.50,arrowlength=3.0,arrowwidth=0.8,arrowinset=0.2,flip](80,0)(20,0,-60)
\Arc[arrow,arrowpos=0.50,arrowlength=3.0,arrowwidth=0.8,arrowinset=0.2,flip](80,0)(20,-60,60)
\Vertex(70,-17.32){1.8} \Vertex(70,17.32){1.8} \Vertex(60,0){1.8}
\Photon[](30,-17.32)(70,-17.32){2}{7.5}
\Photon[](30,17.32)(70,17.32){2}{7.5}
\Photon[](40,0)(60,0){2}{4}
\end{picture}  
& \hspace*{10pt}  -\frac{3(i)^6}{2!} \frac{(-1)^2}{3} \frac{(-1)^2}{3} \; , \nonumber\\
(c) & 
\begin{picture}(100,42.5)(0,-4)
\Arc[arrow,arrowpos=0.50,arrowlength=3.0,arrowwidth=0.8,arrowinset=0.2](20,0)(20,60,-60)
\Arc[arrow,arrowpos=0.50,arrowlength=3.0,arrowwidth=0.8,arrowinset=0.2](20,0)(20,-60,0)
\Arc[arrow,arrowpos=0.50,arrowlength=3.0,arrowwidth=0.8,arrowinset=0.2](20,0)(20,0,60)
\Vertex(30,-17.32){1.8} \Vertex(30,17.32){1.8} \Vertex(40,0){1.8}
\Arc[arrow,arrowpos=0.50,arrowlength=3.0,arrowwidth=0.8,arrowinset=0.2](80,0)(20,60,0)
\Arc[arrow,arrowpos=0.50,arrowlength=3.0,arrowwidth=0.8,arrowinset=0.2](80,0)(20,0,-60)
\Arc[arrow,arrowpos=0.50,arrowlength=3.0,arrowwidth=0.8,arrowinset=0.2](80,0)(20,-60,60)
\Vertex(70,-17.32){1.8} \Vertex(70,17.32){1.8} \Vertex(60,0){1.8}
\Photon[](30,-17.32)(70,-17.32){2}{7.5}
\Photon[](30,17.32)(70,17.32){2}{7.5}
\Photon[](40,0)(60,0){2}{4}
\end{picture}  
& \hspace*{10pt}  -\frac{3(i)^6}{2!} \frac{(-1)^2}{3} \frac{(-1)^2}{3}  \; . \nonumber 
\eea  
\phantom{12345}\vspace*{10pt}\\
The first diagram(a)  among the three will be discarded since one can cut a $\tilde {\mathcal D}_0$ line
 and then separate it into two diagrams.  
Let us display below the diagram corresponding to $J_1 \4di D_0 \4di J_1$,
\[
J_1 \4di D_0 \4di J_1 = (-J_1) \4di D_0 \4di (- J_1) = 
\begin{picture}(48,20)(18,0)
\Line[double](48,2.5)(65,2.5)
\Arc[arrow,arrowpos=0.50,arrowlength=3.0,arrowwidth=0.8,arrowinset=0.2,clock](34,2.5)(14,-90,90)
\Arc[arrow,arrowpos=0.50,arrowlength=3.0,arrowwidth=0.8,arrowinset=0.2,clock](34,2.5)(14,90,0)
\Arc[arrow,arrowpos=0.50,arrowlength=3.0,arrowwidth=0.8,arrowinset=0.2,clock](34,2.5)(14,0,-90)
\Vertex(48,2.5){1.0}
\Vertex(65,2.5){1.0}
\Vertex(34,16.5){1.8}
\Vertex(34,-11.5){1.8}
\Photon[](34,-11.5)(34,16.5){2}{6}
\end{picture} \hspace*{5pt}
\left( \begin{picture}(17,8)(0,0)
\Line[double](0,2.5)(17,2.5)
\Vertex(0,2.5){1.0} \Vertex(17,2.5){1.0} 
\end{picture}\right)^{-1}
\begin{picture}(48,20)(0,0)
\Line[double](3,2.5)(20,2.5)
\Arc[arrow,arrowpos=0.50,arrowlength=3.0,arrowwidth=0.8,arrowinset=0.2](34,2.5)(14,-90,90)
\Arc[arrow,arrowpos=0.50,arrowlength=3.0,arrowwidth=0.8,arrowinset=0.2](34,2.5)(14,90,0)
\Arc[arrow,arrowpos=0.50,arrowlength=3.0,arrowwidth=0.8,arrowinset=0.2](34,2.5)(14,0,-90)
\Vertex(3,2.5){1.0}
\Vertex(20,2.5){1.0}
\Vertex(34,16.5){1.8}
\Vertex(34,-11.5){1.8}
\Photon[](34,-11.5)(34,16.5){2}{6}
\end{picture} \; 
= \; \; 
\begin{picture}(76,20)(0,0)
\Arc[arrow,arrowpos=0.50,arrowlength=3.0,arrowwidth=0.8,arrowinset=0.2,clock](17,2.5)(14,-90,90)
\Arc[arrow,arrowpos=0.50,arrowlength=3.0,arrowwidth=0.8,arrowinset=0.2,clock](17,2.5)(14,90,0)
\Arc[arrow,arrowpos=0.50,arrowlength=3.0,arrowwidth=0.8,arrowinset=0.2,clock](17,2.5)(14,0,-90)
\Photon[](17,-11.5)(17,16.5){2}{6}
\Vertex(17,16.5){1.8}
\Vertex(17,-11.5){1.8}
\Line[double](31,2.5)(48,2.5)
\Arc[arrow,arrowpos=0.50,arrowlength=3.0,arrowwidth=0.8,arrowinset=0.2](62,2.5)(14,-90,90)
\Arc[arrow,arrowpos=0.50,arrowlength=3.0,arrowwidth=0.8,arrowinset=0.2](62,2.5)(14,90,0)
\Arc[arrow,arrowpos=0.50,arrowlength=3.0,arrowwidth=0.8,arrowinset=0.2](62,2.5)(14,0,-90)
\Vertex(31,2.5){1.0}
\Vertex(48,2.5){1.0}
\Photon[](62,-11.5)(62,16.5){2}{6}
\Vertex(62,16.5){1.8}
\Vertex(62,-11.5){1.8}
\end{picture} \; \; .
\]

Therefore, the diagrammatic expression for $\Gamma_2[n]$ is given by 
\begin{figure}[!h]
\bea
\Gamma_2[n] &=& \frac{1}{4} \; \begin{picture}(40,20)(0,-4)
\Arc[arrow,arrowpos=0.50,arrowlength=3.0,arrowwidth=0.8,arrowinset=0.2](20,0)(20,0,90)
\Arc[arrow,arrowpos=0.50,arrowlength=3.0,arrowwidth=0.8,arrowinset=0.2](20,0)(20,90,180)
\Arc[arrow,arrowpos=0.50,arrowlength=3.0,arrowwidth=0.8,arrowinset=0.2](20,0)(20,180,270)
\Arc[arrow,arrowpos=0.50,arrowlength=3.0,arrowwidth=0.8,arrowinset=0.2](20,0)(20,270,360)
\Vertex(20,-20){1.8}
\Vertex(20,20){1.8}
\Vertex(0,0){1.8}
\Vertex(40,0){1.8} 
\Photon[](20,20)(20,-20){2}{6}
\Photon[](0,0)(40,0){2}{6} 
\end{picture} \; \; + \frac{1}{2} \; \begin{picture}(40,20)(0,-4)
\Arc[arrow,arrowpos=0.50,arrowlength=3.0,arrowwidth=0.8,arrowinset=0.2](20,0)(20,0,90)
\Arc[arrow,arrowpos=0.50,arrowlength=3.0,arrowwidth=0.8,arrowinset=0.2](20,0)(20,90,180)
\Arc[arrow,arrowpos=0.50,arrowlength=3.0,arrowwidth=0.8,arrowinset=0.2](20,0)(20,180,270)
\Arc[arrow,arrowpos=0.50,arrowlength=3.0,arrowwidth=0.8,arrowinset=0.2](20,0)(20,270,360)
\Vertex(20,-20){1.8}
\Vertex(20,20){1.8}
\Vertex(0,0){1.8}
\Vertex(40,0){1.8} 
\Photon[](20,20)(40,0){2}{6}
\Photon[](0,0)(20,-20){2}{6} 
\end{picture} \; \;  - \frac{1}{2} \; \begin{picture}(100,20)(0,-4)
\Arc[arrow,arrowpos=0.50,arrowlength=3.0,arrowwidth=0.8,arrowinset=0.2,clock](20,0)(20,-90,90)
\Arc[arrow,arrowpos=0.50,arrowlength=3.0,arrowwidth=0.8,arrowinset=0.2,clock](20,0)(20,90,0)
\Arc[arrow,arrowpos=0.50,arrowlength=3.0,arrowwidth=0.8,arrowinset=0.2,clock](20,0)(20,0,-90)
\Photon[](20,-20)(20,20){2}{8}
\Vertex(20,-20){1.8} \Vertex(20,20){1.8}
\Line[double](40,0)(60,0)
\Arc[arrow,arrowpos=0.50,arrowlength=3.0,arrowwidth=0.8,arrowinset=0.2](80,0)(20,-90,90)
\Arc[arrow,arrowpos=0.50,arrowlength=3.0,arrowwidth=0.8,arrowinset=0.2](80,0)(20,90,0)
\Arc[arrow,arrowpos=0.50,arrowlength=3.0,arrowwidth=0.8,arrowinset=0.2](80,0)(20,0,-90)
\Vertex(40,0){1.8}
\Vertex(60,0){1.8}
\Photon[](80,-20)(80,20){2}{8}
\Vertex(80,-20){1.8} \Vertex(80,20){1.8}
\end{picture} \nonumber \\
&& + \frac{1}{3!} \; \begin{picture}(100,42.5)(0,-4)
\Arc[arrow,arrowpos=0.50,arrowlength=3.0,arrowwidth=0.8,arrowinset=0.2](20,0)(20,60,-60)
\Arc[arrow,arrowpos=0.50,arrowlength=3.0,arrowwidth=0.8,arrowinset=0.2](20,0)(20,-60,0)
\Arc[arrow,arrowpos=0.50,arrowlength=3.0,arrowwidth=0.8,arrowinset=0.2](20,0)(20,0,60)
\Vertex(30,-17.32){1.5} \Vertex(30,17.32){1.5} \Vertex(40,0){1.5}
\Arc[arrow,arrowpos=0.50,arrowlength=3.0,arrowwidth=0.8,arrowinset=0.2,flip](80,0)(20,60,0)
\Arc[arrow,arrowpos=0.50,arrowlength=3.0,arrowwidth=0.8,arrowinset=0.2,flip](80,0)(20,0,-60)
\Arc[arrow,arrowpos=0.50,arrowlength=3.0,arrowwidth=0.8,arrowinset=0.2,flip](80,0)(20,-60,60)
\Vertex(70,-17.32){1.8} \Vertex(70,17.32){1.8} \Vertex(60,0){1.8}
\Photon[](30,-17.32)(70,-17.32){2}{8}
\Photon[](30,17.32)(70,17.32){2}{8}
\Photon[](40,0)(60,0){2}{4}
\end{picture} \;+ \frac{1}{3!} \;  \begin{picture}(100,42.5)(0,-4)
\Arc[arrow,arrowpos=0.50,arrowlength=3.0,arrowwidth=0.8,arrowinset=0.2](20,0)(20,60,-60)
\Arc[arrow,arrowpos=0.50,arrowlength=3.0,arrowwidth=0.8,arrowinset=0.2](20,0)(20,-60,0)
\Arc[arrow,arrowpos=0.50,arrowlength=3.0,arrowwidth=0.8,arrowinset=0.2](20,0)(20,0,60)
\Vertex(30,-17.32){1.8} \Vertex(30,17.32){1.8} \Vertex(40,0){1.8}
\Arc[arrow,arrowpos=0.50,arrowlength=3.0,arrowwidth=0.8,arrowinset=0.2](80,0)(20,60,0)
\Arc[arrow,arrowpos=0.50,arrowlength=3.0,arrowwidth=0.8,arrowinset=0.2](80,0)(20,0,-60)
\Arc[arrow,arrowpos=0.50,arrowlength=3.0,arrowwidth=0.8,arrowinset=0.2](80,0)(20,-60,60)
\Vertex(70,-17.32){1.8} \Vertex(70,17.32){1.8} \Vertex(60,0){1.8}
\Photon[](30,-17.32)(70,-17.32){2}{7.5}
\Photon[](30,17.32)(70,17.32){2}{7.5}
\Photon[](40,0)(60,0){2}{4}
\end{picture} \; \; \; .\nonumber \\
&& \begin{picture}(20,20)(0,0)
\end{picture} \nonumber
\eea 
\vspace*{-20pt}
\caption{Diagrammatic expression for $\Gamma_2[n]$.}\label{fig:Gamma.2}
\end{figure}

Note that each $\tilde {\mathcal D}_0$ propagator line contains the sum of infinitely many terms
\bea
\tilde {\mathcal D}_0 &=& \left(U^{-1} -D_0 \right)^{-1} = \left( {\mathbf I}- U \4di D_0 \right)^{-1} 
\4di \, U
= U + U\4di D_0 \4di \, U + U\4di D_0\4di \, U\4di D_0\4di \, U + \ldots  \label{tD.exp} \\
\begin{picture}(20,30)(0,-3)
\Photon[](0,0)(20,0){2}{4}
\Vertex(0,0){1.8} \Vertex(20,0){1.8}
\end{picture} 
&=& \begin{picture}(20,30)(0,-3)
\Line[dash,dashsize=1.5](0,0)(20,0)
\Vertex(0,0){1.0} \Vertex(20,0){1.0}
\end{picture} \; \;
+ \; \begin{picture}(60,30)(0,-3)
\Line[dash,dashsize=1.5](0,0)(20,0) \Line[dash,dashsize=1.5](40,0)(60,0)
\Arc[arrow,arrowpos=0.50,arrowlength=3.0,arrowwidth=0.8,arrowinset=0.2](30,0)(10,0,180)
\Arc[arrow,arrowpos=0.50,arrowlength=3.0,arrowwidth=0.8,arrowinset=0.2](30,0)(10,180,360)
\Vertex(0,0){1.0} \Vertex(20,0){1.0}
\Vertex(40,0){1.0} \Vertex(60,0){1.0}
\end{picture} \; \;
+ \;  \begin{picture}(100,30)(0,-3)
\Line[dash,dashsize=1.5](0,0)(20,0) \Line[dash,dashsize=1.5](40,0)(60,0) \Line[dash,dashsize=1.5](80,0)(100,0)
\Arc[arrow,arrowpos=0.50,arrowlength=3.0,arrowwidth=0.8,arrowinset=0.2](30,0)(10,0,180)
\Arc[arrow,arrowpos=0.50,arrowlength=3.0,arrowwidth=0.8,arrowinset=0.2](30,0)(10,180,360)
\Arc[arrow,arrowpos=0.50,arrowlength=3.0,arrowwidth=0.8,arrowinset=0.2](70,0)(10,0,180)
\Arc[arrow,arrowpos=0.50,arrowlength=3.0,arrowwidth=0.8,arrowinset=0.2](70,0)(10,180,360)
\Vertex(0,0){1.0} \Vertex(20,0){1.0}
\Vertex(40,0){1.0} \Vertex(60,0){1.0}
\Vertex(80,0){1.0} \Vertex(100,0){1.0}
\end{picture}  \;\; + \; \cdots 
 \;. \nonumber 
\eea
Since each $U$ carries a factor of $e^2$, the above expansion may be viewed as the $e^2$ expansion of $\tilde {\mathcal D}_0$
 with $U$ being the leading order. Therefore, in the diagrams corresponding to
  $\Gamma_2$ (Fig.~\ref{fig:Gamma.2}), if one were to 
  expand $\tilde {\mathcal D}_0$ in $e^2$, the first three diagrams are of order $e^4$ or higher, 
  while the last two diagrams contains terms of order $e^6$ or higher only.
 
It is now instructive to compare with the perturbative methods which use $e^2$ as  the expansion parameter. 
Among those methods, the approach of Valiev and Fernando~\cite{VF_97_arxiv} is the closest to ours. Let us 
pull out the diagrams contributing to order $e^4$ in $\Gamma_{i\le 2}$ and compare to the results of 
reference~\onlinecite{VF_97_arxiv}.
 From $\Gamma_1[n]$, we see that the second diagram in Fig.~\ref{fig:polarization} corresponding to
 $ -\frac{1}{4} \Tr \left[ \left( D_0 \4di \, U\right)^2 \right]$ will contribute to this order. The first three diagrams 
  representing $\Gamma_2[n]$ will contribute to the same order if we replace the $\tilde {\mathcal D}$ propagator by 
  $U$. Thus, one obtains the diagrams shown in Fig.~\ref{fig:OE4}, 
 which are identical to the results in reference~\onlinecite{VF_97_arxiv}.    
\begin{figure*}[!h]
\begin{center}
\begin{picture}(420,40)(0,0)
\put(1,20){\makebox(0,0){\large $\frac{1}{4}$}}
\put(10,20){\begin{picture}(40,20)(0,0)
\Arc[arrow,arrowpos=0.50,arrowlength=3.0,arrowwidth=0.8,arrowinset=0.2](20,0)(20,0,90)
\Arc[arrow,arrowpos=0.50,arrowlength=3.0,arrowwidth=0.8,arrowinset=0.2](20,0)(20,90,180)
\Arc[arrow,arrowpos=0.50,arrowlength=3.0,arrowwidth=0.8,arrowinset=0.2](20,0)(20,180,270)
\Arc[arrow,arrowpos=0.50,arrowlength=3.0,arrowwidth=0.8,arrowinset=0.2](20,0)(20,270,360)
\Vertex(20,-20){1.0}
\Vertex(20,20){1.0}
\Vertex(0,0){1.0}
\Vertex(40,0){1.0} 
\Line[dash,dashsize=1.5](20,20)(20,-20)
\Line[dash,dashsize=1.5](0,0)(40,0) 
\end{picture}
}
\put(68,20){\makebox(0,0){\large $- \frac{1}{4}$}}
\put(85,20){\begin{picture}(90,20)(0,0)
\Arc[arrow,arrowpos=0.50,arrowlength=3.0,arrowwidth=0.8,arrowinset=0.2,clock](20,0)(20,-90,90)
\Arc[arrow,arrowpos=0.50,arrowlength=3.0,arrowwidth=0.8,arrowinset=0.2,clock](20,0)(20,90,-90)
\Line[dash,dashsize=1.5](20,-20)(70,-20)
\Arc[arrow,arrowpos=0.50,arrowlength=3.0,arrowwidth=0.8,arrowinset=0.2](70,0)(20,-90,90)
\Arc[arrow,arrowpos=0.50,arrowlength=3.0,arrowwidth=0.8,arrowinset=0.2](70,0)(20,90,-90)
\Line[dash,dashsize=1.5](20,20)(70,20)
\Vertex(70,-20){1.0}
\Vertex(70,20){1.0}
\Vertex(20,-20){1.0}
\Vertex(20,20){1.0}
\end{picture}
}
\put(197,20){\makebox(0,0){\large $+ \frac{1}{2}$}}
\put(215,20){\begin{picture}(40,20)(0,0)
\Arc[arrow,arrowpos=0.50,arrowlength=3.0,arrowwidth=0.8,arrowinset=0.2](20,0)(20,0,90)
\Arc[arrow,arrowpos=0.50,arrowlength=3.0,arrowwidth=0.8,arrowinset=0.2](20,0)(20,90,180)
\Arc[arrow,arrowpos=0.50,arrowlength=3.0,arrowwidth=0.8,arrowinset=0.2](20,0)(20,180,270)
\Arc[arrow,arrowpos=0.50,arrowlength=3.0,arrowwidth=0.8,arrowinset=0.2](20,0)(20,270,360)
\Vertex(20,-20){1.0}
\Vertex(20,20){1.0}
\Vertex(0,0){1.0}
\Vertex(40,0){1.0} 
\Line[dash,dashsize=1.5](20,20)(40,0)
\Line[dash,dashsize=1.5](0,0)(20,-20) 
\end{picture}
}
\put(275,20){\makebox(0,0){\large $- \frac{1}{2}$}}
\put(292,20){\begin{picture}(100,20)(0,0)
\Arc[arrow,arrowpos=0.50,arrowlength=3.0,arrowwidth=0.8,arrowinset=0.2,clock](20,0)(20,-90,90)
\Arc[arrow,arrowpos=0.50,arrowlength=3.0,arrowwidth=0.8,arrowinset=0.2,clock](20,0)(20,90,0)
\Arc[arrow,arrowpos=0.50,arrowlength=3.0,arrowwidth=0.8,arrowinset=0.2,clock](20,0)(20,0,-90)
\Line[dash,dashsize=1.5](20,-20)(20,20)
\Line[double](40,0)(60,0)
\Arc[arrow,arrowpos=0.50,arrowlength=3.0,arrowwidth=0.8,arrowinset=0.2](80,0)(20,-90,90)
\Arc[arrow,arrowpos=0.50,arrowlength=3.0,arrowwidth=0.8,arrowinset=0.2](80,0)(20,90,0)
\Arc[arrow,arrowpos=0.50,arrowlength=3.0,arrowwidth=0.8,arrowinset=0.2](80,0)(20,0,-90)
\Vertex(40,0){1.0}
\Vertex(60,0){1.0}
\Line[dash,dashsize=1.5](80,-20)(80,20)
\Vertex(80,-20){1.0}
\Vertex(80,20){1.0}
\Vertex(20,-20){1.0}
\Vertex(20,20){1.0}
\end{picture} 
}
\end{picture}
\end{center}
\caption[]{The Feynman diagrams of order $U^2$ (or $e^4$) of the effective action $\Gamma[n]$. 
The correct symmetry factors are also provided. 
} \label{fig:OE4}
\end{figure*}

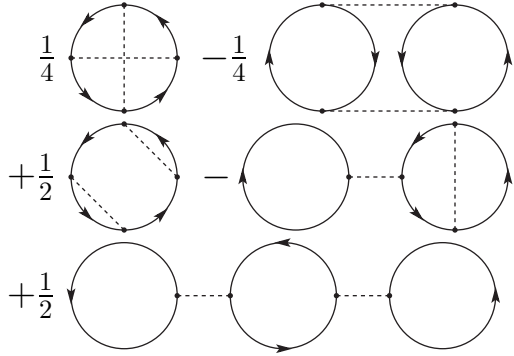
\begin{figure*}[!h]
\begin{center}
\begin{picture}(185,125)(-10,-40)
\put(1,65){\makebox(0,0){\large $\frac{1}{4}$}}
\put(10,65){\begin{picture}(40,20)(0,0)
\Arc[arrow,arrowpos=0.50,arrowlength=3.0,arrowwidth=0.8,arrowinset=0.2](20,0)(20,0,90)
\Arc[arrow,arrowpos=0.50,arrowlength=3.0,arrowwidth=0.8,arrowinset=0.2](20,0)(20,90,180)
\Arc[arrow,arrowpos=0.50,arrowlength=3.0,arrowwidth=0.8,arrowinset=0.2](20,0)(20,180,270)
\Arc[arrow,arrowpos=0.50,arrowlength=3.0,arrowwidth=0.8,arrowinset=0.2](20,0)(20,270,360)
\Vertex(20,-20){1.0}
\Vertex(20,20){1.0}
\Vertex(0,0){1.0}
\Vertex(40,0){1.0} 
\Line[dash,dashsize=1.5](20,20)(20,-20)
\Line[dash,dashsize=1.5](0,0)(40,0) 
\end{picture}
}
\put(68,65){\makebox(0,0){\large $- \frac{1}{4}$}}
\put(85,65){\begin{picture}(90,20)(0,0)
\Arc[arrow,arrowpos=0.50,arrowlength=3.0,arrowwidth=0.8,arrowinset=0.2,clock](20,0)(20,-90,90)
\Arc[arrow,arrowpos=0.50,arrowlength=3.0,arrowwidth=0.8,arrowinset=0.2,clock](20,0)(20,90,-90)
\Line[dash,dashsize=1.5](20,-20)(70,-20)
\Arc[arrow,arrowpos=0.50,arrowlength=3.0,arrowwidth=0.8,arrowinset=0.2](70,0)(20,-90,90)
\Arc[arrow,arrowpos=0.50,arrowlength=3.0,arrowwidth=0.8,arrowinset=0.2](70,0)(20,90,-90)
\Line[dash,dashsize=1.5](20,20)(70,20)
\Vertex(70,-20){1.0}
\Vertex(70,20){1.0}
\Vertex(20,-20){1.0}
\Vertex(20,20){1.0}
\end{picture}
}

\put(-5,20){\makebox(0,0){\large $+ \frac{1}{2}$}}
\put(10,20){\begin{picture}(40,20)(0,0)
\Arc[arrow,arrowpos=0.50,arrowlength=3.0,arrowwidth=0.8,arrowinset=0.2](20,0)(20,0,90)
\Arc[arrow,arrowpos=0.50,arrowlength=3.0,arrowwidth=0.8,arrowinset=0.2](20,0)(20,90,180)
\Arc[arrow,arrowpos=0.50,arrowlength=3.0,arrowwidth=0.8,arrowinset=0.2](20,0)(20,180,270)
\Arc[arrow,arrowpos=0.50,arrowlength=3.0,arrowwidth=0.8,arrowinset=0.2](20,0)(20,270,360)
\Vertex(20,-20){1.0}
\Vertex(20,20){1.0}
\Vertex(0,0){1.0}
\Vertex(40,0){1.0} 
\Line[dash,dashsize=1.5](20,20)(40,0)
\Line[dash,dashsize=1.5](0,0)(20,-20) 
\end{picture}
}

\put(65,20){\makebox(0,0){\large $-$}}
\put(75,20){\begin{picture}(100,20)(0,0)
\Arc[arrow,arrowpos=0.50,arrowlength=3.0,arrowwidth=0.8,arrowinset=0.2,clock](20,0)(20,0,360)
\Line[dash,dashsize=1.5](40,0)(60,0)
\Arc[arrow,arrowpos=0.50,arrowlength=3.0,arrowwidth=0.8,arrowinset=0.2](80,0)(20,-90,90)
\Arc[arrow,arrowpos=0.50,arrowlength=3.0,arrowwidth=0.8,arrowinset=0.2](80,0)(20,90,180)
\Arc[arrow,arrowpos=0.50,arrowlength=3.0,arrowwidth=0.8,arrowinset=0.2](80,0)(20,180,270)
\Line[dash,dashsize=1.5](80,-20)(80,20)
\Vertex(80,-20){1.0}
\Vertex(80,20){1.0}
\Vertex(40,0){1.0}
\Vertex(60,0){1.0}
\end{picture}
}
\put(-5,-25){\makebox(0,0){\large $+ \frac{1}{2}$}}
\put(10,-25){\begin{picture}(160,20)(0,0)
\Arc[arrow,arrowpos=0.50,arrowlength=3.0,arrowwidth=0.8,arrowinset=0.2](20,0)(20,0,360)
\Arc[arrow,arrowpos=0.50,arrowlength=3.0,arrowwidth=0.8,arrowinset=0.2](80,0)(20,0,180)
\Arc[arrow,arrowpos=0.50,arrowlength=3.0,arrowwidth=0.8,arrowinset=0.2](80,0)(20,180,360)
\Arc[arrow,arrowpos=0.50,arrowlength=3.0,arrowwidth=0.8,arrowinset=0.2](140,0)(20,-180,180)
\Vertex(40,0){1.0}
\Vertex(60,0){1.0}
\Vertex(100,0){1.0}
\Vertex(120,0){1.0} 
\Line[dash,dashsize=1.5](40,0)(60,0)
\Line[dash,dashsize=1.5](100,0)(120,0) 
\end{picture}
}
\end{picture}
\end{center}
\caption[]{The Feynman diagrams of order $U^2$ (or $e^4$) of $\beta W[J]$. 
 The correct symmetry factors are also provided. 
} \label{fig:VF_1}
\end{figure*}

We use this example to illustrate the difference between our diagrammatic rules 
 and those of  reference~\onlinecite{VF_97_arxiv}. 
  In order to obtain $e^4$ diagrams, the diagrammatic rules  of reference~\onlinecite{VF_97_arxiv}
 require the generation of all $e^4$ diagrams of $\beta W[J]$, 
  each of which is shown in Fig.~\ref{fig:VF_1}. 

 Then one deletes diagrams that can be cut into two parts by cutting a Coulomb line $u$.
 This means that the last two diagrams above will be deleted from consideration. 
 One then needs to find in the remaining diagrams the two-particle reducible (in the sense of fermion propagator) ones  
  followed by an iterative operation to construct $D_0^{-1}$ lines.\cite{FKYetal_95} In this case, the only 
  two-particle-reducible diagram is the third one and the iterative procedure
   generates exactly the only diagram containing a $D_0^{-1}$ line in $\Gamma_2$, with 
    $\tilde {\mathcal D}$ replaced by $U$. 
 Therefore, the diagrammatic rules  of reference~\onlinecite{VF_97_arxiv} require first one-particle-irreducibility
  of the Coulomb line followed by searching for diagrams that are two-particle-reducible (in 
   fermion propagators sense).

  For our case, when considering a diagram's reducibility, we only consider the $\tilde {\mathcal D}_0$ lines.   
  Let us denote $I^{(k)}[\varphi] \4di \,b_1 \ldots \4di\, b_k $ by ${\mathcal B}^{(k)}$, and 
  call it  blob $k$. The one-particle-irreducible criterion simply means that when one 
   joins any two blobs, say ${\mathcal B}^{(k_1)}$ and ${\mathcal B}^{(k_2)}$, one must make sure
  that at least two or more $\tilde {\mathcal D}_0$ lines are connecting ${\mathcal B}^{(k_1)}$ 
   and ${\mathcal B}^{(k_2)}$ due to contraction of
   $b$ fields. Therefore, it is very easy to find and exclude one-particle-reducible diagrams in our method. 
 For this particular example, only the diagrams surviving in the end are present in our formalism.  
   Since the leading order of $\tilde {\mathcal D}_0$, in terms of expansion of $e^2$, is $U$ itself, 
    our one-particle-irreducibility in $\tilde {\mathcal D}_0$ lines covers the one-particle-irreducibility
  of the Coulomb lines in reference~\onlinecite{VF_97_arxiv}.  
 Although one may show that the rule~\cite{FKYetal_95} of inserting $D_0^{-1}$ for  
two-particle-reducible (in terms of fermion propagators) diagrams still applies , it is not essential to have. 
However, one may choose to use this rule as a tool to ensure correct generation of all distinct diagrams.
 Equipped with the diagrammatic expansion rules, one may construct $\Gamma_{l\,\ge 2}$  
 following the inversion method described in section~\ref{sec:inversion}.

\subsection{Evaluation of $\PH_{0}$ using single particle orbitals}
\label{sec:KS_systematics} 
 To calculate $\PH_0(x,x')$, we define $v(\x) \equiv \upsilon_{\rm ion}(\x) -\mu + J_0(\x)$. 
 Since the evaluation of the Green's function is general, 
  we use the symbol $G(x,x')$ in place of $\PH_0(x,x')$ in the following derivation.

Consider first a generic free fermion Hamiltonian, 
\[
 H [\hpsid,\hpsi] = \int d{\bf x}\; \hpsid(\x) \left( -\frac{\nabla ^2}{2m}
+\upsilon ({\x}) \right) \hpsi (\x) =  \hpsid \3di \hat h \3di \hpsi
\] 
 where 
\begin{equation} \label{single_ptcle_h}
 \hat h(\x,\y) = \left( -\frac{\nabla_{\x}^2}{2m} +\upsilon ({\x}) \right) \delta (\x-\y)
\end{equation}  
and the corresponding Green's function
 (with $Z \equiv \Tr e^{-\beta H} $)
\bea
G(x,y) &=& \la T \hpsi(x) \hpsid (y) \ra = Z^{-1} \Tr \left[ e^{-\beta H} T (\hpsi(x) \hpsid (y) ) \right] \nonumber \\
 &=& \theta(\tau_x - \tau_y-\eta) Z^{-1} \Tr \left[ e^{-\beta H} \hpsi(x) \hpsid (y)\right] - \theta (\tau_y-\tau_x +\eta)
 Z^{-1} \Tr \left[ e^{-\beta H} \hpsid (y) \hpsi(x) \right]   \nonumber \; .
\eea
 The positive infinitesimal quantity $\eta$ is introduced to ensure that in the limit $\tau_x = \tau_y$,
 the time ordered product corresponds to normal ordering (represented by the second term at equal time). 
  Note that $0 < \tau_x,\tau_y < \beta$, $-\beta < \tau_x - \tau_y < \beta$ and  
 $\hpsi (x) = e^{\tau_x H} \hpsi(\x) e^{-\tau_x H}$ and $\hpsid (y) = e^{\tau_y H} \hpsid(\y) e^{-\tau_y H}$. 
 When $\tau_x-\tau_y < 0$, one must have $\tau_x-\tau_y + \beta > 0$. Observe that  
\bea
G(\tau_x-\tau_y < 0) &=& - \Tr \left[ e^{-\beta H} e^{\tau_y H} \hpsid(\y) e^{-\tau_y H} e^{\tau_x H} \hpsi(\x) e^{-\tau_x H} \right]/Z \nonumber \\
&=& -\Tr \left[  e^{\tau_x H} \hpsi(\x) e^{-\tau_x H} e^{-\beta H} e^{\tau_y H} \hpsid(\y) e^{-\tau_y H} \right]/Z \nonumber \\
&=& -\Tr \left[ e^{-\beta H} e^{(\tau_x+\beta) H} \hpsi(\x) e^{-(\tau_x+\beta) H} e^{\tau_y H} \hpsid(\y) e^{-\tau_y H} \right]/Z \nonumber \\
&=& -G(\tau_x-\tau_y+\beta > 0),
\eea
 where the first equality results from $\Tr(AB) = \Tr(BA)$, while the final equality results from the definition
  of the Green's function with positive time argument $\tau_x + \beta -\tau_y> 0$.
 Similarly, one may easily show that
$G(\tau_x - \tau_y > 0 ) = -G(\tau_x-\tau_y-\beta < 0)$. Therefore, the Green's function is antiperiodic in the imaginary time $\tau$. 
 
Letting $M(x,y) \equiv \left( \partial_{\tau_x}  \delta (x-y)  + \hat h(\x, \y) \delta (\tau_x - \tau_y) \right)
 = \left( \partial_{\tau_x} -\frac{\nabla_{\x}^2}{2m} +\upsilon ({\x}) \right) \delta (x-y) $, we find   
\bea
G(x,y) &=& \la T \hpsi(x) \hpsid (y) \ra = - {\delta \over \delta \bar \xi(x)} {\delta \over \delta \xi(y)}\la T e^{\int \bar \xi \hpsi 
 + \hpsid \xi} \ra_{\bar \xi \to 0, \xi \to 0} \nonumber \\
&=& - {\delta \over \delta \bar \xi(x)} {\delta \over \delta \xi(y)} \left.
 \frac {\int {\cal D}[\psi^\dag,\psi] e^{-\psi^\dag \4di M \4di \psi + \bar \xi \4di \psi + \psi^\dag \4di \xi }}
{ \int {\cal D}[\psi^\dag,\psi] e^{-\psi^\dag \4di M  \4di \psi } } \right|_{\bar \xi \to 0, \xi \to 0} \nonumber \\
&=& - {\delta \over \delta \bar \xi(x)} {\delta \over \delta \xi(y)}  
\left[ e^{\bar \xi \4di M^{-1} \4di \xi } \right]_{\bar \xi \to 0, \xi \to 0} = M^{-1}(x,y) \nonumber \; .
\eea
This implies that $\int dx' M(x,x') G(x', y) = \left( \partial_{\tau_x} + \hat h_{\x} \right) G(x,y) =\delta(x-y)$  
 with $\hat h_{\x}$ being a one particle first quantized Hamiltonian $\hat h_{\x} =  -\frac{\nabla_{\x}^2}{2m} +\upsilon ({\x}) $. 
Note that $\delta (x-y) = \delta (\x - \y) \delta (\tau_x - \tau_y)$ and the latter delta function in time
 is defined via $\int_0^\beta g(\tau_x) \delta(\tau_x - \tau_y) d\tau_x = g(\tau_y)$ for any antiperiodic function
 $g(\tau)$. With this understanding, one may express $\delta(x-y)$ in the following way to obtain 
 $G(x,y)$. Observing that 
\bean
(\partial_{\tau_x} + \hat h_{\x}) G(x,y) &=&  \la x | y \ra = 
 \sum_{\omega_n, \alpha} \la x | \omega_n, \alpha \ra \la \omega_n, \alpha | y\ra  \\
&=& \sum_{\omega_n,\alpha } \phi_\alpha(\x) \phi_\alpha^*(\y) {e^{-i\omega_n(\tau_x-\tau_y)} \over \beta}  \; ,
\eean 
 one sees that the action of $(\partial_{\tau_x} + \hat h_{\x})$ may be compensated, leading to
\be 
G(x,y) 
 = \sum_\alpha  \phi_\alpha(\x) \phi_\alpha^*(\y)\left[ {1\over \beta}\sum_{\omega_n}  {e^{-i\omega_n(\tau_x-\tau_y)} \over -i\omega_n + \ve_\alpha - \mu} 
 \right] \; ,
 \label{Green_free}
\ee
where 
\bea
\la x | \omega_n,\alpha \ra &=& \phi_\alpha(\x) {e^{-i\omega_n \tau_x} \over \sqrt{\beta}} \; , \label{1ptle_basis} \\
\omega_n &=& {\pi(2n+1) \over \beta} \; , \label{fermion_freq }\\
{\rm and~~~~}\hat h_{\x} \, \phi_\alpha(\x) = \left[ -\frac{\nabla^2}{2m} + \upsilon (\x) \right] \phi_\alpha(\x)  &=& (\ve_\alpha - \mu) \phi_\alpha(\x) \label{Eigen_1ptle} \; .
\eea
Eq.~(\ref{Eigen_1ptle}) implies that 
\be \label{KS_orbital}
\left[ -\frac{\nabla^2}{2m} + \upsilon_{\rm ion} (\x) + J_0(\x) \right] \phi_\alpha(\x) = \ve_\alpha  \phi_\alpha(\x) \;.
\ee 
Only frequencies of the type ${\pi(2n+1) \over \beta}$ is included in the expansion to ensure the antiperiodicity of 
the fermionic Green's function. 
To proceed further in (\ref{Green_free}), one may sum the frequency by introducing a function $-\beta/(e^{\beta \omega} + 1)$ which
  has poles at $\omega =  {i\pi(2n+1) \over \beta}$ with residue $1$. Evidently, poles for the function $-\beta/(e^{\beta \omega} + 1)$ 
 occur whenever $e^{\beta \omega} + 1 = 0 $. To investigate the strength of each pole, let's rewrite $e^{\beta \omega} + 1$ in
 the limit when $\omega \to i\pi(2n+1)/\beta$ as
\bea
&& \hspace{-0.4in}1+ \exp \left( i\pi (2n+1) + \beta (\omega - {i\pi (2n+1) \over \beta} ) \right)
= 1- \exp \left( \beta (\omega - {i\pi (2n+1) \over \beta} ) \right) \nonumber \\ 
&&  = -\beta \left( \omega - {i\pi (2n+1) \over \beta} \right) 
 + {\cal O} \left[\left( \omega - {i \pi (2n+1) \over \beta} \right)^2 \right] \; . \nonumber 
\eea
 Therefore, $-\beta/(e^{\beta \omega} + 1)$ indeed has residue strength $1$ at each of the allowable 
 frequencies. 

To evaluate ${1\over \beta} \sum_{\omega_n}  {e^{-i\omega_n(\tau_x-\tau_y)} \over -i\omega_n + \ve_\alpha - \mu}$ when $\tau_x < \tau_y$, 
 one integrates over a circular contour (with radius $|\omega| \to \infty$) on  the complex $\omega$-plane 
\[
{1\over \beta} \oint_{|\omega|\to \infty} {e^{-\omega (\tau_x-\tau_y)} \over -\omega + \ve_\alpha - \mu} {-\beta \over e^{\beta \omega} + 1}
 d\omega \; .
\]
Because the line integral along the infinite circle vanishes, the sum of residues must vanish, meaning that
\[ 
{1\over \beta} \sum_{\omega_n} {e^{-i\omega_n(\tau_x-\tau_y)} \over -i\omega_n + \ve_\alpha - \mu}
 + e^{-(\ve_\alpha-\mu)(\tau_x - \tau_y)}{1 \over e^{\beta (\ve_\alpha-\mu)} + 1 } = 0 \; .
\] 
Thus, when $\tau_x \le \tau_y$ (because when $\tau_x = \tau_y$, it must agree with $\tau_x - \tau_y \to 0^{-}$)
 one finds 
\[
G(x,y) = \sum_\alpha  \phi_\alpha(\x) \phi_\alpha^*(\y) e^{-(\ve_\alpha-\mu)(\tau_x - \tau_y)} (-n_\alpha) \; .
\] 
 On the other hand, when $\tau_x > \tau_y$, one considers
\[
{1\over \beta} \oint_{|\omega|\to \infty} {e^{\omega (\tau_x-\tau_y)} \over \omega + \ve_\alpha - \mu} {-\beta \over e^{\beta \omega} + 1}
 d\omega \; .
\] The residue sum then turns into 
\[
{1\over \beta} \sum_{\omega_n} {e^{-i\omega_n(\tau_x-\tau_y)} \over -i\omega_n + \ve_\alpha - \mu}
 - e^{-(\ve_\alpha-\mu)(\tau_x - \tau_y)}{1 \over e^{-\beta (\ve_\alpha-\mu) } + 1 } = 0 \; .
\]
Thus, when $\tau_x > \tau_y$ one  obtains
\[
G(x,y) = \sum_\alpha  \phi_\alpha(\x) \phi_\alpha^*(\y) e^{-(\ve_\alpha-\mu)(\tau_x - \tau_y)} (1-n_\alpha) \; .
\]
Therefore, we have 
\be
G(x,y) = \sum_\alpha \phi_\alpha(\x) \phi_\alpha^*(\y) e^{-(\ve_\alpha-\mu)(\tau_x - \tau_y)} \left\{ \begin{array}{l r}
(-n_\alpha) & {\rm if~} \tau_x \le \tau_y \\
(1-n_\alpha) & {\rm if~} \tau_x > \tau_y \end{array} \right. \; \; ,\label{Green.orbital}
\ee
where $n_\alpha = 1/(e^{\beta (\ve_\alpha-\mu)} + 1)$. Note that in the expression involving $\ve_\alpha$, 
it is always $\ve_\alpha$ minus the chemical potential $\mu$. 

To evaluate $\PH_0(x,y)$, we need to solve the eingensystem (\ref{Eigen_1ptle}).
 It requires evaluations of the RHS of Eq.~(\ref{steepest_descent})
  and self-consistency is reached when $\kappa(x) \to 0$. Of course, one cannot evaluate all the terms 
 and must truncate the series on the RHS of Eq.~(\ref{steepest_descent}) at some stage.   
  This implies that the density profile obtained through the self-consistency condition in this manner depends on the 
   number of terms one includes on the  RHS of Eq.~(\ref{steepest_descent}). Nevertheless, the self-consistent 
    solution obtained when keeping $k$ terms on the RHS of Eq.~(\ref{steepest_descent}) can be used as the starting  
    point when one wishes to include $k+1$ terms on the RHS of Eq.~(\ref{steepest_descent}).

\subsection{Functional derivative of $\PH_0(x,x')$ and $ \int d\tau_y \;\PH_{0}(x,y)\PH_{0}(y,x')$ }
\label{subsec:FDPH0}
 Especially when $J_0$ is time-independent, we need to evaluate
\bea
\frac{\delta \PH_0(x,x')}{\delta  J_0 (\y)} &=& - \int  \PH_0(x,z) \,
\left[\frac{\delta \PH_0^{-1}(z,z')}{\delta J_0(\y)} 
 \right]\,\PH_0(z',x') \; dz dz' \nonumber \\
 &=& - \int  \PH_0(x,z) \,\delta (z-z')  \delta(\y-\z) \, \PH_0(z',x') \; dz dz' \nonumber \\
 &=&  - \int_0^\beta   \PH_0(x,y) \, \PH_0(y,x') \, d\tau_y  , \label{dgdJ}
\eea
where the central expression $ \int_0^\beta   \PH_0(x,y) \, \PH_0(y,x') \, d\tau_y$ may be re-expressed by single-particle orbitals as will be shown below.

From Eq.~(\ref{Green_free}), we see that 
\bea
G(x,y) &=& \frac{1}{\beta} 
\sum_{n,\alpha} \frac{e^{-i\omega_n(\tau_x-\tau_y-\eta)}}{-i\omega_n+\ve_\alpha -\mu } \phi_\alpha(\x)\phi^*_\alpha(\y) \; , \nonumber \\
{\rm and~~~~}G(y,x') &=&  \frac{1}{\beta}
\sum_{n',\rho} \frac{e^{-i\omega_{n'}(\tau_y-\tau_{x'}-\eta)}}{-i\omega_{n'}+\ve_\rho -\mu} \phi_\rho(\y) \phi^*_\rho(\x')  \; , \nonumber 
\eea
and therefore 
\bea
\int_0^\beta d\tau_y \, G(x,y)\,G(y,x') &=& \frac{\beta}{\beta^2} \sum_{n,n',\alpha, \rho}  
\frac{\phi_\alpha(\x)\phi_\alpha^*(\y) \phi_\rho(\y)\phi_\rho^*(\x') }
{(-i\omega_n+\ve_\alpha-\mu)(-i\omega_{n'}+\ve_\rho-\mu)} \delta_{n,n'}\, e^{-i\omega_n(\tau_x-\tau_{x'}-2\eta)} \nonumber \\
&& \hspace*{-30pt} =\sum_{\alpha ,\rho} \left(\frac{1}{\beta} \sum_n \frac{e^{-i\omega_n(\tau_x-\tau_{x'}-2\eta)}}
  {(-i\omega_n+\ve_\alpha-\mu)(-i\omega_{n}+\ve_\rho-\mu)}\right)  \phi_\alpha(\x)\phi_\alpha^*(\y) \phi_\rho(\y)\phi_\rho^*(\x') \nonumber  \; .
\eea
Let us now concentrate on the portion inside the parentheses. Assuming that there is no energy degeneracy, 
we rewrite the denominator of this fctor when $\alpha\ne \rho $
\[
\frac{1}{(-i\omega_n+\ve_\alpha-\mu)(-i\omega_{n}+\ve_\rho-\mu)} = 
 \frac{1}{\ve_\rho-\ve_\alpha} \left( \frac{1}{-i\omega_n+ \ve_\alpha-\mu} - \frac{1}{-i\omega_n+\ve_\rho-\mu}\right)
\]
and when $\alpha = \rho $
\[
\frac{1}{(-i\omega_n+\ve_\alpha-\mu)(-i\omega_{n}+\ve_\rho-\mu)} \to \frac{1}{(-i\omega_n+\ve_\alpha-\mu)^2} =-\frac{\partial}{\partial \ve_\alpha} \left( \frac{1}{-i\omega_n+\ve_\alpha-\mu} \right) \; .
\]
Therefore, we need to evaluate $\sum_n \frac{e^{i\omega_n(\tau_{x'}-\tau_x)}}{-i\omega_n+\ve_\alpha-\mu}$.

We distinguish the two cases: $\tau_x \le \tau_{x'}$ and $\tau_x > \tau_{x'}$.
 When $\tau_x \le \tau_{x'}$, we consider the following integral over the infinite circle
\[
\oint \frac{e^{\omega (\tau_{x'}-\tau_x)}}{-\omega + \ve_\alpha - \mu}
 \frac{-\beta}{e^{\beta \omega} + 1} \frac{d\omega }{2\pi i} = \sum_n \frac{e^{i\omega_n(\tau_{x'}-\tau_x)}}{-i\omega_n+\ve_\alpha-\mu}
 + \frac{\beta e^{(\ve_\alpha-\mu)(\tau_{x'}-\tau_x)}}{e^{\beta (\ve_\alpha - \mu) } + 1} \; .
\]  
Since the integral over the infinite  circle vanishes, we have 
\be
\frac{1}{\beta}\sum_n \frac{e^{i\omega_n(\tau_{x'}-\tau_x})}{-i\omega_n+\ve_\alpha - \mu} 
= - \frac{ e^{-(\ve_\alpha - \mu)(\tau_{x}-\tau_{x'})}}{e^{\beta (\ve_\alpha-\mu) } + 1} = - e^{-(\ve_\alpha-\mu)(\tau_x - \tau_{x'})} n_\alpha \;.
\ee
To evaluate the expression $\frac{1}{\beta} \sum_n \frac{e^{i\omega_n(\tau_{x'}-\tau_x)}}{(-i\omega_n+\ve_\alpha - \mu)^2}$, we consider 
\[
-\frac{\partial}{\partial \ve_\alpha } \left(\frac{1}{\beta} \sum_n \frac{e^{i\omega_n(\tau_{x'}-\tau_x})}{-i\omega_n+\ve_\alpha - \mu} \right)=  e^{-(\ve_\alpha-\mu)(\tau_x-\tau_{x'})} \left[ -\beta n_\alpha (1-n_\alpha) - (\tau_x - \tau_{x'}) n_\alpha \right] \; .
\]

Therefore, when $\tau_x \le \tau_{x'}$, 
\bea
\int_0^\beta \!\! d\tau_y \, G(x,y)\,G(y,x') & = &\sum_\alpha \phi_\alpha(\x)n_\alpha(\y)\phi_\alpha^*(\x')
 e^{-(\ve_\alpha-\mu)(\tau_x-\tau_{x'})} \left[ -\beta n_\alpha (1-n_\alpha) - (\tau_x - \tau_{x'}) n_\alpha \right] \nonumber \\
&& \hspace*{-10pt} + \sum_{\alpha\ne \rho} \phi_\alpha(\x)\phi_\alpha^*(\y) \phi_\rho(\y)\phi_\rho^*(\x') \left[ 
\frac{e^{-\ve_\rho(\tau_x-\tau_{x'})} n_\rho- e^{-\ve_\alpha(\tau_x-\tau_{x'})} n_\alpha}{\ve_\rho-\ve_\alpha}\right] \; , \label{tau_lessthan_taup}
\eea
where $n_\alpha(\y) = \phi_\alpha^*(\y)\phi_\alpha(\y)$.

On the other hand, when $\tau_x > \tau_{x'}$, we consider the integral 
\[
\oint \frac{e^{\omega (\tau_{x}-\tau_{x'})}}{\omega + \ve_\alpha -\mu}
 \frac{-\beta}{e^{\beta \omega} + 1} \frac{d\omega }{2\pi i} = \sum_n \frac{e^{-i\omega_n(\tau_{x}-\tau_{x'})}}{-i\omega_n+\ve_\alpha-\mu}
 - \frac{\beta e^{-(\ve_\alpha-\mu)(\tau_{x}-\tau_{x'})}}{e^{-\beta (\ve_\alpha-\mu)} + 1} \; ,
\]  
and obtain (since the integral over the infinite circle vanishes)
\be
\frac{1}{\beta}\sum_n \frac{e^{-i\omega_n(\tau_{x}-\tau_{x'})}}{-i\omega_n+\ve_\alpha-\mu}
= \frac{ e^{-(\ve_\alpha-\mu)(\tau_{x}-\tau_{x'})}}{e^{-\beta (\ve_\alpha-\mu)} + 1}
 = e^{-(\ve_\alpha-\mu)(\tau_x - \tau_{x'})}(1- n_\alpha ) \; .
\ee
Similarly, 
\[
-\frac{\partial}{\partial \ve_\alpha} \left( \frac{1}{\beta}\sum_n \frac{e^{-i\omega_n(\tau_{x}-\tau_{x'}})}{-i\omega_n+\ve_\alpha-\mu} \right)= 
  e^{-(\ve_\alpha-\mu)(\tau_x-\tau_{x'})} \left[ -\beta n_\alpha (1-n_\alpha) + (\tau_x - \tau_{x'}) (1- n_\alpha) \right] \; .
\]
Therefore, when $\tau_x > \tau_{x'}$, 
\bea
\int_0^\beta \!\! d\tau_y \, G(x,y)\,G(y,x') & = &\sum_\alpha \phi_\alpha(\x)n_\alpha(\y)\phi_\alpha^*(\x')
 e^{-(\ve_\alpha-\mu)(\tau_x-\tau_{x'})} \left[ -\beta n_\alpha (1-n_\alpha) + (\tau_x - \tau_{x'}) (1-n_\alpha) \right] \nonumber \\
&& \hspace*{-24pt}+ \sum_{\alpha\ne \rho} \phi_\alpha(\x)\phi_\alpha^*(\y) \phi_\rho(\y)\phi_\rho^*(\x')  \left[ 
\frac{ e^{-(\ve_\rho-\mu)(\tau_x-\tau_{x'})} n_\rho- e^{-(\ve_\alpha-\mu)(\tau_x-\tau_{x'})}n_\alpha}{\ve_\rho-\ve_\alpha}\right] \; .\label{tau_greaterthan_taup}
\eea

We may now write down the full expression for $\delta \PH_0(x,x')/\delta J_0(\y)$ as  
\bea
\frac{\delta \PH_0(x,x')}{\delta J_0(\y)} &=& - \int_0^\beta   \PH_0(x,y) \, \PH_0(y,x') \, d\tau_y  \nonumber \\
&=& \sum_\alpha \phi_\alpha(\x)\, n_\alpha(\y)\, \phi_\alpha^*(\x')\, 
e^{-(\ve_\alpha-\mu)(\tau_x-\tau_{x'})} \left[ \beta n_\alpha (1-n_\alpha) + (\tau_x - \tau_{x'}) n_\alpha \right] \nonumber \\
&& - (\tau_x - \tau_{x'})\, \theta(\tau_x - \tau_{x'}-\eta) \sum_\alpha \phi_\alpha(\x)\, n_\alpha(\y)\, \phi_\alpha^*(\x')
\, e^{-(\ve_\alpha-\mu)(\tau_x-\tau_{x'})} \nonumber \\
&&  - \sum_{\alpha\ne \rho} \phi_\alpha(\x)\phi_\alpha^*(\y) \phi_\rho(\y)\phi_\rho^*(\x')  \left[ 
\frac{e^{-(\ve_\rho-\mu)(\tau_x-\tau_{x'})} n_\rho - e^{-(\ve_\alpha-\mu)(\tau_x-\tau_{x'})} n_\alpha}{\ve_\rho-\ve_\alpha}\right]
\; .  \label{dPHdJ}
\eea

In the absence of time-dependence, we have
\[
n(\x) = \frac{1}{\beta} \frac{\delta (\beta W[J_0])}{\delta J_0(\x)} 
 = - \frac{1}{\beta} \!\int\!\! dz dy \,\PH_0(z,y) \delta (\y-\x) \delta (y-z) = -\PH_0(x,x) \; .
\]
Therefore, using Eq.~(\ref{dPHdJ}) we have 
\bea
D_0(\x,\y) = \frac{\delta n(\x)}{\delta J_0(\y)}  &=&  \sum_{\alpha\ne \rho} \phi_\alpha(\x)\phi_\alpha^*(\y)
 \phi_\rho(\y)\phi_\rho^*(\x)  \left[ 
\frac{n_\rho-n_\alpha}{\ve_\rho-\ve_\alpha}\right] \nonumber \\
&& + \sum_\alpha n_\alpha(\x)\, n_\alpha(\y)\, 
 \left[ \beta n_\alpha (1-n_\alpha) \right] \nonumber\;.
\eea
Note that the expression $\beta n_\alpha(1-n_\alpha)$ vanishes as $\beta \to \infty$,
 because $n_\alpha(1-n_\alpha)$ decays exponentially with $\beta$ when $\mu \ne \ve_\alpha$. 
 That is, at the zero temperature limit, one has 
 \[
D_0(\x,\y) \rightarrow  \sum_{\alpha\ne \rho} \phi_\alpha(\x)\phi_\alpha^*(\y)
 \phi_\rho(\y)\phi_\rho^*(\x)  \left[ 
\frac{n_\rho-n_\alpha}{\ve_\rho-\ve_\alpha}\right]
 \] 
as long as $\mu$ is not equal to any eigenenergy of the orbital.

\subsection{The Computational Procedure}
\label{subsec:recipe}
The recipe for computation goes as follows. Starting with a reasonably guessed $J_0(\x)$, one first 
obtains single particle wave functions $\phi_\alpha$ and energies $\ve_\alpha$ through (\ref{KS_orbital}). 
 Note that the occupation number of state $\alpha$ is given by 
\[
 n_\alpha = \frac{1}{e^{\beta (\ve_\alpha - \mu)} + 1 } 
\]
 and the chemical potential is chosen such that 
\[
g_s \sum_\alpha n_\alpha = N_e \; ,
\] 
where $g_s$ denotes the spin degeneracy ($g_s = 2$ for spin $1/2$ electrons).

Through eqs.~(\ref{Green.orbital}), (\ref{D0}), and (\ref{tDI}), 
 one constructs respectively the $\PH_0$, $D_0^{-1}$, and
 $\tilde {\mathcal D}_0$ propagators. Combined with their differentiation rules (\ref{dgdJ.tdep}-\ref{dDdJ.tdep}) 
 with respect to $J_0$, these propagators are used 
to compute $\delta \Gamma /\delta J_0$. 

As the third step, one obtains a new estimate of $J_0$ given by $J_0 - \varsigma \, \delta \Gamma/\delta J_0$
  with $\varsigma >0$ being the step size. See eq.~(\ref{steepest_descent}) and the text nearby for details.

Finally, one starts again with the new $J_0$
   and goes through the other steps, iteratively until the convergence condition $\delta \Gamma/ \delta J_0 = 0$ is reached.
 This simultaneously determines $J_0$ (sum of the Hartree potential and the KS potential), the ground state charge density 
  and the ground state energy, as well as the KS orbitals and energies.

\section{Case Studies} \label{sec:case_studies}
\subsection{The Universal Functional ${\mathcal F}[n]$ at Arbitrary Temperature}
 There are vast discussions~\cite{FKSY_94,VF_97_arxiv,PS_02} 
  on the equivalence between $E_\upsilon[n]$ in (\ref{enfunc}-\ref{first.func}) and 
 $\lim_{\beta \to \infty} \frac{1}{\beta}\Gamma[n]$. However, not much attention was given to
  the emergence of the universal functional ${\mathcal F}[n]$ (at any given temperature) resulting 
   from the effective action formalism. We address here for the first time 
   how ${\mathcal F}[n]$ arises naturally from our effective action formulation.

Fukuda {\it et al.}~\cite{FKSY_94} showed that it is possible to eliminate the $\upsilon$ dependence in the
 functional to obtain (when translated into our terms)
 \[
 \Gamma_\upsilon [i\varphi] = \Gamma_{\upsilon = 0}[i\varphi] + \upsilon \4di (i\varphi) + 
 \frac{1}{2}\upsilon \4di \, U^{-1} \4di \upsilon \; .
 \]
 They interpret $i\varphi$ as some sort of electron density since it coincides with the real electron density
  when $J=0$. The appearance of the quadratic term in $\upsilon$, however, disagrees with (\ref{first.func})
   where it is clearly stated that in addition to a term that is linear in 
   $\upsilon$, the remainder should be $\upsilon$-independent.  We settle this discrepancy below by showing 
   that the appearance of the quadratic $\upsilon$ dependence is due to the fact that reference~\onlinecite{FKSY_94} 
   did not use the system's electron density as the natural variable. 
   Once the system's electron density is used as the natural variable, the universal functional
    ${\mathcal F}[n]$ emerges naturally.

Note that (\ref{H_J}) indicates that one may view $\upsilon_{\rm ion} (\x)$ as a nonvanishing source term. 
 In the change of variable $\phi \to \phi + i U^{-1} \4di J$ right after~(\ref{GJ}), if we make 
  $\phi \to \phi + i U^{-1} \4di (J + \upsilon)$ instead, we see that $\upsilon$ is then bonded with $J$ from that point on.
  That said, we may view $W[J] \equiv W'[J+\upsilon]$. That is, the generating functional in the presence of the 
  one-body potential $\upsilon$ with source $J$  is equivalent to the generating functional of a system without a 
  one-body potential but with source $J + \upsilon$. Following the algebra in Eqs.~(\ref{genfunc.3}-\ref{W.varphi.3}), one 
   sees that 
 \be \label{J+v} 
 \beta W[J] = \beta W'_\phi[J+\upsilon]  - \frac{1}{2} (J + \upsilon) \4di \, U^{-1} (J+\upsilon) \; , 
 \ee   
where 
\bea
\beta W'_\phi [J+\upsilon] &=&  \frac{1}{2} \varphi \4di \, U \4di \varphi
- \Tr \ln \left( \bar G_{\varphi}^{-1} \right) + i (J+\upsilon) \4di \varphi \nonumber \\
&& + \frac{1}{2} \Tr \ln \left( \tilde {\mathcal D}^{-1} \4di \, U \right) 
 - \sum_{n=1}^\infty \frac{1}{n!} \la \left[ \sum_{k=3}^\infty
 I^{(k)}[\varphi] \4di \,b_1 \ldots \4di\, b_k  \right]^n \ra_{\rm 1PI,~conn.} \;  \nonumber \\
\bar G_\varphi^{-1}(x,x') &=& \left( \partial_\tau -\frac{\nabla ^2}{2m}
 -\mu  + i (U \4di \varphi)_x  \right)\delta(x-x') \nonumber 
 \; .
\eea

Upon using the following variant of (\ref{n2phiJ4d})  
\[
n = i\varphi - U^{-1} \4di (J + \upsilon)  \; ,
\] 
the quadratic term in $J+\upsilon$ in (\ref{J+v}) gets absorbed into the density of the electron 
and one arrives at the following variant of (\ref{W.varphi.2})
\bea
\beta W[J] = \beta W' [ J+\upsilon ]  &=& 
- \Tr \ln \left( \bar G_{\varphi}^{-1} \right)   - \frac{1}{2} n \4di \, U \4di n 
+ \frac{1}{2} \Tr \ln \left( \tilde{\mathcal D}^{-1} \4di \, U \right) \nonumber \\
&& \hspace*{15pt}  - \sum_{n=1}^\infty \frac{1}{n!} \la \left[ \sum_{k=3}^\infty
 I^{(k)}[\varphi] \4di \,b_1 \ldots \4di\, b_k  \right]^n \ra_{\rm 1PI,~conn.} \;.
 \nonumber
\eea
Note that here $n = n_J = n'_{J+\upsilon}$ represents the electron density of the system. 
 
The effective action $\Gamma[n] = \beta W[J] - J \4di n$ can thus be rewritten as
\[
\Gamma[n] = \beta W'[J+\upsilon] - J \4di n = \upsilon \4di n + \beta W'[J+\upsilon] - (J+\upsilon)\4di n \; ,
\]
where the first term on the RHS represents $\beta \int n(\x) \upsilon(\x) d\x$, and the last two terms represent
 the effective action, $\Gamma'[n] = \beta W' [J+\upsilon] - (J+\upsilon) \4di n$, 
 of a system in the absence of one-body potential.  We thus identify ${\mathcal F}[n]$ as
 \[
{\mathcal F}[n]= \frac{1}{\beta} \left\{ \beta W'[J+\upsilon] - (J+\upsilon) \4di n \right\} \; .
 \] 
This also serves as an alternative exposition of what Mermin proved.\cite{Mermin_65}

\subsection{Effective Potential near Zero Temperature}
\label{sec:zero.T}
The effective potential divided by $\beta$ 
 is the ground state energy plus $\mu N_e$ in the $T \to 0$ limit. 
 Below, we will show how the Hartree-Fock terms appear in this formulation. 

Equations~(\ref{W.loop}-\ref{J2}) provide a systematic expansion for calculating the effective potential. 
 In the expression (\ref{tGamma.0.def}), the term $- \Tr \ln (\PH_0^{-1})$ is equivalent to
 $- \ln [\det (\PH_0^{-1})]$. There are many ways to obtain $\det(\PH_0^{-1})$: one may obtain the results directly through
 the definition of $e^{-\beta W_0[J_0]}$ or one may express  $e^{-\beta W_0[J_0]}$ as a path integral to obtain the determinant in
 discrete time. We shall denote $-\Tr \ln (\PH_0^{-1})$ by $\beta W_0[J_0]$ with
\bean
e^{-\beta W_0[J_0]} &=& \Tr \left\{ e^{-\beta \left[ \hpsid \3di (\hat h + J_0) \3di \hpsi \right]} \right\}
 = \int {\cal D}[\psi^\dag, \psi] e^{- \int dx \psi^\dag(x) \left[ \partial_\tau + \hat h + J_0(\x) \right] \psi(x)}  \\
&=& \int {\cal D}[\psi^\dag, \psi] e^{- \psi^\dag \4di \PH_0^{-1} \4di \psi } = \det (\PH_0^{-1}) \; .
\eean
Direct evaluation using the trace definition leads to 
\[
\det (\PH_0^{-1}) = \prod_\alpha  \left( 1+ e^{-\beta (\ve_\alpha-\mu) } \right) \; ,
\]
and consequently 
\[
W_0[J_0] =  -{1\over \beta} \Tr \ln \left( \PH_0^{-1} \right) =  -{1\over \beta} \sum_\alpha \ln \left( 1+ e^{-\beta (\ve_\alpha-\mu) } \right)  = {1\over \beta} \sum_\alpha \ln \left( 1-n_\alpha \right) \; .
\]

Note that the energy is measured with respect to the chemical potential $\mu$. Therefore, in the limit of $\beta \to \infty$, we have
\be \label{W0.low.T} 
\lim_{\beta \to \infty}W_0 [J_0] = - \lim_{\beta \to \infty} {1\over \beta} \Tr \ln \left( \PH_0^{-1} \right)
 = \sum_{\alpha = 1}^{N_e} (\ve_\alpha - \mu) \; ,
\ee
with $\ve_\alpha \le \mu$ for $1\le \alpha \le N_e$.

Using Eqs.~(\ref{Gn.1PI}) and (\ref{W0.low.T}), we obtain the low temperature 
limit of the effective action 
\bea 
\lim_{\beta \to \infty} \left( {1\over \beta} \Gamma [n] \right) & = & 
 \sum_{\alpha = 1}^{N_e} (\ve_\alpha - \mu) - J_0 \3di n + \frac{1}{2} n \3di \, U \3di n  \nonumber \\
&& \hspace*{12pt} 
 + \lim_{\beta \to \infty} \left[ \frac{1}{2\beta }\Tr \ln \left( \tilde {\cal D}_0^{-1} \4di \, U \right)
 + \frac{1}{\beta}\sum_{i=2}^\infty \Gamma_i [n] \right] \; . \label{Gamma.low.T.0}
\eea
The first term in (\ref{Gamma.low.T.0}) can be expressed in a different way 
 if we multiply both sides of Eq.~(\ref{Eigen_1ptle}) by $\phi_\alpha^*(\x)$, sum
 $\alpha$ over the lowest $N_e$ states, and integrate over $\x$. Upon doing this, we
  obtain 
\bea 
\sum_{\alpha  = 1}^{N_e} (\ve_\alpha - \mu) &=&
\sum_{\alpha =1}^{N_e} \int d\x \phi_\alpha^*(\x) \left[ -\frac{\nabla^2}{2m} + \upsilon_{\rm ion}(\x) 
 - \mu + J_0(\x) \right] \phi_\alpha(\x) \nonumber \\
& \equiv& T_0[n] -\mu N_e + \int d\x \left[ \upsilon_{\rm ion}(\x) + J_0 (\x) \right]\, n(\x)  \; , \nonumber
\eea  
where 
\[
n(\x) = \sum_{\alpha =1}^{N_e} \phi_\alpha^*(\x) \phi_\alpha(\x)  \; ,
\]
and 
\[
T_0[n] \equiv \sum_{\alpha =1}^{N_e} \int d\x \phi_\alpha^*(\x) \left[ -\frac{\nabla^2}{2m}  \right] \phi_\alpha(\x) \; .
\]
Therefore, the first two terms in (\ref{Gamma.low.T.0}) give us 
\[
\sum_{\alpha = 1}^{N_e} (\ve_\alpha - \mu) - J_0 \3di n = T_0[n] -\mu N_e + \int d\x \upsilon_{\rm ion}(\x)\, n(\x) \; .
\]
Evidently, the third term in (\ref{Gamma.low.T.0}) is nothing but the Hartree energy
\[
\frac{1}{2} n \3di \, U \3di n = \frac{1}{2} \iint d\x d\y \, n(\x) \frac{e^2 }{|\x-\y|} n (\y) \; .
\]
Therefore, we have 
\bea  
\lim_{\beta \to \infty} \left( {1\over \beta} \Gamma [n] \right)&=& T_0[n] + \int \upsilon_{\rm ion}(\x) \, n(\x) \,  d\x
 - \mu N_e  + \frac{1}{2} \iint d\x d\y \, n(\x) \frac{e^2 }{|\x-\y|} n (\y) \nonumber \\
 && + \lim_{\beta \to \infty}   \frac{1}{2\beta}\Tr \ln \left( {\mathbf I} - D_0 \4di \, U \right) 
 + \lim_{\beta \to \infty}  \left[ \frac{1}{\beta}\sum_{i=2}^\infty \Gamma_i [n] \right] 
 \label{Gamma.low.T} \; ,
\eea
with the last two terms combined to form the exchange-correlation energy functional when compared with
 the Kohn-Sham decomposition (\ref{KSfunctional}).

If $D_0 \4di \, U$ (or $e^2$) may be treated as a small quantity, one may expand  
 ${1\over 2\beta}\Tr \ln ( {\mathbf I} - D_0 \4di \, U) $ as 
\be \label{e_weak_exp}
{1\over 2\beta} \Tr \ln \left( {\mathbf I} - D_0 \4di \, U \right) 
 = -{1\over 2\beta} \sum_{n=1}^\infty \frac{ \Tr \; \left[ (D_0 \4di \, U)^n \right]}{n} \; ,
\ee
 with the leading term (in the static limit)
\[
{-1\over 2\beta} \int dx dy D_0(x,y) u(y,x) = {-e^2 \over 2} \int d\x d\y {n(\x,\y) n(\y,\x) \over |\x-\y|}  \; ,
\]
where $n(\x,\y) = \sum_{m=1}^{N_e} \phi_\alpha(\x)\phi_\alpha^*(\y) = -\PH_0(\x, \tau; \y, \tau)$. 
 Consequently, when $D_0\4di \, U$ (or $e^2$) can be treated as a small quantity, 
 we may write the leading terms of the effective potential as 
\bea
\lim_{\beta \to \infty} {1\over \beta} \Gamma [n]  &=&
  T_0[n] + \int \upsilon_{\rm ion} (\x) n(\x) d\x  -\mu N_e + {e^2\over 2} \int d\x d\y {n(\x) n(\y) - n(\x,\y)n(\y,\x) \over |\x-\y|} \nonumber \\
&& +  \lim_{\beta \to \infty}  \left[  {-1\over 2\beta} \sum_{n=2}^\infty \frac{1}{n} \Tr \; 
\left[ (D_0\4di \, U)^n \right] + \frac{1}{\beta}\sum_{i=2}^\infty \Gamma_i [n] \right] \; ,  \label{G.low.T.weak.e}
\eea 
where the fourth term is nothing but the Hartree-Fock term. The higher-order terms inside the square brackets
encode the remaining contributions of exchange and correlation. In the case when the Coulomb interaction is  strong,   
 one may choose not to use the expansion in Eq.~(\ref{e_weak_exp}), but to use (\ref{Gamma.low.T}) for the zero temperature limit and (\ref{Gn.1PI}) for finite temperature. 

\subsection{Single Electron at Zero Temperature}
\label{one_ptcle_limit}
When the system contains only one electron and when the energy difference 
 between the first excited state and the ground state is much larger than 
 $k_BT$, there will be no particle-hole pairs. Consequently, there should be no 
 exchange-correlation energy as well as the Hartree energy. 
 When one employs empirical density functionals, this feature
   is unlikely to be preserved -- an issue known as the self-interaction problem. 
 It is customary to define the exchange-correlation energy $E_{xc}$ as the sum
  of the Fock exchange energy $E_x$ and the correlation energy $E_c$. 
  Since it is easy to show that the exchange energy $E_x$ 
   exactly cancels the Hartree energy in the case of one electron, one easily concludes 
   that $E_c=0$ for the one electron case. This argument has been used, for example, by
  Perdew and Zunger.\cite{PZ_81} However, from the diagrammatic expansion
   point of view, the Hartree term and the Fock exchange term correspond only to the first
    order (in terms of $e^2$) diagrams. The cancellation of the first order terms does not imply
   that the higher order diagrams, making up $E_c$, will give no contribution. 
   That is, although the fact that $E_c = 0$ for one electron system can be argued, 
 a formal diagrammatic exposition incorporating all higher orders 
  is needed to justify the asymptotic exactness of the proposed functional.
   
  The purpose of this section is to illustrate how 
   $E_c=0$ can be derived formally within our formalism. It should be noted, however, that
  the self-interaction will remain if one truncate the series in eq.~(\ref{Gn.1PI}). 
  While the exchange-only functional will have no self-interaction problem, as we will
  show below it is not because the exchange-only functional is an approach with more 
   correct physics, but because the simplification it makes is equivalent to 
    completely ignoring the correlation energy.

When $N_e = 1$ and when $T \to 0$, we find from Eq.~(\ref{Green.orbital}) that 
the Green's function $\PH_0(x,y)$ takes the following form
\be
\PH_0(x,y) = \left\{ \begin{array}{l r} \phi_1(\x) \phi_1^*(\y) e^{-(\ve_1-\mu)(\tau_x - \tau_y)} 
(-1) & {\rm if~} \tau_x \le \tau_y \\
\sum\limits_{\alpha \ge 2} \phi_\alpha(\x) \phi_\alpha^*(\y) e^{-(\ve_\alpha-\mu)(\tau_x - \tau_y)}
 & {\rm if~} \tau_x > \tau_y \end{array} \right. \; , \label{Green.one.ptcle}
\ee
where $\ve_1 < \mu < \ve_{\alpha \ge 2}$ and
 $\phi_{1(\rho)}(\x)$ is the ground ($\rho^{\rm th}$) state wave function of the single particle Hamiltonian
\[
\hat h(\x) = -\frac{\nabla^2}{2m} + \upsilon_{\rm ion}(\x) + J_0(\x) - \mu \; 
\] 
with eigenvalue $\ve_{1(\rho)}-\mu$.

The disappearance of the Hartree energy and the exchange-correlation energy is best seen
 by grouping terms with the same number of Coulomb lines $U$. We will explicitly show the
 first few calculations followed by a sketch of the general proof.
  
Let us first show that the first term ($n=1$) on the RHS of (\ref{e_weak_exp})
cancels the Hartree term exactly. When there is only one electron, 
 the Hartree term becomes 
\[
\frac{e^2}{2}\int d\x d\y \frac{n(\x)n(\y)}{|\x-\y|}
=  \frac{e^2}{2}\int d\x d\y 
\frac{\phi_1(\x)\phi_1^*(\x)\phi_1(\y)\phi_1^*(\y)}{|\x-\y|} \;.
\] 
The Fock term ($n=1$ term of order $U$ in (\ref{e_weak_exp})),  
when there is only one electron, reads, 
\[
{-1\over 2\beta} \Tr \left[ D_0 \4di \, U \right] = 
{-1\over 2\beta} \int dx dy D_0(x,y) U(y,x) = {-e^2 \over 2} \int d\x d\y {\phi_1(\x)\phi_1^*(\x)\phi_1(\y)\phi_1^*(\y) \over |\x-\y|}  \; .
\]
The cancellation between the Fock term and the Hartree term is thus apparent. 
If one were to approximate the exchange-correlation functional by the Fock exchange
 functional, correlation energy can never be accounted for. That is, this 
  approximation coincidently leads to the expected result $E_c = 0$ at the single 
  electron limit simply because it never takes $E_c$ into account.

In terms of diagrammatic expression, the Hartree term is given by 
 diagram (a) of Fig.~\ref{fig:HF}, while the lowest order exchange term
  ($n=1$ term of (\ref{e_weak_exp}) ) corresponds to diagram (b) of Fig.~\ref{fig:HF}.
 That is, the order $U$ diagrams in $\frac{1}{\beta}\Gamma[n]$ are identical to the order $U$ diagrams in
  $W[J_0]$ at zero temperature. 
 The perfect cancellation between the Hartree term and the lowest order exchange term
 implies that the sum of these two terms remains zero when one makes a derivative 
  with respect to either $J_0$ or $n$. Diagrammatically speaking, this means  that
  when  the system contains only one electron, we have 
 \be \label{1e.HF} 
 \left\{ \begin{array} {l} 
 0 = \frac{\delta 0}{\delta J_0 }  \\
 0 = \frac{\delta 0}{\delta n } \end{array} \right.
  =   \left\{ \begin{array} {l} 
 \frac{\delta}{\delta J_0 }  \\
 \frac{\delta}{\delta n } \end{array} \right. \left[ \begin{picture}(136,20)(0,17)
\Text(0,20)[lc]{\large $\frac{1}{2}$}
\Arc[arrow,arrowpos=0.50,arrowlength=3.0,arrowwidth=0.8,arrowinset=0.2](25,20)(12,0,360)
\Line[dash,dashsize=1.5](37,20)(50,20)
\Vertex(37,20){1.0} 
\Vertex(50,20){1.0}
\Arc[arrow,arrowpos=0.50,arrowlength=3.0,arrowwidth=0.8,arrowinset=0.2](62,20)(12,-180,180)
\Text(84,20)[lc]{\large $-\frac{1}{2}$} 
\Arc[arrow,arrowpos=0.50,arrowlength=3.0,arrowwidth=0.8,arrowinset=0.2](120,20)(14,-90,90)
\Arc[arrow,arrowpos=0.50,arrowlength=3.0,arrowwidth=0.8,arrowinset=0.2](120,20)(14,90,-90)
\Line[dash,dashsize=1.5](120,6)(120,34)
\Vertex(120,6){1.0} 
\Vertex(120,34){1.0}
\end{picture} \right] \; .
 \ee

Terms of the next order in $U$ are contained in $\Gamma_{i\le 2}$. See Fig.~\ref{fig:OE4} for a diagrammatic expression 
 of $\Gamma_{i\le 2}$ of order $U^2$ (or of order $e^4$). We show below that in Fig.~\ref{fig:OE4} the first two 
 graphs cancel each other and the last two diagrams also cancel each other. For illustration, we will work out
  the explicit cancellation of the first two graphs by labelling the space-time points. The cancellation of the  
 the last two graphs require additional operations which we will turn to later. The first two graphs in Fig.~\ref{fig:OE4}
  with the space-time points labelled appear to be
\bea
&& 
\begin{picture}(200,55)(-20,15)
\put(-20,45){\makebox(0,0){\large $\frac{1}{4}$}}
\put(-7,45){\makebox(0,0){$x_1$}}
\put(49,45){\makebox(0,0){$x_2$}}
\put(21,70){\makebox(0,0){$y_1$}}
\put(21,20){\makebox(0,0){$y_2$}}
\put(0,45){\begin{picture}(40,20)(0,0)
\Arc[arrow,arrowpos=0.50,arrowlength=3.0,arrowwidth=0.8,arrowinset=0.2](20,0)(20,0,90)
\Arc[arrow,arrowpos=0.50,arrowlength=3.0,arrowwidth=0.8,arrowinset=0.2](20,0)(20,90,180)
\Arc[arrow,arrowpos=0.50,arrowlength=3.0,arrowwidth=0.8,arrowinset=0.2](20,0)(20,180,270)
\Arc[arrow,arrowpos=0.50,arrowlength=3.0,arrowwidth=0.8,arrowinset=0.2](20,0)(20,270,360)
\Vertex(20,-20){1.0}
\Vertex(20,20){1.0}
\Vertex(0,0){1.0}
\Vertex(40,0){1.0} 
\Line[dash,dashsize=1.5](20,20)(20,-20)
\Line[dash,dashsize=1.5](0,0)(40,0) 
\end{picture}
}
\put(70,45){\makebox(0,0){\large $- \frac{1}{4}$}}
\put(106,20){\makebox(0,0){$x_1$}}
\put(106,70){\makebox(0,0){$y_1$}}
\put(156,20){\makebox(0,0){$x_2$}}
\put(156,70){\makebox(0,0){$y_2$}}
\put(85,45){\begin{picture}(90,20)(0,0)
\Arc[arrow,arrowpos=0.50,arrowlength=3.0,arrowwidth=0.8,arrowinset=0.2,clock](20,0)(20,-90,90)
\Arc[arrow,arrowpos=0.50,arrowlength=3.0,arrowwidth=0.8,arrowinset=0.2,clock](20,0)(20,90,-90)
\Line[dash,dashsize=1.5](20,-20)(70,-20)
\Arc[arrow,arrowpos=0.50,arrowlength=3.0,arrowwidth=0.8,arrowinset=0.2](70,0)(20,-90,90)
\Arc[arrow,arrowpos=0.50,arrowlength=3.0,arrowwidth=0.8,arrowinset=0.2](70,0)(20,90,-90)
\Line[dash,dashsize=1.5](20,20)(70,20)
\Vertex(70,-20){1.0}
\Vertex(70,20){1.0}
\Vertex(20,-20){1.0}
\Vertex(20,20){1.0}
\end{picture}
}
\end{picture}
\nonumber \\
&& \hspace*{2pt} = \int\!\! d\tau_x\! d\tau_y \! \prod_{i=1}^2 d\x_i d\y_i 
\frac{\PH_0(x_1,y_1)(\PH_0(x_2,y_2)}{4|\x_1-\x_2| |\y_1-\y_2|}  \times \nonumber \\
&& \hspace*{10pt} \times
\large[ \PH_0(y_1,x_2)\PH_0(y_2,x_1) -\PH_0(y_1,x_1)\PH_0(y_2,x_2) \large]  \; . 
\label{e4.cancel.1}
\eea  
We have used $\tau_x$ to denote the time associated with both $\x_1$ and $\x_2$, while 
 denoting by $\tau_y$ the time associated with both $\y_1$ and $\y_2$.
When $\tau_y \le \tau_x$, the quantity inside the square brackets of Eq.~(\ref{e4.cancel.1})   
 vanishes because
 \[
 \PH_0(y_i,x_j) \propto \phi_1(\y_i)\phi^*_1(\x_j) \; , 
 \]
and thus $\PH_0(y_1,x_2)\PH_0(y_2,x_1) =  \PH_0(y_1,x_1)\PH_0(y_2,x_2) $.  When $\tau_y > \tau_x$,
 we simply swap the dummy variable $y_1$ and $y_2$ in the second graph above to arrive at
\[
\int\!\! d\tau_x\! d\tau_y \! \prod_{i=1}^2 d\x_i d\y_i 
\frac{\PH_0(y_1,x_2)(\PH_0(y_2,x_1)}{4|\x_1-\x_2| |\y_1-\y_2|}
\large[ \PH_0(x_1,y_1)\PH_0(x_2,y_1) -\PH_0(x_1,y_2)\PH_0(x_2,y_1) \large]  \; . 
\]
And  again the quantity inside the square brackets vanishes due to the same reason as before. 
As for the cancellation of the last two graphs in Fig.~\ref{fig:OE4}, we first use the bottom portion
 of (\ref{1e.HF}) to obtain
\[
0 = \frac{\delta}{\delta n}
\left[ \begin{picture}(136,20)(0,17)
\Text(0,20)[lc]{\large $\frac{1}{2}$}
\Arc[arrow,arrowpos=0.50,arrowlength=3.0,arrowwidth=0.8,arrowinset=0.2](25,20)(12,0,360)
\Line[dash,dashsize=1.5](37,20)(50,20)
\Vertex(37,20){1.0} 
\Vertex(50,20){1.0}
\Arc[arrow,arrowpos=0.50,arrowlength=3.0,arrowwidth=0.8,arrowinset=0.2](62,20)(12,-180,180)
\Text(84,20)[lc]{\large $-\frac{1}{2}$} 
\Arc[arrow,arrowpos=0.50,arrowlength=3.0,arrowwidth=0.8,arrowinset=0.2](120,20)(14,-90,90)
\Arc[arrow,arrowpos=0.50,arrowlength=3.0,arrowwidth=0.8,arrowinset=0.2](120,20)(14,90,-90)
\Line[dash,dashsize=1.5](120,6)(120,34)
\Vertex(120,6){1.0} 
\Vertex(120,34){1.0}
\end{picture} \right] = 
\begin{picture}(116,20)(0,17)
\Text(0,20)[lc]{\large $-$}
\Line[dash,dashsize=1.5](17,20)(30,20)
\Vertex(17,20){1.8} 
\Vertex(30,20){1.0}
\Arc[arrow,arrowpos=0.50,arrowlength=3.0,arrowwidth=0.8,arrowinset=0.2](42,20)(12,-180,180)
\Text(60,20)[lc]{\large $+$} 
\Line[double](77,20)(90,20)
\Arc[arrow,arrowpos=0.50,arrowlength=3.0,arrowwidth=0.8,arrowinset=0.2](102,20)(12,-90,90)
\Arc[arrow,arrowpos=0.50,arrowlength=3.0,arrowwidth=0.8,arrowinset=0.2](102,20)(12,90,180)
\Arc[arrow,arrowpos=0.50,arrowlength=3.0,arrowwidth=0.8,arrowinset=0.2](102,20)(12,-180,-90)
\Line[dash,dashsize=1.5](102,8)(102,32)
\Vertex(102,8){1.0} 
\Vertex(102,32){1.0}
\Vertex(77,20){1.8}
\Vertex(90,20){1.0}
\end{picture}  \; ,
\] 
and then
\[
0 = \frac{1}{2}\left[
\begin{picture}(116,20)(0,17)
\Text(0,20)[lc]{\large $-$}
\Line[dash,dashsize=1.5](41,20)(54,20)
\Vertex(54,20){1.8} 
\Vertex(41,20){1.0}
\Arc[arrow,arrowpos=0.50,arrowlength=3.0,arrowwidth=0.8,arrowinset=0.2,flip](29,20)(12,360,0)
\Text(60,20)[lc]{\large $+$} 
\Line[double](101,20)(114,20)
\Arc[arrow,arrowpos=0.50,arrowlength=3.0,arrowwidth=0.8,arrowinset=0.2,clock](89,20)(12,-90,90)
\Arc[arrow,arrowpos=0.50,arrowlength=3.0,arrowwidth=0.8,arrowinset=0.2,clock](89,20)(12,90,0)
\Arc[arrow,arrowpos=0.50,arrowlength=3.0,arrowwidth=0.8,arrowinset=0.2,clock](89,20)(12,0,-90)
\Line[dash,dashsize=1.5](89,8)(89,32)
\Vertex(89,8){1.0} 
\Vertex(89,32){1.0}
\Vertex(114,20){1.8}
\Vertex(101,20){1.0}
\end{picture}
\right] 
\left( \begin{picture}(17,8)(0,0)
\Line[double](0,2.5)(17,2.5)
\Vertex(0,2.5){1.0} \Vertex(17,2.5){1.0} 
\end{picture}\right)^{-1}
\left[
\begin{picture}(116,20)(0,17)
\Text(0,20)[lc]{\large $-$}
\Line[dash,dashsize=1.5](17,20)(30,20)
\Vertex(17,20){1.8} 
\Vertex(30,20){1.0}
\Arc[arrow,arrowpos=0.50,arrowlength=3.0,arrowwidth=0.8,arrowinset=0.2](42,20)(12,-180,180)
\Text(60,20)[lc]{\large $+$} 
\Line[double](77,20)(90,20)
\Arc[arrow,arrowpos=0.50,arrowlength=3.0,arrowwidth=0.8,arrowinset=0.2](102,20)(12,-90,90)
\Arc[arrow,arrowpos=0.50,arrowlength=3.0,arrowwidth=0.8,arrowinset=0.2](102,20)(12,90,180)
\Arc[arrow,arrowpos=0.50,arrowlength=3.0,arrowwidth=0.8,arrowinset=0.2](102,20)(12,-180,-90)
\Line[dash,dashsize=1.5](102,8)(102,32)
\Vertex(102,8){1.0} 
\Vertex(102,32){1.0}
\Vertex(77,20){1.8}
\Vertex(90,20){1.0}
\end{picture} 
\right] \; ,
\] 
or
\be \label{1e.Id.1}
- \frac{1}{2} \; \; 
\begin{picture}(60,20)(78,17)
\Line[double](102,20)(114,20)
\Arc[arrow,arrowpos=0.50,arrowlength=3.0,arrowwidth=0.8,arrowinset=0.2,clock](90,20)(12,-90,90)
\Arc[arrow,arrowpos=0.50,arrowlength=3.0,arrowwidth=0.8,arrowinset=0.2,clock](90,20)(12,90,0)
\Arc[arrow,arrowpos=0.50,arrowlength=3.0,arrowwidth=0.8,arrowinset=0.2,clock](90,20)(12,0,-90)
\Line[dash,dashsize=1.5](90,8)(90,32)
\Vertex(90,8){1.0} 
\Vertex(90,32){1.0}
\Vertex(114,20){1.0}
\Vertex(102,20){1.0}
\Arc[arrow,arrowpos=0.50,arrowlength=3.0,arrowwidth=0.8,arrowinset=0.2](126,20)(12,-90,90)
\Arc[arrow,arrowpos=0.50,arrowlength=3.0,arrowwidth=0.8,arrowinset=0.2](126,20)(12,90,180)
\Arc[arrow,arrowpos=0.50,arrowlength=3.0,arrowwidth=0.8,arrowinset=0.2](126,20)(12,-180,-90)
\Line[dash,dashsize=1.5](126,8)(126,32)
\Vertex(126,8){1.0} 
\Vertex(126,32){1.0}
\end{picture} \;\;
=\;  \frac{1}{2}\;\; \begin{picture}(96,20)(0,-3)
\Arc[arrow,arrowpos=0.50,arrowlength=3.0,arrowwidth=0.8,arrowinset=0.2](12,0)(12,0,360)
\Arc[arrow,arrowpos=0.50,arrowlength=3.0,arrowwidth=0.8,arrowinset=0.2](48,0)(12,0,180)
\Arc[arrow,arrowpos=0.50,arrowlength=3.0,arrowwidth=0.8,arrowinset=0.2](48,0)(12,180,360)
\Arc[arrow,arrowpos=0.50,arrowlength=3.0,arrowwidth=0.8,arrowinset=0.2](84,0)(12,-180,180)
\Vertex(24,0){1.0}
\Vertex(36,0){1.0}
\Vertex(60,0){1.0}
\Vertex(72,0){1.0} 
\Line[dash,dashsize=1.5](24,0)(36,0)
\Line[dash,dashsize=1.5](60,0)(72,0) 
\end{picture} \; \; -\;
\begin{picture}(60,20)(0,-3)
\Arc[arrow,arrowpos=0.50,arrowlength=3.0,arrowwidth=0.8,arrowinset=0.2,clock](12,0)(12,0,360)
\Arc[arrow,arrowpos=0.50,arrowlength=3.0,arrowwidth=0.8,arrowinset=0.2](48,0)(12,-90,90)
\Arc[arrow,arrowpos=0.50,arrowlength=3.0,arrowwidth=0.8,arrowinset=0.2](48,0)(12,90,180)
\Arc[arrow,arrowpos=0.50,arrowlength=3.0,arrowwidth=0.8,arrowinset=0.2](48,0)(12,180,270)
\Line[dash,dashsize=1.5](48,-12)(48,12)
\Vertex(48,-12){1.0}
\Vertex(48,12){1.0}
\Vertex(24,0){1.0}
\Vertex(36,0){1.0}
\Line[dash,dashsize=1.5](24,0)(36,0)
\end{picture} \; ,  
\ee
when there is only one electron in the system.

Equation (\ref{1e.Id.1}) means that at the single electron limit, the last two graphs in Fig.~\ref{fig:OE4} 
 are equivalent to the last three graphs in Fig.~\ref{fig:VF_1}. In other words, at the single electron limit and
  at zero temperature, the set of order $U^2$ diagrams in $\Gamma[n]$ is equivalent to the set of order $U^2$ diagrams in
   $\beta W[J_0]$. We shall pause at this point and elucidate the general
  situation by looking at these cancellations via Hugenholtz diagrams.~\cite{Negele_Orland_book}

With a two-body interaction term such as the Coulomb interaction, 
 typical Feynman diagrams treat the direct (Coulomb) and exchange matrix element separately. 
  It is not surprising that one may simplify the calculation 
  by combining the direct and exchange matrix elements into a single antisymmetrized matrix element.
 The basic idea is to combine the following two scenarios into one
\bea
\begin{picture}(50,40)(-5,15)
\Line[arrow,arrowpos=0.50,arrowlength=3.0,arrowwidth=0.8,arrowinset=0.2](0,0)(10,20)
\Line[arrow,arrowpos=0.50,arrowlength=3.0,arrowwidth=0.8,arrowinset=0.2](10,20)(0,40)
\Line[arrow,arrowpos=0.50,arrowlength=3.0,arrowwidth=0.8,arrowinset=0.2](40,0)(30,20)
\Line[arrow,arrowpos=0.50,arrowlength=3.0,arrowwidth=0.8,arrowinset=0.2](30,20)(40,40)
\Line[dash,dashsize=1.5](10,20)(30,20)
\Text(-5,-5)[lc]{$\alpha$} 
\Text(45,-5)[rc]{$\rho$}
\Text(-5,45)[lc]{$\gamma$} 
\Text(45,45)[rc]{$\delta$}
\end{picture}
-
\begin{picture}(50,40)(-5,15)
\Line[arrow,arrowpos=0.50,arrowlength=3.0,arrowwidth=0.8,arrowinset=0.2](0,0)(30,20)
\Line[arrow,arrowpos=0.50,arrowlength=3.0,arrowwidth=0.8,arrowinset=0.2](10,20)(0,40)
\Line[arrow,arrowpos=0.50,arrowlength=3.0,arrowwidth=0.8,arrowinset=0.2](40,0)(21.5,12.333)
\Line[](18.5,14.33)(10,20)
\Line[arrow,arrowpos=0.50,arrowlength=3.0,arrowwidth=0.8,arrowinset=0.2](30,20)(40,40)
\Line[dash,dashsize=1.5](10,20)(30,20)
\Text(-5,-5)[lc]{$\alpha$} 
\Text(45,-5)[rc]{$\rho$}
\Text(-5,45)[lc]{$\gamma$} 
\Text(45,45)[rc]{$\delta$}
\end{picture}
&\equiv &
\begin{picture}(50,40)(-5,15)
\Line[arrow,arrowpos=0.50,arrowlength=3.0,arrowwidth=0.8,arrowinset=0.2](0,0)(20,20)
\Line[arrow,arrowpos=0.50,arrowlength=3.0,arrowwidth=0.8,arrowinset=0.2](20,20)(0,40)
\Line[arrow,arrowpos=0.50,arrowlength=3.0,arrowwidth=0.8,arrowinset=0.2](40,0)(20,20)
\Line[arrow,arrowpos=0.50,arrowlength=3.0,arrowwidth=0.8,arrowinset=0.2](20,20)(40,40)
\Vertex(20,20){3.0}
\Text(-5,-5)[lc]{$\alpha$} 
\Text(45,-5)[rc]{$\rho$}
\Text(-5,45)[lc]{$\gamma$} 
\Text(45,45)[rc]{$\delta$}
\end{picture} \nonumber \\
&& \nonumber \\
\left(\gamma \delta \vert \upsilon \vert \alpha \rho \right) - \left(\gamma \delta \vert \upsilon \vert \rho \alpha \right)
& \equiv & \{ \gamma \delta \vert \upsilon \vert \alpha \rho \}\; ,
\eea 
where 
\[
\left(\gamma \delta \vert \upsilon \vert \alpha \rho \right) \equiv 
\int d\tau_x d\tau_y d\x d\y \phi_\gamma^*(\x) \phi^*_\delta(\y) \upsilon(x,y) \phi_\alpha (\x) \phi_\rho (\y) \; , 
\] 
and
\[
\left\{ \gamma \delta \vert \upsilon \vert \alpha \rho \right\} \equiv 
\int d\tau_x d\tau_y d\x d\y \phi_\gamma^*(\x) \phi^*_\delta(\y) \upsilon(x,y) \left[ \phi_\alpha (\x) \phi_\rho (\y)
- \phi_\rho (\x) \phi_\alpha (\y)
\right]   \; , 
\]  
with $\phi_\alpha (\x)$ being the single-particle wave function described earlier in Eq.~(\ref{Eigen_1ptle}).
 The  resulting diagrams with bullet dots as the new vertices are called Hugenholtz diagrams. 
The appearance of those vertex matrix elements comes from the following. When one evaluates a 
 Feynman diagram, a propagator $\PH_0(x,x')$ going from $x'$ to $x$ connects two vertices located at $x'$ and $x$ 
  respectively. From Eq.~(\ref{Green.orbital}), we know that 
  \[
  \PH_0(x,x') = \sum_\alpha \phi_\alpha(\x) \phi_\alpha^*(\x') e^{-(\ve_\alpha-\mu)(\tau_x - \tau_{x'})} \left\{ \begin{array}{l r}
(-n_\alpha) & {\rm if~} \tau_x \le \tau_{x'} \\
(1-n_\alpha) & {\rm if~} \tau_x > \tau_{x'} \end{array} \right.   \; .
  \] 
Upon integration of the space-time coordinates, we see that $\phi_\alpha(\x)$ will be integrated with 
vertex $x$, whither the propagator leads, while  $\phi^*_\alpha(\x')$ will be integrated with 
vertex $x'$, whence the propagator originates. Therefore, when evaluating a Feynman diagram,
 one may associate each propagator going from time $\tau'$ to time $\tau$  with a state index $\alpha$ with 
\[ 
 \PH_0(\alpha, \tau-\tau') = e^{-(\ve_\alpha - \mu) (\tau-\tau') } \left[ (1-n_\alpha) \theta (\tau-\tau'-\eta)
 - n_\alpha \theta(\tau'-\tau+\eta) \right]  \; .
\] 
Each vertex will contribute a numerical factor equivalent to its vertex matrix element.
Note that each vertex carries a time index. At zero temperature, for a vertex with time index $\tau$,  
if the two incoming lines originate from vertices with times larger than or equal to $\tau$, 
 the vertex matrix element associated with $\tau$ becomes zero in the single electron limit. 
 This is because both incoming lines each carry only the $\alpha = 1$ index. 
 Upon antisymmetrization, the Hugenholtz vertex matrix element becomes zero. 
  
For an arbitrary Hugenholtz diagram with $n$ vertices, one may always name their time
   index such that 
  $\tau_1 \le \tau_2 \le \tau_3 \le \cdots \le \tau_n$. In this case, the vertex associated with $\tau_1$ gives 
   zero matrix element since its two incoming lines must come from two other time indices that 
    are larger than or equal to $\tau_1$. Consequently, each Hugenholtz diagram yields value zero at zero 
    temperature when there is only one electron present.    
 As a matter of fact, the order $U$ diagrams of $W[J_0]$ are represented by the Hugenholtz diagram
 $
 \begin{picture}(24,6)(0,-3)
\Arc[arrow,arrowpos=0.50,arrowlength=3.0,arrowwidth=0.8,arrowinset=0.2](6,0)(6,0,360)
\Arc[arrow,arrowpos=0.0,arrowlength=3.0,arrowwidth=0.8,arrowinset=0.2](18,0)(6,0,360)
\Vertex(12,0){2.0}
\end{picture}
 $    .
 On the other hand, the first two graphs of Fig.~\ref{fig:OE4} (order $U^2$ terms of $\Gamma[n]$) 
 correspond to the Hugenholtz diagram
 $
 \begin{picture}(20,5)(0,-3)
\Arc[arrow,arrowpos=0.50,arrowlength=3.0,arrowwidth=0.8,arrowinset=0.2](10,0)(10,-90,90)
\Arc[arrow,arrowpos=0.50,arrowlength=3.0,arrowwidth=0.8,arrowinset=0.2,clock](10,0)(10,-90,90)
\Arc[arrow,arrowpos=0.50,arrowlength=3.0,arrowwidth=0.8,arrowinset=0.2,flip](0,0)(14.142,-45,45)
\Arc[arrow,arrowpos=0.50,arrowlength=3.0,arrowwidth=0.8,arrowinset=0.2,clock,flip](20,0)(14.142,-135,135)
\Vertex(10,10){2.0}
\Vertex(10,-10){2.0}
\end{picture}
$\; ,
 while the last two graphs of Fig.~\ref{fig:OE4} (order $U^2$ terms of $\Gamma[n]$)
  are equivalent to the last three graphs of Fig.~\ref{fig:VF_1} (order $U^2$ terms of $\beta W[J_0]$)
  and correspond to the Hugenholtz diagram 
 $ 
\begin{picture}(38,10)(0,-3)
\Arc[arrow,arrowpos=0.50,arrowlength=3.0,arrowwidth=0.8,arrowinset=0.2](6,0)(6,0,360)
\Arc[arrow,arrowpos=0.0,arrowlength=3.0,arrowwidth=0.8,arrowinset=0.2](38,0)(6,0,360)
\Arc[arrow,arrowpos=0.50,arrowlength=3.0,arrowwidth=0.8,arrowinset=0.2,flip](22,-10)(14.142,45,135)
\Arc[arrow,arrowpos=0.50,arrowlength=3.0,arrowwidth=0.8,arrowinset=0.2,flip](22,10)(14.142,-135,-45)
\Vertex(12,0){2.0}
\Vertex(32,0){2.0}
\end{picture}
$ \;  (see reference~\onlinecite{Negele_Orland_book}).
 Therefore, we see that all the order $U^2$ terms cancel each other out in the effective action $\Gamma[n]$. 

If one were to expand the effective action (\ref{Gn.1PI}) in powers of $U$, our approach reduces to performing the inversion
 method using $e^2$ as the expansion parameter.\cite{VF_97_arxiv,FKYetal_95} In this case, $W_{l\ge 1}[J_0]$ 
 in Eqs.~(\ref{W.loop}-\ref{J2}) contains exactly all the order $U^l$ diagrams in $W[J_0]$ and can
  be expressed as Hugenholtz diagrams of order $U^l$ as well. Since all the Hugenholtz diagrams give value zero,
 all the derivatives of $W_l$ (all the order $u^l$ diagrams) with respect to the density vanish as well. 
  This implies that all the $J_{l\ge 1}$ vanish and consequently in the effective action the sum of Hartree energy and
   the exchange-correlation energy equals zero at zero temperature when there is only one electron present.     
  
The perfect cancellation of the Hartree energy and exchange-correlation energy means that one is the negative of the other.
 This means that $\sum_{i=1}^\infty \Gamma_i = -\frac{1}{2} n \4di \, U \4di n$. Employing Eq.~(\ref{ex.general.1})
  for the most general case, we obtain
 \[
\tilde J_0 - J  = \frac{\delta \left( \sum_{i=1} \Gamma_i[n] \right) } {\delta n} 
=  -\frac{1}{2}\frac{\delta \left( n \4di \, U \4di n \right) } {\delta n } = - U\4di n \; .
 \]
Since $J_0 = \tilde J_0 + U\4di n$, we find that $J_0 =  J $ at the single electron limit at zero temperature.    
The fact that $J_0 = J $ means that the final effective potential $\ve_1-\mu - J \3di n $ is nothing but the lowest eigenenergy 
 of the following single-particle Hamiltonian
\[
\hat h(\x) = -\frac{\nabla^2}{2m} + \upsilon_{\rm ion}(\x) - \mu + J  
\] 
less the expectation value of $J$, which is exactly what one expected.

\subsection{Screening of Coulomb Interaction}
\label{sec:screening}
In the physical limit, $n = i\varphi$. In our formulation, lumps of charge fluctuation 
 around  the configuration $i\varphi$ interact with each other via 
\[
U\4di \tilde {\cal D}_0^{-1} \4di \, U =  \left( U - U \4di D_0 \4di \, U\right)\;.
\]
As will be described in the discussion section, 
 $U\4di (i \phi ) = ib$ plays the role of the photon field here. Therefore,  
\[
-\phi \4di \left( U - U \4di D_0 \4di \, U\right) \4di \phi = (i\phi \4di \, U) \4di \tilde {\mathcal D}_0^{-1} \4di (U\4di i\phi) 
 = (ib) \4di \left( U^{-1} - D_0 \right) \4di (ib) \; .
\]
Since $U(x,y)= \delta (\tau_x - \tau_y) U(\x,\y)$ and $\PH_0(x,y)$ only depends on the 
 relative time difference $\tau_x - \tau_y $, we expect $D_0(x,y)$ 
 to depend only on $\tau \equiv \tau_x - \tau_y$. Furthermore, since $\PH_0(x,y)$ is antiperiodic
 in $\tau_x$, $\tau_y$, and $\tau_x-\tau_y$, one expects that $\PH_0(x,y) \PH_0(y,x)$ to 
 be periodic in $\tau_x-\tau_y$.

 Introducing the spatial momenta $\p$ and $\q$ conjugate to
 the spatial variables $\x$ and $\y$, we may write 
\[
\left[U^{-1} - D_0 \right](\nu_n,\p,\q) = \int e^{i\p \x + i\q \y} d\x d\y \int_0^\beta \, d \tau 
\;  e^{i\nu_n \tau} \left[ U^{-1}(\x,\y) \delta (\tau) - \PH_0(x,y) \PH_0(y,x) \right] \; .
\]

When $U(\x,\y) = {e^2 \over |\x-\y|}$, we find that $U^{-1}(x,y) = -{1\over 4\pi e^2} 
 \delta (\tau_x-\tau_y) \nabla^2_\x \delta (\x-\y)$ and 
\[ 
U^{-1}(\nu_n,\p,\q) = {(2\pi)^3\over 4\pi e^2} \q^2 \delta(\p+\q) = {L^3\over 4\pi e^2} \q^2 \delta_{\p+\q}\; ,
\]
where $L^3$ is the spatial volume of the system.  
Let us now write $D_0(\nu_n,\p,\q)$
 as
\[
D_0(\nu_n,\p,\q) = \int e^{i\p \x + i\q \y} d\x d\y \int_0^\beta \, d \tau 
\;  e^{i\nu_n \tau} \, \PH_0(x,y) \PH_0(y,x) \; .
\]

In momentum space, $U^{-1} \propto {\q}^2 \delta (\p+\q)$ and it is well known that this type of Coulomb interaction
 leads to infrared divergence (that occurs near ${\q} \to 0$). In the presence of $-D_0$, one may wish to calculate the zero momentum contribution of $-D_0$. Let us define  
 $D_0(\nu_n) \equiv D_0(\nu_n,\p=0,\q=0)$. Using Eq.~(\ref{Green.orbital}), we find
\be
D_0(\nu_n) =  - \sum_{\alpha,\alpha'} \delta_{\alpha,\alpha'} \int_0^\beta d\tau e^{i\nu_n \tau} \left[ n_\alpha (1-n_{\alpha'}) \right]
 = -\beta \delta_{\nu_n,0} \sum_\alpha n_\alpha(1-n_\alpha) \; .
\ee
From Eq.~(\ref{dPHdJ}), we see that when $x=x'$,  
\[
\beta \sum_\alpha n_\alpha (1-n_\alpha) = -\int d\x \frac{\partial \PH_0(x,x)}{\partial \mu} = \frac{\partial N_e}{\partial \mu} \; .
\]
We therefore have 
\[
D_0(\nu_n) = - \delta_{\nu_n,0} {L^3} \frac{\partial \bar n}{\partial \mu} \; ,
\]
where $\bar n = N_e/L^3$ is the average electron density.

Therefore at the stationary limit, where $\nu_n = 0$, the inverse propagator with nearly zero momentum  
 behaves as
\[
L^3  \left[ \frac{q^2}{4\pi e^2}  + \frac{\partial \bar n}{\partial \mu} \right] 
 \Rightarrow \frac{L^3}{4\pi e^2}\left[ q^2 +  4\pi e^2 \frac{\partial \bar n}{\partial \mu} \right] \; ,
\]
analogous to the Thomas-Fermi results for electric charge screening.  

Except during the intermediate steps of computing the excitation spectrum, one deals with
 the time independent system.  Note that $ 4\pi e^2 \partial \bar n/\partial \mu$  plays the role of $m^2$
  in the screened potential $e^{-mr}/r$.  This shows that the static interaction between charge fluctuations 
   in the long wave length limit is a screened one instead of the bare Coulomb interaction. 
   It should be noted that the use of this screened propagator dates back half a century. 
 DuBois~\cite{DuBois_59a} replaced the bare Coulomb interaction by the screened interaction 
  in his study of electron interactions. Hedin~\cite{Hedin_65} also used it to replace the
     bare Coulomb interaction in the expansion of the Luttinger-Ward functional,~\cite{LW_60}
   resulting in the so-called GW approximation. The differences among the three mentioned approaches should
  be described. In both DuBois's and Hedin's formalism, they use the {\it full} polarization of the
  interacting system. The difference between their formalism is in the electron propagators employed.
   In DuBois's approach, the free electron propagator is used, while in Hedin's method, 
  similar to the Luttinger-Ward functional,~\cite{LW_60} the {\it full}
   electron propagator is employed. In our approach, it is the polarization of the KS system that
  enters the calculation and the electron propagator entering the 
   diagrammatic calculation is the noninteracting KS Green's function.

\subsection{Homogeneous Electron Gas} 

Being the foundation for the LDA of the DFT, 
the homogeneous electron gas (HEG) is also a simplified model 
for 
a metal or a plasma. The HEG system is an interacting electron gas placed in
  a uniformly distributed positive background chosen to ensure that the 
   total system is neutral. Due to the translational symmetry of the HEG system, 
   the one-particle propagator functions only depend on the coordinate difference between  
  two variables (space-time points)  instead of on both variables. 
  Consequently, each propagator (Green's function) carries a definite momentum 
  even if the Coulomb interaction among electrons is fully taken into 
   consideration. 
   
 Investigation of the HEG system dates back to 1930s by 
 Wigner,\cite{Wigner_34} who coined the term ``correlation energy" to represent the 
 ground state energy per electron after subtracting away the average kinetic
  and exchange energy. (The Hartree energy is cancelled exactly by the interaction between 
  electrons and the positive background, and by the Coulomb energy among the uniform positive charges.)  
  Apparently, after choosing the Rydberg (the negative of the Hydrogen atom
  ground state energy) as the energy unit, one may express either the correlation energy
   or the ground state energy of the HEG system 
  in terms of the dimensionless quantity $r_s \equiv r_0/r_B$,
  where ${4\pi \over 3} r_0^3 = \frac{V}{N_e}$, $V$ being the volume of the system considered, $N_e$ 
   being the total number of electrons, and $r_B = \frac{\hbar^2}{m e^2}$ representing the Bohr radius.  
 In fact, efforts have been invested to express the correlation energy as a power
 series in $r_s$ (and $r_s \ln r_s$ as well) at both the high density and low density limits.     
 Lacking {\it real} experimental results to compare with, however, 
 it is hard to assess 
 how much improvement each increasing order of $r_s$ (or $1/r_s$) can bring.

Since $r_s \propto 1/r_B \propto e^2$, a small $r_s$ expansion is most naturally performed
 by treating the Coulomb interaction as a perturbation.  
 Equivalent to the method of computing the 
 vacuum amplitude,\cite{FW_71,JM_73,Negele_Orland_book}  
  the time-ordered perturbative expansion developed by  
 Goldstone~\cite{Goldstone_57} formalized the Brueckner theory~\cite{BL_55} and
 made a direct connection to diagrammatic expansion in calculation of
  the ground state energy. The electron propagator in diagrams under this formalism is 
  that of the noninteracting electrons.
  An alternative formalism to compute the ground state energy (or the grand potential) is
  to utilize the full propagator (i.e., the one with self-energy included)   
 in diagrammatic calculation. Under this alternative formalism, 
 only two-particle irreducible diagrams contribute. 
 Indeed, the work of Luttinger and Ward~\cite{LW_60} (as well as that of Klein~\cite{Klein_61}) 
  was exactly along this line. Since the self-energy is not known {\it a priori}, this type of
   energetic expression is a functional of the self-energy (or the full Green's function), which must be  
   determined via a stationary condition.\cite{LW_60}

 An equivalent of the Brueckner-Goldstone formalism~\cite{Goldstone_57,BG_58}
  was used by Gell-Mann and Brueckner~\cite{GB_57} to compute
  the correlation energy of the HEG system. Given the long-range nature of the Coulomb interaction, 
  it is not surprising that Gell-Mann and Brueckner identified the occurrence of divergence as early as in 
  the second order of $e^2$-based perturbative calculation. To circumvent this unphysical divergence, 
  they summed an infinite subset of diagrams to arrive at a contribution proportional to $\ln r_s$.  
To eliminate unphysical divergence, DuBois~\cite{DuBois_59a} replaced the bare Coulomb interaction
  by the screened one.  Starting with calculating the vacuum amplitude,
   he expressed the ground state energy in terms of integration of the {\it full} 
   electronic polarization over the Coulomb coupling strength, using a version of the Feynman-Hellmann theorem. 
  While the full polarization is used, the electron propagator entering DuBois's formalism
   is the free electron propagator instead of the full propagator. 
  Unfortunately, DuBois made a mistake in extracting the higher order terms  
 of the ground state energy at the high density (small $r_s$) limit. 
  Although this error was later corrected by Carr and Maradudin,\cite{CM_64} according to
  Hedin,\cite{Hedin_65} this high density expansion violated   
  at moderate $r_s$ values Ferrell's condition,\cite{Ferrell_58} which is based on the simple 
  fact that the second order perturbation contribution
   to the ground state energy is always negative.   
   
Except for using the screened Coulomb interaction to replace the bare one and working directly 
at zero temperature, Hedin's approach is largely similar to the finite-temperature formalism 
of Luttinger and Ward~\cite{LW_60} in the sense that the {\it full} electron propagator is used 
as the fundamental variable. As a consequence, the higher order diagrams contributing to the full  
   polarization appear different in DuBois's work and in Hedin's formalism. 
    Instead of solving his own self-consistent equations, however, Hedin approximated the 
   {\it full} Green's function within his formalism by the {\it non-interacting} Green's function 
   to compute the energetics of the HEG system. Nevertheless, it is still possible to
  maintain the exactness within Hedin's approach if two-particle reducible diagrams are  
   properly included. 
  We will provide an example of such a two-particle reducible diagram 
  that would not be included in Hedin's approximate calculation but should be incorporated
  for theoretical soundness. 
 
Another important issue in obtaining the ground state energy of a many-electron system
 has to do with whether the finite temperature formalism (setting ${V \to \infty}$ first followed by ${T\to 0}$)
  or the zero temperature formalism (setting $T \to 0$ and then followed by $V \to \infty$) is used.    
  In terms of diagrammatic expansion, there exist diagrams (termed {\it anomalous} diagrams 
  by Luttinger {\it et al.}\cite{KL_60,LW_60})  that are present (giving finite contribution) 
  within the finite temperature formalism but are 
  absent (giving zero contribution) within the zero temperature formalism.   
 To be specific, a diagram is anomalous if within it 
 there exist two electron propagators linking  
 two different times (say $\tau_1$, $\tau_2$ and of course $\tau_1 \ne \tau_2$)  
  in the opposite orders: $\PH(\x,\tau_1,\y,\tau_2) \PH({\mathbf w},\tau_2, {\mathbf z},\tau_1)$.   
  
 Zero temperature methods typically rely on the Gell-Mann and Low theorem,~\cite{GL_51} which assures that 
  adiabatic transformation of the noninteracting ground state by gradually switching on the 
 interaction leads to an eigenstate of the interacting system.\cite{FW_71,Negele_Orland_book} 
 The energetic difference of this eigenstate and the noninteracting ground state is what
   the zero temperature formalism obtains. This approach therefore makes sense only if  
  the adiabatically transformed state is the ground state of the interacting system, which 
   occurs only if the noninteracting Fermi surface is identical to that of the interacting system.~\cite{JM_73}
 For the HEG system, the perturbed Fermi surface remains spherical (identical to that of 
 the unperturbed one) since the Coulomb interaction respects spherical symmetry and 
  the background positive charge distribution also has spherical symmetry. 
 Consequently, for the HEG system, the anomalous diagrams should end up giving no contribution. Indeed,  
    Luttinger {\it et al.}~\cite{KL_60,LW_60} illustrated how the contributions 
 from the anomalous diagrams in this case are cancelled by the chemical potential shift.
  Luttinger and Ward~\cite{LW_60} also showed how one may avoid anomalous contributions from appearing
  by expressing the grand potential as a sum of {\it all} possible linked diagrams (including two-particle reducible ones).
  The key step there is to subtract from each self-energy part a number, which is given by
   that self-energy part evaluated at the Fermi surface. 
  As we will illustrate later, under our finite temperature formalism it is not necessary to implement
  such an elaborate subtraction scheme because each two-particle reducible diagram is automatically  
   accompanied by another appropriate diagrammatic subtraction.   
   
What sets our self-consistent equation (\ref{OEP}) apart from that of Luttinger and Ward 
and of Hedin is the variable to be solved for. 
It is the KS potential, instead of the physical Green's function, that enters
 our exact, self-consistent equation. In general, one needs to first solve $\tilde J_0$ 
  self-consistently using eq.~(\ref{OEP}) prior to the evaluation of the 
  grand potential (or the ground state energy).  
 When limiting to a homogeneous system with constant electron density, however, 
 $\tilde J_0 = {\rm const.}$ becomes the only possibility. Whether or not such a choice 
  satisfies the self-consistent equation (\ref{OEP}) depends on whether 
  a uniform electron density truly represent the lowest energy configuration. 
  For the present purpose of considering a high density HEG, 
  we assume a constant $\tilde J_0$ to proceed. 
 Because a constant $\tilde J_0$ can be easily absorbed into the chemical potential, 
 the resulting KS Green's function carries a corresponding 
 energy that consists of kinetic energy only. That is, rather than being an approximation as in Hedin's case, 
  the single-particle Green's function carrying only kinetic energy represents exactly the self-consistent
   KS Green's function under our formalism. Consequently, for the HEG system, the
    grand potential (or the ground state energy) may be calculated using  eq.~(\ref{Gn.1PI}). 
In the remaining part of this section, we will first show how our formalism naturally avoids 
 divergence and how one may use it to obtain the ground state energy of the HEG
  as a series in $r_s$ (and $\ln r_s$). We will then illustrate with an explicit example 
   how the anomalous contributions are cancelled within our formalism, followed by 
  a brief description of diagrams that would be missed within Hedin's approximation~\cite{Hedin_65}
  in computing the ground state energy of the HEG system. To provide an easier comparison with
   existing results, we restore in this section the electron spins that have been suppressed 
   thus far to simplify the exposition. 
    
In our definition of the energy functional $E_\upsilon$, see eq.~(\ref{KSfunctional}),  
the $-\mu N_e$ term is included but the interaction among
 background charges is not included. Since most zero temperature formalism does not
  include the $-\mu N_e$ term and for the HEG system the interaction between background
 charges is also included, the ground state energy in the literature will
  correspond to
 \[
 \lim_{\beta \to \infty} \left[ \frac{1}{\beta} \Gamma [n] + \mu N_e + \frac{1}{2}
 n_{\rm bg} \3di \, U \3di n_{\rm bg} \right]   
 \]    
 in our formalism.   Eq.~(\ref{Gamma.low.T.0}) can then be used to arrive at
\bea 
E_g &=&  \lim_{\beta \to \infty} \left[ \frac{1}{\beta} \Gamma [n] + \mu N_e + \frac{1}{2}
 n_{\rm bg} \3di \, U \3di n_{\rm bg} \right]  \nonumber  \\
& = &
 \sum_{\alpha = 1}^{N_e} \ve_\alpha - J_0 \3di n + \frac{1}{2} n\3di \, U \3di n 
 + \frac{1}{2} n_{\rm bg} \3di \, U \3di n_{\rm bg} 
 + \lim_{\beta \to \infty} \left[ \frac{1}{2\beta }\Tr \ln \left( \tilde {\cal D}_0^{-1} \4di \, U \right)
 + \frac{1}{\beta}\sum_{i=2}^\infty \Gamma_i [n] \right] \nonumber \\
 &=&  \sum_{\alpha = 1}^{N_e} (\ve_\alpha - \tilde J_0) + 
 \lim_{\beta \to \infty} \left[ \frac{1}{2\beta }\Tr \ln \left( \tilde {\cal D}_0^{-1} \4di \, U \right)
 + \frac{1}{\beta}\sum_{i=2}^\infty \Gamma_i [n] \right]\; ,
\eea   
where $J_0 = \tilde J_0 + U \3di n$ is used and $n_{\rm bg} = n$ for the HEG system is also employed.  
 Note that the state label $\alpha = (\p, \sigma)$ includes both momentum and spin.  
 For the HEG system, $\upsilon_{\rm ion} + U \3di n = 0$ and thus eq.~(\ref{KS_orbital}) 
  leads to $\ve_\alpha = \p^2/2m + \tilde J_0$. Therefore, the ground state energy of the HEG system 
   may be written as
\be \label{Eg.HEG}
E_g = 2 \sum_{\p}^{|\p| \le p_F} \frac{\p^2}{2m} 
 + \lim_{\beta \to \infty} \frac{1}{\beta} \left[ \frac{1}{2}\Tr \ln \left( \tilde {\cal D}_0^{-1}
  \4di \, U \right) + \sum_{i=2}^\infty \Gamma_i [n] \right] \; ,
\ee   
where $p_F$ indicates the Fermi momentum and 
the factor of $2$ in the first term of the right hand side comes from noting that there are two
 spin states associated with each momentum. Furthermore, the KS Green's function 
  in this case may be written as
\be \label{Green.HEG} 
\PH_0(x,\sigma ;y, \sigma') = \frac{1}{\beta V } \sum_{\p,n} 
\frac{e^{-i\omega_n(\tau_x-\tau_y)} e^{-i\p\cdot(\x-\y)}}{-i\omega_n + \ve_\p + \tilde J_0 -\mu}
\, \delta_{\sigma,\sigma'} \, ,
\ee
where $\ve_\p = \p^2/2m$ is the kinetic energy of an electron carrying momentum $\p$.
 By absorbing the constant $\tilde J_0$ into the chemical potential in (\ref{Green.HEG}), 
 one sees that the KS Green's function in the HEG system is indeed the  
  free electron propagator and at the zero temperature, the new chemical potential
   (the original one subtracted by $\tilde J_0$) 
  simply becomes $p_F^2/2m$. 
 
The diagrammatic expression of our first correction term $\Gamma_1[n] = 
\frac{1}{2}\Tr \ln (\tilde {\mathcal{D}_0}^{-1} \4di \, U ) 
=\frac{1}{2} \Tr \ln \left[ {\mathbf I} - D_0\4di \, U \right] $  
 is shown in  Fig.~\ref{fig:polarization}. It can be seen that, 
 upon being divided by the inverse temperature $\beta$, 
  our $\Gamma_1[n]$ contains the exchange energy $\epsilon_x$ and  all the ring-like correlation energy $\epsilon'$ 
 discussed by Gell-Mann and Brueckner~\cite{GB_57} via the relation
  $\Gamma_1[n]/\beta = N_e(\epsilon_x + \epsilon')$, with $N_e$ being the total number of electrons.  
Since both $D_0$ and $U$ are diagonal in momentum space for the HEG system,  
  we may write 
\be   \label{Gamma.1.q}
\frac{1}{\beta} \Gamma_1[n] =  \frac{1}{2 \beta} \sum_{\q,\nu_n} \ln \left[ 1 - D_0(q,\nu_n)\, U(q) \right]
 = \frac{V}{2\beta} \int \frac{d\q}{(2\pi)^3} \sum_{\nu_n} \ln \left[ 1 - D_0(q,\nu_n) \, U(q) \right]   \; . 
\ee 
By comparing this equation to a derived result (eq.~(30.16) of reference~\onlinecite{FW_71}), 
 it is also evident that our $\Gamma_1[n]/(\beta N_e) $ indeed gives $\epsilon_x + \epsilon'$ of 
 reference~\onlinecite{GB_57} as $\beta \to \infty$. 
 In Fig.~\ref{fig:polarization}, except for the first diagram, all other diagrams
  when evaluated individually exhibit
  infrared divergence due to the piling up of $1/\q^2$ propagators.\cite{GB_57} 
  To obtain the leading contribution, one must pay particular attention to the small $|\q|$ region.   
 
The polarization $D_0 \equiv \delta n(x)/\delta J_0(y)$ is defined in eq.~(\ref{dndJ.tdep}). 
When the electron spins are included, $n(x) =  - \sum_\sigma \PH_0(x,\sigma;x,\sigma)$. Because
 the Coulomb interaction does not flip spins, the KS propagator is diagonal in the electron spins
 and thus 
\be \label{D_0.spin} 
D_0(x,y) = \sum_{\sigma} \PH_0(x,\sigma; y,\sigma)\PH_0(y,\sigma; x,\sigma)
 =\, 2 \;\PH_0(x,y)\PH_0(y,x) \; .
\ee
We therefore have 
\bea
D_0(\q,\nu_n) &\equiv& 2 \int d(\x-\y) d(\tau_x-\tau_y) e^{i\q.(\x-\y)} d^{i\nu_n(\tau_x-\tau_y)} 
  \left[ \PH_0(x,y) \PH_0(y,x) \right] \nonumber \\
  &=& \frac{2}{\beta} \sum_{n'} \int \frac{d\p}{(2\pi)^3} 
  \frac{1}{(-i\omega_{n'}-i\nu_n + \ve_{\p+\q} - \mu)(-i\omega_{n'}+\ve_\p -\mu)} \nonumber \\
  &=& 2 \int \frac{d\p}{(2\pi)^3} \frac{n_{\p+\q}-n_{\p}}{-i\nu_n + \ve_{\p+\q} - \ve_{\p} } \nonumber \\
  &=& 
  -2 \int \frac{d\p}{(2\pi)^3} \frac{n_\p(1- n_{\p+\q})-n_{\p+\q}(1-n_{\p})}{-i\nu_n + \ve_{\p+\q} - \ve_{\p} } \nonumber \\
&=& -2 \int \frac{d\p}{(2\pi)^3} n_\p(1- n_{\p+\q}) \left[ 
\frac{1}{-i\nu_n + \ve_{\p+\q} - \ve_{\p}}
 + \frac{1}{i\nu_n + \ve_{\p+\q} - \ve_{\p}}
\right]  
  \label{D_0.HEG} \\
  &=& -2 \int \frac{d\p}{(2\pi)^3}\; n_\p \left[ 
\frac{1}{-i\nu_n + \ve_{\p+\q} - \ve_{\p}}
 + \frac{1}{i\nu_n + \ve_{\p+\q} - \ve_{\p}}
\right]  \nonumber \\
  &=& - 4 \int \frac{d\p}{(2\pi)^3} \frac{n_\p(1- n_{\p+\q})(\ve_{\p+\q}-\ve_\p)}{(\ve_{\p+\q}-\ve_\p)^2+\nu_n^2} = - 4 \int \frac{d\p}{(2\pi)^3} \frac{n_\p (\ve_{\p+\q}-\ve_\p)}{(\ve_{\p+\q}-\ve_\p)^2+\nu_n^2}
  \nonumber
\eea  
where the third line of the above equation is obtained using the technique described in 
section~\ref{sec:KS_systematics}, 
 the fourth line is obtained by rewriting the numerator of
  the third line as $- n_\p (1-n_{\p+\q}) + n_{\p+\q} (1-n_\p)$, the fifth line is obtained by 
  a change of variable ($\p+\q \to -\p$) in the second half and assuming the 
   spherical property of $n_\p$, and the 
   sixth line comes from the fact that $n_\p n_{\p+\q}$ is symmetric with respect to $(\p+\q) \leftrightarrow \p$
    while the quantity inside the square bracket is antisymmetric with respect to $(\p+\q) \leftrightarrow \p$.
  Since $\ve_{\p} \propto |\p|^2$ is a monotonic increasing function of $|\p|$,
 while both $n_{\p} = 1/(e^{\beta (\ve_\p-\mu)}+1)$ and $1-n_\p$ are nonnegative, 
  we see that $D_0(\q,\nu_n)$ is a negative definite quantity. 
  
It turns out that for the HEG system one can further simplify the expression of $D_0(q,\nu)$ by
 introducing a new variable $u$ via $\nu \equiv u|\q|/m$. We then have (by choosing $\hat q$ as the
  $\hat z$ direction in $\p$ and using the fact that $n_\p = n(|\p|) = n(p)$)
\bea
 D_0(q,\frac{uq}{m}) &=&  -2 \int \frac{d\p}{(2\pi)^3}\; n(p) \left[ 
\frac{1}{-iuq/m + \ve_{\p+\q} - \ve_{\p}}
 + \frac{1}{iuq/m + \ve_{\p+\q} - \ve_{\p}}
\right]  \nonumber \\
&=& -2 \frac{m}{q} \int_0^\infty \frac{ n (p) \, p^2\, dp}{(2\pi)^2}\; \int_{-1}^1 d\cos \theta \left[ 
\frac{1}{-iu+ q/2 + p\cos \theta }  + \frac{1}{iu + q/2 + p\cos \theta }
\right]  \nonumber \\
&=& -2 \frac{m}{q} \int_0^\infty \frac{ n (p) \, dp}{(2\pi)^2}\; 
 p \left[ \ln \left(\frac{q}{2}+p+iu \right) + \ln \left( \frac{q}{2} + p - iu \right) \right. \nonumber \\
 && \hspace*{90pt}
 \left. - \ln \left(\frac{q}{2}- p+iu \right) - \ln \left( \frac{q}{2} - p - iu \right)
\right]  \nonumber \\
&=& \frac{2m}{(2\pi)^2}\int_0^\infty \frac{\partial n(p)}{\partial p} \, dp\, \left[ p 
+ \frac{1}{2q} (p^2+u^2-\frac{q^2}{4} )\ln \frac{(p+\frac{q}{2})^2 + u^2}{(p-\frac{q}{2})^2 + u^2} 
 \right.  \nonumber \\
&& \hspace*{115pt} \left. 
 - u \tan^{-1}\frac{\frac{q}{2}+p}{u} + u\tan^{-1}\frac{\frac{q}{2}-p}{u} \right] \; .
\eea 
As $\beta \to \infty$, $n(p) = \theta(p_F-p)$ and therefore 
$\partial n(p)/\partial p = -\delta (p-p_F)$, leading to 
\bea 
D_0(q,\frac{uq}{m}) & \xrightarrow[\beta \to \infty]{}& -\frac{2m}{(2\pi)^2}  \left[ p_F
+ \frac{1}{2q} (p_F^2+u^2-\frac{q^2}{4} )\ln \frac{(p_F+\frac{q}{2})^2 + u^2}{(p_F-\frac{q}{2})^2 + u^2} 
 \right.  \nonumber \\
&& \hspace*{75pt} \left. 
 - u \tan^{-1}\frac{\frac{q}{2}+p_F}{u} + u\tan^{-1}\frac{\frac{q}{2}-p_F}{u} \right] \; . \label{D_0.HEG.zT}
\eea 
The exact expression (\ref{D_0.HEG.zT}) allows one to extract the limits of $q \to \infty$ and $q \to 0$, 
  both of which are important for determining the convergence properties of the energy expansion. 
 We have 
\bea
D_0(q \gg 1,\frac{uq}{m}) &\xrightarrow[\beta \to \infty]{}& - \frac{4m \,p_F^3}{3\pi^2} \frac{1}{q^2 + 4u^2} 
 + {\cal O} \left( (q^2+4u^2)^{-2}\right) \; ,
\label{D_0.zT.bq} \\ 
D_0(q \ll 1,\frac{uq}{m}) &\xrightarrow[\beta \to \infty]{}& -\frac{m}{\pi^2} \left[
 R_0(u) + R_1(u)\, q^2  + R_2 (u)\, q^4 \right] +   {\cal O} \left( q^{6}\right) \label{D_0.zT.sq} \; ,
\eea  
where
\bea
R_0(u) &\equiv &  p_F -u \tan^{-1} \frac{p_F}{u}  \; ,\label{R_0.def}\\
R_1(u) &=&  - \frac{p_F^3}{12(p_F^2+u^2)^2} \; , \label{R_1.def} \\
R_2(u) &=& - \frac{p_F^3(p_F^2-5u^2)}{240 (p_F^2+u^2)^4}  \; . \label{R_2.def}
\eea  
With these asymptotic behaviors, we see from eq.~(\ref{Gamma.1.q}) that 
 $\Gamma_1[n]$ is finite. Methods for extracting coefficients associated with the $e^2$-based
  expansion for $\epsilon'$ can be found in references \onlinecite{GB_57}
   and  \onlinecite{CM_64}. The $\epsilon' = \lim_{\beta \to \infty} \Gamma_1[n]/(N_e\,\beta) 
   - \epsilon_x$ part contains in general 
   $r_s^{n\ge 0} \ln r_s$ and $r_s^{(n\ge 0)}$ terms when the energy is expressed in Rydbergs and
     expanded in power of $r_s$ (or $e^2$).

We now turn our attention to  $\Gamma_2[n]$, whose diagrammatic expression is given in 
Figure~\ref{fig:Gamma.2}. As mentioned earlier, within our formalism, $\Gamma_{i\ge 2}[n]$ 
correspond to diagrams containing only $\tilde {\cal D}_0$ propagator and the KS electron propagator. 
Since the $\tilde {\cal D}_0$ propagator retains a finite value even when its associated momentum
 approaches zero, each of the diagrams corresponding to $\Gamma_{i\ge 2}[n]$ takes a finite value. 
 The absence of divergence no longer holds when one perform high density ( small $r_s$) expansion, 
 as we will illustrate explicitly later. To make easy comparison with existing HEG studies, 
one employs Dyson's equation,   
\bea 
\tilde {\cal D}_0 & = & U + U \4di D_0 \4di \tilde{\cal D}_0 \label{tD.Dyson} \\ 
\begin{picture}(20,30)(0,-3)
\Photon[](0,0)(20,0){2}{4}
\Vertex(0,0){1.2} \Vertex(20,0){1.2}
\end{picture} \;\;
&=& \;\; \begin{picture}(20,30)(0,-3)
\Line[dash,dashsize=1.5](0,0)(20,0)
\Vertex(0,0){1.2} \Vertex(20,0){1.2}
\end{picture} \; \;
+ \; \begin{picture}(60,30)(0,-3)
\Line[dash,dashsize=1.5](0,0)(20,0) \Photon[](40,0)(60,0){2}{4}
\Arc[arrow,arrowpos=0.50,arrowlength=3.0,arrowwidth=0.8,arrowinset=0.2](30,0)(10,0,180)
\Arc[arrow,arrowpos=0.50,arrowlength=3.0,arrowwidth=0.8,arrowinset=0.2](30,0)(10,180,360)
\Vertex(0,0){1.2} \Vertex(20,0){1.2}
\Vertex(40,0){1.2} \Vertex(60,0){1.2}
\end{picture}  \;\;   \nonumber 
\eea   
to decompose the propagator $\tilde {\cal D}_0$ (of order $e^2$ and higher) 
 into the sum of a bare Coulomb propagator $U$
 (of order $e^2$) and $U\4di D_0 \4di \tilde{\cal D}_0$ (of order $e^4$ and higher). 
 The reason that one should not expand further and write 
 $\tilde {\cal D}_0 = U + U \4di D_0 \4di \, U + U \4di D_0 \4di \, U \4di D_0 \4di \tilde {\cal D}_0$ 
 is because the $U \4di D_0 \4di \, U$ term causes infrared divergence 
  due to the momentum integral $\int d\q/q^4$. In fact, this is exactly what causes
   (in the second diagram of Figure~\ref{fig:polarization}) the infrared 
   divergence, motivating the summation of the ring diagrams.

 To illustrate the main points, let us begin by examining the first three diagrams  
 of $\Gamma_2[n]$ and employ the decomposition rule mentioned above:  
\bea 
\frac{1}{4}\;\;\; \;\; \; \; \; \; \; \; \; \; \begin{picture}(80,20)(0,-4)
\put(-16,15){\makebox(0,0){$\p_1+\q$}}
\put(45,15){\makebox(0,0){$\p_1$}}
\put(-8,-15){\makebox(0,0){$-\p_2$}}
\put(58,-15){\makebox(0,0){$-\p_2-\q$}}
\Arc[arrow,arrowpos=0.50,arrowlength=3.0,arrowwidth=0.8,arrowinset=0.2](20,0)(20,0,90)
\Arc[arrow,arrowpos=0.50,arrowlength=3.0,arrowwidth=0.8,arrowinset=0.2](20,0)(20,90,180)
\Arc[arrow,arrowpos=0.50,arrowlength=3.0,arrowwidth=0.8,arrowinset=0.2](20,0)(20,180,270)
\Arc[arrow,arrowpos=0.50,arrowlength=3.0,arrowwidth=0.8,arrowinset=0.2](20,0)(20,270,360)
\Vertex(20,-20){1.8}
\Vertex(20,20){1.8}
\Vertex(0,0){1.8}
\Vertex(40,0){1.8} 
\Photon[](20,20)(20,-20){2}{6}
\Photon[](0,0)(40,0){2}{6} 
\end{picture} \; \; 
&=& \;\;
\frac{1}{4} \; \begin{picture}(40,20)(0,-4)
\Arc[arrow,arrowpos=0.50,arrowlength=3.0,arrowwidth=0.8,arrowinset=0.2](20,0)(20,0,90)
\Arc[arrow,arrowpos=0.50,arrowlength=3.0,arrowwidth=0.8,arrowinset=0.2](20,0)(20,90,180)
\Arc[arrow,arrowpos=0.50,arrowlength=3.0,arrowwidth=0.8,arrowinset=0.2](20,0)(20,180,270)
\Arc[arrow,arrowpos=0.50,arrowlength=3.0,arrowwidth=0.8,arrowinset=0.2](20,0)(20,270,360)
\Vertex(20,-20){1.8}
\Vertex(20,20){1.8}
\Vertex(0,0){1.8}
\Vertex(40,0){1.8} 
\Line[dash,dashsize=1.5](20,20)(20,-20)
\Line[dash,dashsize=1.5](0,0)(40,0) 
\Text(20,-34)[cc]{$(a)$} 
\end{picture} \; \; 
  + \frac{1}{2} \;  \begin{picture}(50,25)(0,-4)
\Arc[arrow,arrowpos=0.50,arrowlength=3.0,arrowwidth=0.8,arrowinset=0.2](25,0)(25,0,90)
\Arc[arrow,arrowpos=0.50,arrowlength=3.0,arrowwidth=0.8,arrowinset=0.2](25,0)(25,90,180)
\Arc[arrow,arrowpos=0.50,arrowlength=3.0,arrowwidth=0.8,arrowinset=0.2](25,0)(25,180,270)
\Arc[arrow,arrowpos=0.50,arrowlength=3.0,arrowwidth=0.8,arrowinset=0.2](25,0)(25,270,360)
\Vertex(25,-25){1.8}
\Vertex(25,25){1.8}
\Vertex(0,0){1.8}
\Vertex(50,0){1.8} 
\Line[dash,dashsize=1.5](25,25)(25,16)
\Arc[](25,11)(5,0,360)
\Photon[](25,-25)(25,6){2}{5}
\Line[dash,dashsize=1.5](0,0)(50,0) 
\Text(25,-34)[cc]{$(b)$}
\end{picture} \; \;   
+ \frac{1}{4} \; \begin{picture}(50,25)(0,-4)
\Arc[arrow,arrowpos=0.50,arrowlength=3.0,arrowwidth=0.8,arrowinset=0.2](25,0)(25,0,90)
\Arc[arrow,arrowpos=0.50,arrowlength=3.0,arrowwidth=0.8,arrowinset=0.2](25,0)(25,90,180)
\Arc[arrow,arrowpos=0.50,arrowlength=3.0,arrowwidth=0.8,arrowinset=0.2](25,0)(25,180,270)
\Arc[arrow,arrowpos=0.50,arrowlength=3.0,arrowwidth=0.8,arrowinset=0.2](25,0)(25,270,360)
\Vertex(25,-25){1.8}
\Vertex(25,25){1.8}
\Vertex(0,0){1.8}
\Vertex(50,0){1.8} 
\Line[dash,dashsize=1.5](25,25)(25,16)
\Arc[](25,11)(5,0,360)
\Photon[](25,-25)(25,6){2}{5}
\Line[dash,dashsize=1.5](0,0)(9,0)
\Arc[](14,0)(5,0,360)
\Photon[](19,0)(50,0){2}{5} 
\Text(25,-34)[cc]{$(c)$}
\end{picture} \label{Gamma.2.1.decomp}  \\
\frac{1}{2} \;\;\;\; \;\; \; \; \; \; \; \; \;  \begin{picture}(80,60)(0,-4)
\put(-17.5,19){\makebox(0,0){$-\p_1-\q$}}
\put(50,15){\makebox(0,0){$-\p_1$}}
\put(-16,-19){\makebox(0,0){$-\p_2-\q$}}
\put(60,-15){\makebox(0,0){$-\p_1-\q$}}
\Arc[arrow,arrowpos=0.50,arrowlength=3.0,arrowwidth=0.8,arrowinset=0.2](20,0)(20,0,90)
\Arc[arrow,arrowpos=0.50,arrowlength=3.0,arrowwidth=0.8,arrowinset=0.2](20,0)(20,90,180)
\Arc[arrow,arrowpos=0.50,arrowlength=3.0,arrowwidth=0.8,arrowinset=0.2](20,0)(20,180,270)
\Arc[arrow,arrowpos=0.50,arrowlength=3.0,arrowwidth=0.8,arrowinset=0.2](20,0)(20,270,360)
\Vertex(20,-20){1.8}
\Vertex(20,20){1.8}
\Vertex(0,0){1.8}
\Vertex(40,0){1.8} 
\Photon[](20,20)(40,0){2}{6}
\Photon[](0,0)(20,-20){2}{6} 
\end{picture}  \; \; 
&=& \;\;
\frac{1}{2} \; \begin{picture}(40,20)(0,-4)
\Arc[arrow,arrowpos=0.50,arrowlength=3.0,arrowwidth=0.8,arrowinset=0.2](20,0)(20,0,90)
\Arc[arrow,arrowpos=0.50,arrowlength=3.0,arrowwidth=0.8,arrowinset=0.2](20,0)(20,90,180)
\Arc[arrow,arrowpos=0.50,arrowlength=3.0,arrowwidth=0.8,arrowinset=0.2](20,0)(20,180,270)
\Arc[arrow,arrowpos=0.50,arrowlength=3.0,arrowwidth=0.8,arrowinset=0.2](20,0)(20,270,360)
\Vertex(20,-20){1.8}
\Vertex(20,20){1.8}
\Vertex(0,0){1.8}
\Vertex(40,0){1.8} 
\Line[dash,dashsize=1.5](20,20)(40,0)
\Line[dash,dashsize=1.5](0,0)(20,-20) 
\Text(20,-34)[cc]{$(a)$}
\end{picture}\; \; 
  +  \; \; \; \; \begin{picture}(50,25)(0,-4)
\Arc[arrow,arrowpos=0.50,arrowlength=3.0,arrowwidth=0.8,arrowinset=0.2](25,0)(25,0,90)
\Arc[arrow,arrowpos=0.50,arrowlength=3.0,arrowwidth=0.8,arrowinset=0.2](25,0)(25,90,180)
\Arc[arrow,arrowpos=0.50,arrowlength=3.0,arrowwidth=0.8,arrowinset=0.2](25,0)(25,180,270)
\Arc[arrow,arrowpos=0.50,arrowlength=3.0,arrowwidth=0.8,arrowinset=0.2](25,0)(25,270,360)
\Vertex(25,-25){1.8}
\Vertex(25,25){1.8}
\Vertex(0,0){1.8}
\Vertex(50,0){1.8} 
\Line[dash,dashsize=1.5](25,25)(32.071,17.929)
\Arc[](35.606,14.3934)(5,0,360)
\Photon[](39.142,10.858)(50,0){2}{4}
\Line[dash,dashsize=1.5](0,0)(25,-25) 
\Text(25,-34)[cc]{$(b)$}
\end{picture} \; \;   
+ \frac{1}{2} \; \begin{picture}(50,25)(0,-4)
\Arc[arrow,arrowpos=0.50,arrowlength=3.0,arrowwidth=0.8,arrowinset=0.2](25,0)(25,0,90)
\Arc[arrow,arrowpos=0.50,arrowlength=3.0,arrowwidth=0.8,arrowinset=0.2](25,0)(25,90,180)
\Arc[arrow,arrowpos=0.50,arrowlength=3.0,arrowwidth=0.8,arrowinset=0.2](25,0)(25,180,270)
\Arc[arrow,arrowpos=0.50,arrowlength=3.0,arrowwidth=0.8,arrowinset=0.2](25,0)(25,270,360)
\Vertex(25,-25){1.8}
\Vertex(25,25){1.8}
\Vertex(0,0){1.8}
\Vertex(50,0){1.8} 
\Line[dash,dashsize=1.5](25,25)(32.071,17.929)
\Arc[](35.606,14.3934)(5,0,360)
\Photon[](39.142,10.858)(50,0){2}{4}
\Line[dash,dashsize=1.5](0,0)(7.071,-7.071)
\Arc[](10.606,-10.606)(5,0,360) 
\Photon[](14.142,-14.142)(25,-25){2}{4}
\Text(25,-34)[cc]{$(c)$}
\end{picture} \label{Gamma.2.2.decomp} \\
 - \frac{1}{2} \; \; \; \; \; \;\; \begin{picture}(108,60)(0,-4)
 \put(-5.5,15){\makebox(0,0){$\p_2$}}
 \put(-6.5,8){\makebox(0,0){$+$}}
 \put(-6.5,2){\makebox(0,0){$\q'$}}
\put(43,18){\makebox(0,0){$\p_2$}}
\put(43,-18){\makebox(0,0){$\p_2$}}
\put(59,18){\makebox(0,0){$\p_1$}}
\put(60,-18){\makebox(0,0){$\p_1$}}
 \put(106,15){\makebox(0,0){$\p_1$}}
 \put(107,8){\makebox(0,0){$+$}}
 \put(107,1){\makebox(0,0){$\q$}}
\Arc[arrow,arrowpos=0.50,arrowlength=3.0,arrowwidth=0.8,arrowinset=0.2,clock](20,0)(20,-90,90)
\Arc[arrow,arrowpos=0.50,arrowlength=3.0,arrowwidth=0.8,arrowinset=0.2,clock](20,0)(20,90,0)
\Arc[arrow,arrowpos=0.50,arrowlength=3.0,arrowwidth=0.8,arrowinset=0.2,clock](20,0)(20,0,-90)
\Photon[](20,-20)(20,20){2}{8}
\Vertex(20,-20){1.8} \Vertex(20,20){1.8}
\Line[double](40,0)(60,0)
\Arc[arrow,arrowpos=0.50,arrowlength=3.0,arrowwidth=0.8,arrowinset=0.2](80,0)(20,-90,90)
\Arc[arrow,arrowpos=0.50,arrowlength=3.0,arrowwidth=0.8,arrowinset=0.2](80,0)(20,90,0)
\Arc[arrow,arrowpos=0.50,arrowlength=3.0,arrowwidth=0.8,arrowinset=0.2](80,0)(20,0,-90)
\Vertex(40,0){1.8}
\Vertex(60,0){1.8}
\Photon[](80,-20)(80,20){2}{8}
\Vertex(80,-20){1.8} \Vertex(80,20){1.8}
\end{picture}\; \;  &=& \; \; 
- \frac{1}{2} \; \begin{picture}(100,60)(0,-4)
\Arc[arrow,arrowpos=0.50,arrowlength=3.0,arrowwidth=0.8,arrowinset=0.2,clock](20,0)(20,-90,90)
\Arc[arrow,arrowpos=0.50,arrowlength=3.0,arrowwidth=0.8,arrowinset=0.2,clock](20,0)(20,90,0)
\Arc[arrow,arrowpos=0.50,arrowlength=3.0,arrowwidth=0.8,arrowinset=0.2,clock](20,0)(20,0,-90)
\Line[dash,dashsize=1.5](20,-20)(20,20)
\Vertex(20,-20){1.8} \Vertex(20,20){1.8}
\Line[double](40,0)(60,0)
\Arc[arrow,arrowpos=0.50,arrowlength=3.0,arrowwidth=0.8,arrowinset=0.2](80,0)(20,-90,90)
\Arc[arrow,arrowpos=0.50,arrowlength=3.0,arrowwidth=0.8,arrowinset=0.2](80,0)(20,90,0)
\Arc[arrow,arrowpos=0.50,arrowlength=3.0,arrowwidth=0.8,arrowinset=0.2](80,0)(20,0,-90)
\Vertex(40,0){1.8}
\Vertex(60,0){1.8}
\Line[dash,dashsize=1.5](80,-20)(80,20)
\Vertex(80,-20){1.8} \Vertex(80,20){1.8}
\Text(50,-20)[cc]{$(a)$}
\end{picture}  \; \; - \; \;  
\begin{picture}(100,60)(0,-4)
\Arc[arrow,arrowpos=0.50,arrowlength=3.0,arrowwidth=0.8,arrowinset=0.2,clock](20,0)(20,-90,90)
\Arc[arrow,arrowpos=0.50,arrowlength=3.0,arrowwidth=0.8,arrowinset=0.2,clock](20,0)(20,90,0)
\Arc[arrow,arrowpos=0.50,arrowlength=3.0,arrowwidth=0.8,arrowinset=0.2,clock](20,0)(20,0,-90)
\Line[dash,dashsize=1.5](20,-20)(20,20)
\Vertex(20,-20){1.8} \Vertex(20,20){1.8}
\Line[double](40,0)(60,0)
\Arc[arrow,arrowpos=0.50,arrowlength=3.0,arrowwidth=0.8,arrowinset=0.2](80,0)(20,-90,90)
\Arc[arrow,arrowpos=0.50,arrowlength=3.0,arrowwidth=0.8,arrowinset=0.2](80,0)(20,90,0)
\Arc[arrow,arrowpos=0.50,arrowlength=3.0,arrowwidth=0.8,arrowinset=0.2](80,0)(20,0,-90)
\Vertex(40,0){1.8}
\Vertex(60,0){1.8}
\Line[dash,dashsize=1.5](80,20)(80,10)
\Arc[](80,4)(6,0,360)
\Photon[](80,-2)(80,-20){2}{4}
\Vertex(80,-20){1.8} \Vertex(80,20){1.8}
\Text(50,-20)[cc]{$(b)$}
\end{picture}\; \nonumber \\
&& \; \; 
- \frac{1}{2} \; \; \begin{picture}(100,50)(0,-4)
\Arc[arrow,arrowpos=0.50,arrowlength=3.0,arrowwidth=0.8,arrowinset=0.2,clock](20,0)(20,-90,90)
\Arc[arrow,arrowpos=0.50,arrowlength=3.0,arrowwidth=0.8,arrowinset=0.2,clock](20,0)(20,90,0)
\Arc[arrow,arrowpos=0.50,arrowlength=3.0,arrowwidth=0.8,arrowinset=0.2,clock](20,0)(20,0,-90)
\Line[dash,dashsize=1.5](20,20)(20,10)
\Arc[](20,4)(6,0,360)
\Photon[](20,-2)(20,-20){2}{4}
\Vertex(20,-20){1.8} \Vertex(20,20){1.8}
\Line[double](40,0)(60,0)
\Arc[arrow,arrowpos=0.50,arrowlength=3.0,arrowwidth=0.8,arrowinset=0.2](80,0)(20,-90,90)
\Arc[arrow,arrowpos=0.50,arrowlength=3.0,arrowwidth=0.8,arrowinset=0.2](80,0)(20,90,0)
\Arc[arrow,arrowpos=0.50,arrowlength=3.0,arrowwidth=0.8,arrowinset=0.2](80,0)(20,0,-90)
\Vertex(40,0){1.8}
\Vertex(60,0){1.8}
\Line[dash,dashsize=1.5](80,20)(80,10)
\Arc[](80,4)(6,0,360)
\Photon[](80,-2)(80,-20){2}{4}
\Vertex(80,-20){1.8} \Vertex(80,20){1.8}
\Text(50,-20)[cc]{$(c)$}
\end{picture}
 \label{Gamma.2.3.decomp}\\
&& \nonumber
\eea
Let us start with the diagrams to the left of the equal signs in (\ref{Gamma.2.1.decomp}-\ref{Gamma.2.3.decomp}). 
 We will for now keep the bosonic propagators general, that is, allowing them to carry frequencies.  
  After that, we will discuss the (a) diagrams in (\ref{Gamma.2.1.decomp}-\ref{Gamma.2.3.decomp}),
   followed by the (b) diagrams and then the (c) diagrams.  

In (\ref{Gamma.2.1.decomp}), let the vertical boson propagator, denoted by $B_1$,  
carry momentum $\q$ (upward) 
and frequency $\nu$. Let the horizontal boson propagator, denoted by $B_2$, carry momentum $\q' \equiv -(\p_1+\p_2+\q)$ (leftward) and frequency $\nu'$. 
 Using techniques described in section~\ref{sec:KS_systematics} for frequency summation, one 
 obtains the following generic result 
\bea 
&& \hspace*{-20pt}\frac{\beta V}{4} \frac{2}{\beta^2}\sum_{\nu,\nu'} \int \frac{d\p_1}{(2\pi)^3}\frac{d\p_2}{(2\pi)^3}
\frac{d\q}{(2\pi)^3} \frac{B_1(\q,\nu)}{-i\nu+\ve_{\p_1+\q}-\ve_{\p_1}} 
\frac{B_2(\q',\nu')}{i\nu+\ve_{\p_2+\q}-\ve_{\p_2}}  
\left\{ \frac{n_{\p_2} - n_{\p_1}}{\ve_{\p_2}-\ve_{\p_1}-i(\nu+\nu') } \right.
 \nonumber \\
&& \left. \hspace*{15pt} + \frac{n_{\p_1} - n_{\p_1+\q'}}{\ve_{\p_1+\q'}-\ve_{\p_1}-i\nu' }
 + \frac{n_{\p_2} - n_{\p_2+\q'}}{\ve_{\p_2+\q'}-\ve_{\p_2}+ i\nu' }
 + \frac{n_{\p_2+\q} - n_{\p_1+\q}}{\ve_{\p_2+\q}-\ve_{\p_1+\q} + i(\nu - \nu') } \right\} \; , \label{Gamma.2.1.v1}
\eea   
 or equivalently,
\bea 
&& \hspace*{-20pt}\frac{\beta V}{4} \frac{2}{\beta^2}\sum_{\nu,\nu'} \int \frac{d\p_1}{(2\pi)^3}\frac{d\p_2}{(2\pi)^3}
\frac{d\q}{(2\pi)^3} \frac{B_1(\q,\nu)}{-i\nu'+\ve_{\p_1+\q'}-\ve_{\p_1}} 
\frac{B_2(\q',\nu')}{i\nu'+\ve_{\p_2+\q'}-\ve_{\p_2}}  
\left\{ \frac{n_{\p_2} - n_{\p_1}}{\ve_{\p_2}-\ve_{\p_1}-i(\nu+\nu') } \right.
 \nonumber \\
&& \left. \hspace*{15pt} + \frac{n_{\p_1} - n_{\p_1+\q}}{\ve_{\p_1+\q}-\ve_{\p_1}-i\nu }
 + \frac{n_{\p_2} - n_{\p_2+\q}}{\ve_{\p_2+\q}-\ve_{\p_2}+ i\nu }
 + \frac{n_{\p_2+\q} - n_{\p_1+\q}}{\ve_{\p_2+\q}-\ve_{\p_1+\q} + i(\nu - \nu') } \right\} \; . \label{Gamma.2.1.v2} 
\eea     
 Note that the factor $2$ associated with $\frac{2}{\beta^2}$ comes from the two possible
  spin states for an electron. 
  
In (\ref{Gamma.2.2.decomp}), let the boson propagator on top, denoted by $B_1$,  
carry momentum $\q$ (towards lower-right) 
and frequency $\nu$. Let the boson propagator at the bottom, denoted by $B_2$, carry momentum $\q' \equiv (\p_1-\p_2)$ (towards upper-left) and frequency $\nu'$. 
 Using techniques described in section~\ref{sec:KS_systematics} for frequency summation, one 
 obtains the following generic result   
\bea 
&& \hspace*{-20pt}\frac{\beta V}{2} \frac{2}{\beta^2}\sum_{\nu,\nu'} \int \frac{d\p_1}{(2\pi)^3}\frac{d\p_2}{(2\pi)^3}
\frac{d\q}{(2\pi)^3} \frac{B_1(\q,\nu)}{-i\nu+\ve_{\p_1+\q}-\ve_{\p_1}} 
\frac{B_2(\q',\nu')}{-i\nu'+\ve_{\p_1+\q}-\ve_{\p_2+\q}}  
\left\{ -\beta n_{\p_1+\q} (1-n_{\p_1+\q})\right.  \nonumber \\
&& \left. \hspace*{15pt} - 
\frac{n_{\p_1+\q} - n_{\p_2 +\q}}{-i\nu'+ \ve_{\p_1+\q}-\ve_{\p_2+\q} } 
 + \frac{n_{\p_1} - n_{\p_1+\q}}{-i\nu + \ve_{\p_1+\q}-\ve_{\p_1} }
 + \frac{n_{\p_1} - n_{\p_2+\q}}{i(\nu -\nu')+ \ve_{\p_1}-\ve_{\p_2+\q}  } \right\} \; . \label{Gamma.2.2.v1}
\eea     
  
In (\ref{Gamma.2.3.decomp}), let the boson propagator at the right hand side, denoted by $B_1$,  
carry momentum $\q$ (downwards) and frequency $\nu$. Let the boson propagator 
at the left hand side, denoted by $B_2$, carry momentum $\q'$ (also downwards) and frequency $\nu'$. 
 Using techniques described in section~\ref{sec:KS_systematics} for frequency summation, one 
 obtains the following generic result   
\bea 
&& \hspace*{-25pt} -\frac{\beta V}{2} \frac{2^2}{\beta^2} 
\frac{-1}{2\beta \int \frac{d\p}{(2\pi)^3} n_{\p}(1-n_{\p})} 
\sum_{\nu,\nu'} \int \frac{d\p_1}{(2\pi)^3}\frac{d\p_2}{(2\pi)^3}
\frac{d\q}{(2\pi)^3} \frac{d\q'}{(2\pi)^3} \frac{B_1(\q,\nu)}{-i\nu+\ve_{\p_1+\q}-\ve_{\p_1}} 
\frac{B_2(\q',\nu')}{-i\nu'+\ve_{\p_2+\q'}-\ve_{\p_2}}  \nonumber \\
&&  \hspace*{15pt}\left\{ \beta^2 n_{\p_1}(1-n_{\p_1}) n_{\p_2} (1-n_{\p_2}) +\beta n_{\p_1}(1-n_{\p_1}) 
\frac{n_{\p_2+\q'} - n_{\p_2}}{-i\nu'+ \ve_{\p_2+\q'}-\ve_{\p_2} } \right. \nonumber \\
&& \hspace*{20pt} \left. +\beta n_{\p_2}(1-n_{\p_2}) 
\frac{n_{\p_1+\q} - n_{\p_1}}{-i\nu+ \ve_{\p_1+\q}-\ve_{\p_1} } 
+ \frac{n_{\p_2+\q'} - n_{\p_2}}{-i\nu'+ \ve_{\p_2+\q'}-\ve_{\p_2} }
\frac{n_{\p_1+\q} - n_{\p_1}}{-i\nu+ \ve_{\p_1+\q}-\ve_{\p_1} }
 \right\} \; , \label{Gamma.2.3.v1}
\eea     
where the factor $-1/\left\{ 2\beta \int \frac{d\p}{(2\pi)^3} n_\p (1-n_\p) \right\} = 
 D_0^{-1}(\q=0,\nu=0)$ arises from the fact that in a translationally invariant system,
  momentum conservation at each vertex demands that the inverse density correlator must 
carry zero momentum and frequency.

For the diagram (a) on the right hand side of eq.~(\ref{Gamma.2.1.decomp}), 
 $B_1(\q,\nu) = 4\pi e^2/\q^2$ and $B_2(\q',\nu') = 4\pi e^2 /(\p_1+\p_2+\q)^2$ are both
  frequency independent. Thus, one may sum over both $\nu$ and $\nu'$. 
 And upon doing so, we obtain
\be 
\frac{1}{4\beta } \; \begin{picture}(40,20)(0,-4)
\Arc[arrow,arrowpos=0.50,arrowlength=3.0,arrowwidth=0.8,arrowinset=0.2](20,0)(20,0,90)
\Arc[arrow,arrowpos=0.50,arrowlength=3.0,arrowwidth=0.8,arrowinset=0.2](20,0)(20,90,180)
\Arc[arrow,arrowpos=0.50,arrowlength=3.0,arrowwidth=0.8,arrowinset=0.2](20,0)(20,180,270)
\Arc[arrow,arrowpos=0.50,arrowlength=3.0,arrowwidth=0.8,arrowinset=0.2](20,0)(20,270,360)
\Vertex(20,-20){1.8}
\Vertex(20,20){1.8}
\Vertex(0,0){1.8}
\Vertex(40,0){1.8} 
\Line[dash,dashsize=1.5](20,20)(20,-20)
\Line[dash,dashsize=1.5](0,0)(40,0) 
\end{picture} \; =\;  {V} \int \frac{d\p_1}{(2\pi)^3}\frac{d\p_2}{(2\pi)^3}
\frac{d\q}{(2\pi)^3} \frac{(4\pi e^2)^2}{\q^2 (\p_1+\p_2+\q)^2} 
\frac{n_{\p_1}n_{\p_2}(1-n_{\p_1+\q})(1-n_{\p_2+\q})}{\ve_{\p_1+\q}+\ve_{\p_2+\q}-\ve_{\p_1}-\ve_{\p_2}}
\; . \vspace*{8pt}\nonumber  
\ee   
If we divide the quantity above by $N_e $ and then take the zero temperature limit, it gives 
 rise to (using $p_F$ as the unit for momentum) 
\bea
&& \hspace*{-40pt}\frac{m p_F^3 V }{N_e (2\pi)^9} (4\pi e^2)^2 \int_{|\p_i|<1,\p_i+\q|> 1}\; d\q \; d\p_1 \; d\p_2 
\frac{1}{\q^2 (\p_1+\p_2+\q)^2} \frac{1}{\q^2 + \q \cdot (\p_1+\p_2)} \nonumber \\
&& = \left[ \frac{3}{16 \pi^5} \int_{|\p_i|<1,\p_i+\q|> 1}\; d\q \; d\p_1 \; d\p_2 
\frac{1}{\q^2 (\p_1+\p_2+\q)^2} \frac{1}{\q^2 + \q \cdot (\p_1+\p_2)}\right] \; \; {\rm Rydberg},\nonumber
\eea 
 which is exactly the $\epsilon_b^{(2)}$ term of Gell-Mann and Brueckner.\cite{GB_57} 
 
Diagram (a) on the right hand side of eq.~(\ref{Gamma.2.2.decomp}) is one of the  
 {\it anomalous} diagrams\cite{KL_60,LW_60} that give rise to finite contribution 
  as the $T \to 0$ limit is taken within finite temperature formalism but are absent within 
   zero temperature formalism. For this diagram, 
 $B_1(\q,\nu) = 4\pi e^2/\q^2$ and $B_2(\q',\nu') = 4\pi e^2 /(\p_1-\p_2)^2$ are both
  frequency independent. Thus, one may evaluate its zero temperature contribution
   by summing over both $\nu$ and $\nu'$. 
 Upon doing so, one obtains from diagram (a) of (\ref{Gamma.2.2.decomp}) the 
 following anomalous contribution 
\bea 
\frac{1}{2\beta } \; \begin{picture}(40,20)(0,-4)
\Arc[arrow,arrowpos=0.50,arrowlength=3.0,arrowwidth=0.8,arrowinset=0.2](20,0)(20,0,90)
\Arc[arrow,arrowpos=0.50,arrowlength=3.0,arrowwidth=0.8,arrowinset=0.2](20,0)(20,90,180)
\Arc[arrow,arrowpos=0.50,arrowlength=3.0,arrowwidth=0.8,arrowinset=0.2](20,0)(20,180,270)
\Arc[arrow,arrowpos=0.50,arrowlength=3.0,arrowwidth=0.8,arrowinset=0.2](20,0)(20,270,360)
\Vertex(20,-20){1.8}
\Vertex(20,20){1.8}
\Vertex(0,0){1.8}
\Vertex(40,0){1.8} 
\Line[dash,dashsize=1.5](20,20)(40,0)
\Line[dash,dashsize=1.5](0,0)(20,-20) 
\end{picture}\; \; &=& -(V)\int \frac{d\p_1\, d\p_2\, d\q}{(2\pi)^9} 
\frac{4\pi e^2}{\q^2} \frac{4\pi e^2}{(\p_1-\p_2)^2} \, \beta n_{\p_1+\q} (1-n_{\p_1+\q}) 
 n_{\p_2+\q} n_{\p_1} \nonumber \\
&=& -(V)\int \frac{d\p_1\, d\q\, d\q'}{(2\pi)^9} 
\frac{4\pi e^2}{\q^2} \frac{4\pi e^2}{(\q')^2} \, \beta n_{\p_1} (1-n_{\p_1}) 
 n_{\p_1+\q'} n_{\p_1+\q} \; , \label{anomalous.1.0}
\eea  
where the last expression is obtained via the following change of 
 variables: $\p_{1(2)} \to  -\p_{1(2)}-\q$, eliminating $\p_2$ by $\q' \equiv \p_1-\p_2$, 
 and $\q' \to -\q'$. This expression can be further simplified via the following definition
\[
f(\p) \equiv \int \frac{d\q}{(2\pi)^3} \frac{1}{\q^2} n_{\p +\q} \;,
\] 
leading to 
\bea 
\frac{1}{2\beta } \; \begin{picture}(40,20)(0,-4)
\Arc[arrow,arrowpos=0.50,arrowlength=3.0,arrowwidth=0.8,arrowinset=0.2](20,0)(20,0,90)
\Arc[arrow,arrowpos=0.50,arrowlength=3.0,arrowwidth=0.8,arrowinset=0.2](20,0)(20,90,180)
\Arc[arrow,arrowpos=0.50,arrowlength=3.0,arrowwidth=0.8,arrowinset=0.2](20,0)(20,180,270)
\Arc[arrow,arrowpos=0.50,arrowlength=3.0,arrowwidth=0.8,arrowinset=0.2](20,0)(20,270,360)
\Vertex(20,-20){1.8}
\Vertex(20,20){1.8}
\Vertex(0,0){1.8}
\Vertex(40,0){1.8} 
\Line[dash,dashsize=1.5](20,20)(40,0)
\Line[dash,dashsize=1.5](0,0)(20,-20) 
\end{picture}\; \; 
&=& - V (4\pi e^2)^2\int \frac{d\p_1}{(2\pi)^3} 
 \beta n_{\p_1} (1-n_{\p_1}) f^2(\p_1) \nonumber \\
 & = & - V (4\pi e^2)^2\int \frac{d\p_1}{(2\pi)^3} 
 \frac{\partial  n_{\p_1}}{\partial \mu} f^2(\p_1) \; . \vspace*{18pt}\label{anomalous.1.1}
\eea   
 
As mentioned earlier, each two-particle reducible diagram within our formalism is accompanied by
 a corresponding diagram that will eliminate anomalous contribution when applicable.
Diagrams in eq.~(\ref{Gamma.2.3.decomp}), appearing only under the effective 
 action formalism, are such diagrams. They are neither
 present in the zero temperature formalism of Goldstone~\cite{Goldstone_57} 
 and Brueckner~\cite{BL_55} nor in the finite temperature formalism
  of Luttinger and Ward.\cite{LW_60}
Diagram (a) on the right hand side of (\ref{Gamma.2.3.decomp}) corresponds to the case
 $B_1(\q,\nu) = 4\pi e^2/\q^2$ and $B_2(\q',\nu') = 4\pi e^2 /(\q')^2$, both being
  frequency independent. Thus, upon summing over both $\nu$ and $\nu'$,  
 one obtains 
\bea 
- \frac{1}{2 \beta} \; \begin{picture}(100,40)(0,-4)
\Arc[arrow,arrowpos=0.50,arrowlength=3.0,arrowwidth=0.8,arrowinset=0.2,clock](20,0)(20,-90,90)
\Arc[arrow,arrowpos=0.50,arrowlength=3.0,arrowwidth=0.8,arrowinset=0.2,clock](20,0)(20,90,0)
\Arc[arrow,arrowpos=0.50,arrowlength=3.0,arrowwidth=0.8,arrowinset=0.2,clock](20,0)(20,0,-90)
\Line[dash,dashsize=1.5](20,-20)(20,20)
\Vertex(20,-20){1.8} \Vertex(20,20){1.8}
\Line[double](40,0)(60,0)
\Arc[arrow,arrowpos=0.50,arrowlength=3.0,arrowwidth=0.8,arrowinset=0.2](80,0)(20,-90,90)
\Arc[arrow,arrowpos=0.50,arrowlength=3.0,arrowwidth=0.8,arrowinset=0.2](80,0)(20,90,0)
\Arc[arrow,arrowpos=0.50,arrowlength=3.0,arrowwidth=0.8,arrowinset=0.2](80,0)(20,0,-90)
\Vertex(40,0){1.8}
\Vertex(60,0){1.8}
\Line[dash,dashsize=1.5](80,-20)(80,20)
\Vertex(80,-20){1.8} \Vertex(80,20){1.8}
\end{picture} 
&=& \frac{2 (-1)^2 V (4\pi e^2)^2}{2\beta \int \frac{d\p}{(2\pi)^3} n_\p (1-n_\p)}
 \int \frac{d\p_1\, d\p_2\, d\q \, d\q'}{(2\pi)^{12}} 
\frac{4\pi e^2}{q^2} \frac{4\pi e^2}{{q'}^2}\times \nonumber \\
&& \times \beta^2 n_{\p_1} (1-n_{\p_1}) n_{\p_2} (1-n_{\p_2})
 n_{\p_1+\q} n_{\p_2+\q'}  \label{counter_anomalous.1.0} \\
 &=& \frac{V (4\pi e^2)^2}{\int \frac{d\p}{(2\pi)^3} \frac{\partial n_\p}{\partial \mu}}
 \left[\int \frac{d\p}{(2\pi)^3} \frac{\partial n_\p}{\partial \mu} f(\p) \right]^2 \; .
  \label{counter_anomalous.1.1}
\eea    
When combined with eq.~(\ref{anomalous.1.1}), one obtains
\be
\frac{1}{2\beta } \; \begin{picture}(40,20)(0,-4)
\Arc[arrow,arrowpos=0.50,arrowlength=3.0,arrowwidth=0.8,arrowinset=0.2](20,0)(20,0,90)
\Arc[arrow,arrowpos=0.50,arrowlength=3.0,arrowwidth=0.8,arrowinset=0.2](20,0)(20,90,180)
\Arc[arrow,arrowpos=0.50,arrowlength=3.0,arrowwidth=0.8,arrowinset=0.2](20,0)(20,180,270)
\Arc[arrow,arrowpos=0.50,arrowlength=3.0,arrowwidth=0.8,arrowinset=0.2](20,0)(20,270,360)
\Vertex(20,-20){1.8}
\Vertex(20,20){1.8}
\Vertex(0,0){1.8}
\Vertex(40,0){1.8} 
\Line[dash,dashsize=1.5](20,20)(40,0)
\Line[dash,dashsize=1.5](0,0)(20,-20) 
\end{picture}\; \; - \frac{1}{2 \beta} \; \begin{picture}(100,30)(0,-4)
\Arc[arrow,arrowpos=0.50,arrowlength=3.0,arrowwidth=0.8,arrowinset=0.2,clock](20,0)(20,-90,90)
\Arc[arrow,arrowpos=0.50,arrowlength=3.0,arrowwidth=0.8,arrowinset=0.2,clock](20,0)(20,90,0)
\Arc[arrow,arrowpos=0.50,arrowlength=3.0,arrowwidth=0.8,arrowinset=0.2,clock](20,0)(20,0,-90)
\Line[dash,dashsize=1.5](20,-20)(20,20)
\Vertex(20,-20){1.8} \Vertex(20,20){1.8}
\Line[double](40,0)(60,0)
\Arc[arrow,arrowpos=0.50,arrowlength=3.0,arrowwidth=0.8,arrowinset=0.2](80,0)(20,-90,90)
\Arc[arrow,arrowpos=0.50,arrowlength=3.0,arrowwidth=0.8,arrowinset=0.2](80,0)(20,90,0)
\Arc[arrow,arrowpos=0.50,arrowlength=3.0,arrowwidth=0.8,arrowinset=0.2](80,0)(20,0,-90)
\Vertex(40,0){1.8}
\Vertex(60,0){1.8}
\Line[dash,dashsize=1.5](80,-20)(80,20)
\Vertex(80,-20){1.8} \Vertex(80,20){1.8}
\end{picture} 
= -V (4\pi e^2)^2 \left( \int \frac{d\p}{(2\pi)^3} \frac{\partial n_\p}{\partial \mu} \right)
\overline{\left(f - \overline{f}\right)^2} \; , \label{Gamma.2.2-3.1} \vspace*{10pt}
\ee  
where the overline symbol is defined as  
$ \overline{f} \equiv 
\int \frac{d\p}{(2\pi)^3} \frac{\partial n_\p}{\partial \mu} f(\p)$. Apparently,
 the expression (\ref{Gamma.2.2-3.1}) is in general negative unless $f = \overline{f}$. 
 We will show that this happens only at zero temperature and only 
 if the system has spherical symmetry.  

In order to have $f(\p) = \overline{f}$, $\p$ can only have support at
 a constant $f(\p)$ surface. This is achieved when $T\to 0$, where 
  $n_\p \to \theta (\mu - \ve_\p)$ and $\partial n_\p /\partial \mu = \delta (\mu - \ve_\p)$
   forcing $\p$ to lie on a constant $\ve_\p$ surface. Furthermore, $f(\p) = \overline{f}$
    demands that $f(\p)$ depends only on
   the magnitude of $\p$, i.e., $f(\p) = f(p)$. This can only be achieved if 
 $n_\p$ depends only on $|\p| = p$. For the HEG system, $\ve_\p = \p^2/2m$, thus 
 $n_\p = n(p)$ and  $\mu = p_F^2/2m$.  We therefore have 
 $\partial n_\p /\partial \mu = m\delta(p-p_F)/p_F $, fixing the length of $\p$.  
 Now $n_{\p+\q} = \theta (\mu - (\p+\q)^2/2m)|_{|\p|=p_F} = \theta (-\cos \vt - \frac{q}{2p_F})$,
  with $\vt$ being the angle between $\q$ and $\p$. Thus, in the integral defining
  $f(\p)$, although $\hat \p$ defines the $\hat z$ direction of vector $\q$ the integral
  is independent of the choice of $\hat \p$. Therefore, the anomalous contribution (\ref{anomalous.1.1})
 may be written as (when $T\to 0$) 
\be
-(V)\left[ \int \frac{d\p_1}{(2\pi)^3}  \frac{m}{p_F}
\delta (p_1-p_F)  \right]
\left[ \int_{-1}^1 dx  \int_0^\infty \frac{q^2 \, dq}{(2\pi)^2}    
\frac{4\pi e^2}{q^2} \; \theta \big( -x - \frac{q}{2p_F} \big) \right]^2 \;,  
 \label{anomalous.1.2} 
\ee  
 while the corresponding subtraction term (\ref{counter_anomalous.1.1}) may be written as  
\bea 
&&\hspace*{-10pt}\frac{V}{\int \frac{d\p_1}{(2\pi)^3}  \frac{m}{p_F}
\delta (p_1-p_F)} \left[ \int \frac{d\p_1}{(2\pi)^3}  \frac{m}{p_F}
\delta (p_1-p_F)\right]^2 \left[ \int_{-1}^1 dx  \int_0^\infty \frac{q^2 \, dq}{(2\pi)^2}    
\frac{4\pi e^2}{q^2} \; \theta \big( -x - \frac{q}{2p_F} \big) \right]^2 \; \nonumber \\
&& = V \left[ \int \frac{d\p_1}{(2\pi)^3}  \frac{m}{p_F}
\delta (p_1-p_F)  \right]
\left[ \int_{-1}^1 dx  \int_0^\infty \frac{q^2 \, dq}{(2\pi)^2}    
\frac{4\pi e^2}{q^2} \; \theta \big( -x - \frac{q}{2p_F} \big) \right]^2 \; ,  
 \label{counter_anomalous.1.2} 
\eea  
cancelling exactly the anomalous contribution (\ref{anomalous.1.2}) in the HEG case.

Within the framework of Luttinger {\it et al.},\cite{KL_60,LW_60} 
the anomalous contribution is cancelled by the chemical potential shift,
 the difference between $\mu (T \to 0)$, under finite temperature formalism, 
 and $\mu (T=0) = p_F^2/2m$, under zero temperature formalism.  
Within the current finite temperature framework, however, the $\mu(T\to 0)$ is identical to
 $\mu(T=0)$ and the cancellation of anomalous contribution is explicit. 
 As we will show later, however, diagram (b) of (\ref{Gamma.2.2.decomp}) can't be 
  cancelled by diagram (b) of (\ref{Gamma.2.3.decomp}). This is because diagram (b) of (\ref{Gamma.2.2.decomp}) is not an anomalous diagram. 
 
Before moving onto (b) diagrams in (\ref{Gamma.2.1.decomp}-\ref{Gamma.2.3.decomp}), 
 let us remark that diagrams within our formalism, i.e.,
 diagrams on the left hand side of (\ref{Gamma.2.1.decomp}-\ref{Gamma.2.3.decomp})
  with $B_{1(2)} \to \tilde {\cal D}_0$ yield
 only finite contribution. The decomposition made in (\ref{Gamma.2.1.decomp}-\ref{Gamma.2.3.decomp}),
  however, may introduce divergence as we will illustrate using the (b) diagrams
 that correspond to the main results of DuBois.\cite{DuBois_59a}

For the (b) diagram of (\ref{Gamma.2.1.decomp}), 
$B_2(\q',\nu') = 4\pi e^2 /(\p_1+\p_2+\q)^2$ 
 while $B_1(\q,\nu) = (4\pi e^2)^2 D_0(\q,\nu)/\left\{\q^2[\q^2- 4\pi e^2 D_0(\q,\nu)] \right\}$. 
Therefore, one may sum over $\nu'$ to simplify the expression. Using again the methods described in 
section~\ref{sec:KS_systematics} for frequency summation, one obtains 
\bea   
\frac{1}{2\beta}\; \;  \begin{picture}(50,25)(0,-4)
\Arc[arrow,arrowpos=0.50,arrowlength=3.0,arrowwidth=0.8,arrowinset=0.2](25,0)(25,0,90)
\Arc[arrow,arrowpos=0.50,arrowlength=3.0,arrowwidth=0.8,arrowinset=0.2](25,0)(25,90,180)
\Arc[arrow,arrowpos=0.50,arrowlength=3.0,arrowwidth=0.8,arrowinset=0.2](25,0)(25,180,270)
\Arc[arrow,arrowpos=0.50,arrowlength=3.0,arrowwidth=0.8,arrowinset=0.2](25,0)(25,270,360)
\Vertex(25,-25){1.8}
\Vertex(25,25){1.8}
\Vertex(0,0){1.8}
\Vertex(50,0){1.8} 
\Line[dash,dashsize=1.5](25,25)(25,16)
\Arc[](25,11)(5,0,360)
\Photon[](25,-25)(25,6){2}{5}
\Line[dash,dashsize=1.5](0,0)(50,0) 
\end{picture} \; &=& \; 
2 \frac{V}{4} \frac{(4\pi e^2)^3}{(2\pi)^9}\frac{2}{\beta}\sum_{\nu} \int d\p_1 d\p_2 d\q 
\frac{D_0(\q,\nu)}{\q^2(\q^2-4\pi e^2 D_0(\q,\nu))} \frac{1}{(\p_1+\p_2+\q)^2} \times \nonumber \\
&& \hspace*{60pt} \times \frac{n_{\p_1+\q} - n_{\p_1}}{-i\nu+\ve_{\p_1+\q} -\ve_{\p_1}}
\, \frac{n_{\p_2+\q} - n_{\p_2}}{i\nu+\ve_{\p_2+\q} -\ve_{\p_2}} \nonumber \\
&=& \; 
\frac{V}{\beta}\frac{(4\pi e^2)^3}{(2\pi)^9}\sum_{\nu} \int d\p_1 d\p_2 d\q 
\frac{D_0(\q,\nu)}{\q^2(\q^2-4\pi e^2 D_0(\q,\nu))} \frac{1}{(\p_1-\p_2)^2} \times \nonumber \\
&& \hspace*{60pt} \times \frac{n_{\p_1+\q} - n_{\p_1}}{-i\nu+\ve_{\p_1+\q} -\ve_{\p_1}}
\, \frac{n_{\p_2+\q} - n_{\p_2}}{-i\nu+\ve_{\p_2+\q} -\ve_{\p_2}} \; , \label{Gamma.2.1.b.0}
\eea
where the last expression is obtained by changing variables: 
$\p_2+\q \to -\p_2'$ followed by $\p_2' \to \p_2$.
For the high density expansion, where $e^2$ is treated as a small parameter, 
the major contribution in (\ref{Gamma.2.1.b.0}) comes from the region $\q \to 0$. One thus writes 
\[
n_{\p+\q} - n_\p \xrightarrow[|\q| \to 0]{} (\ve_{\p+q} - \ve_{\p}) \frac{\partial n_{\p}}{\partial \ve_{\p}}
 = -\beta  n_{\p}(1-n_{\p}) (\ve_{\p+\q}-\ve_{\p}) \xrightarrow[T \to 0]{} 
 - q \cos \vt \, \delta(p-p_F) \; ,
\]
where $\vt$ is  the angle between $\p$ and $\q$. 
As $T\to 0$, $\frac{1}{\beta} \sum_n F(\nu_n) \to 
\int_{-\infty}^\infty \frac{d\nu}{2\pi} F(\nu)$ if $F(\nu)$ does not have pole strength greater
 than one. 
Making a change of variable  $\nu \equiv u |\q|/m$ 
 and treating $q$ as a small quantity, the major contribution 
  of the (b) diagram of (\ref{Gamma.2.1.decomp}) is given by
\bea 
\frac{1}{2\beta}\; \;  \begin{picture}(50,25)(0,-4)
\Arc[arrow,arrowpos=0.50,arrowlength=3.0,arrowwidth=0.8,arrowinset=0.2](25,0)(25,0,90)
\Arc[arrow,arrowpos=0.50,arrowlength=3.0,arrowwidth=0.8,arrowinset=0.2](25,0)(25,90,180)
\Arc[arrow,arrowpos=0.50,arrowlength=3.0,arrowwidth=0.8,arrowinset=0.2](25,0)(25,180,270)
\Arc[arrow,arrowpos=0.50,arrowlength=3.0,arrowwidth=0.8,arrowinset=0.2](25,0)(25,270,360)
\Vertex(25,-25){1.8}
\Vertex(25,25){1.8}
\Vertex(0,0){1.8}
\Vertex(50,0){1.8} 
\Line[dash,dashsize=1.5](25,25)(25,16)
\Arc[](25,11)(5,0,360)
\Photon[](25,-25)(25,6){2}{5}
\Line[dash,dashsize=1.5](0,0)(50,0) 
\end{picture} & \approx &
 \frac{V}{m} \frac{(4\pi e^2)^3}{(2\pi)^9}\int_{-\infty}^\infty \frac{du}{2\pi} 
 \int d\p_1\,d\p_2 \frac{ m \delta(p_1-p_F)\cos \vt_1 }{-iu + p_F\cos \vt_1} 
 \frac{ m \delta(p_2-p_F)\cos \vt_2 }{iu + p_F\cos \vt_2}  \times  \nonumber \\
&& \hspace*{10pt} \frac{1}{(\p_1-\p_2)^2} \int_0^{q_c} 4\pi q dq \frac{D_0(q,uq/m)}
{q^2-4\pi e^2 D_0(q,uq/m)} \nonumber  \\
&\approx & V m \, \frac{(4\pi e^2)^3}{(2\pi)^9}\int_{-\infty}^\infty \frac{du}{2\pi}
\int_{-1}^1 \frac{\cos \vt_1\, d\cos \vt_1}{-iu+p_F\cos \vt_1} 
\, \frac{\cos \vt_2\, d\cos \vt_2}{-iu+p_F\cos \vt_2}
  \times \nonumber \\ 
 &&\hspace*{10pt} \frac{(2\pi)^2 p_F^2/2}{|\cos \vt_1 - \cos \vt_2|} 
 \int 4\pi q dq \frac{D_0(q,uq/m)}{q^2-4\pi e^2 D_0(q,uq/m)} \; .
 \label{Gamma.2.1.b.1}
\eea 
Apparently, the $1/|\cos \vt_1 - \cos \vt_2|$ factor in the integrand causes 
 an undesirable divergence in the $\vt$ integral that is absent if 
 one does not decompose the diagrams using (\ref{tD.Dyson}). 
 A finite result emerges only when one combines (\ref{Gamma.2.1.b.1}) 
  with the (b) diagram of (\ref{Gamma.2.2.decomp}), which we soon turn to.

Before elaborating on the evaluation of the (b) diagram of (\ref{Gamma.2.2.decomp}), 
 we wish to point out that this is the type of diagram (two-particle reducible ones) 
 that was missed in Hedin's approximation~\cite{Hedin_65} but should have been included
   for the exactness of the theory. 
For this diagram, $B_2(\q',\nu') = 4\pi e^2 /(\p_1-\p_2)^2$ 
 while $B_1(\q,\nu) = (4\pi e^2)^2 D_0(\q,\nu)/\left\{\q^2[\q^2- 4\pi e^2 D_0(\q,\nu)] \right\}$. 
Summing over $\nu'$ via methods described in 
section~\ref{sec:KS_systematics}, one obtains 
\bea 
\frac{1}{\beta}\; \begin{picture}(50,25)(0,-4)
\Arc[arrow,arrowpos=0.50,arrowlength=3.0,arrowwidth=0.8,arrowinset=0.2](25,0)(25,0,90)
\Arc[arrow,arrowpos=0.50,arrowlength=3.0,arrowwidth=0.8,arrowinset=0.2](25,0)(25,90,180)
\Arc[arrow,arrowpos=0.50,arrowlength=3.0,arrowwidth=0.8,arrowinset=0.2](25,0)(25,180,270)
\Arc[arrow,arrowpos=0.50,arrowlength=3.0,arrowwidth=0.8,arrowinset=0.2](25,0)(25,270,360)
\Vertex(25,-25){1.8}
\Vertex(25,25){1.8}
\Vertex(0,0){1.8}
\Vertex(50,0){1.8} 
\Line[dash,dashsize=1.5](25,25)(32.071,17.929)
\Arc[](35.606,14.3934)(5,0,360)
\Photon[](39.142,10.858)(50,0){2}{4}
\Line[dash,dashsize=1.5](0,0)(25,-25) 
\end{picture} &=& V \frac{(4\pi e^2)^3}{(2\pi)^9} \frac{2}{\beta} \sum_\nu \int d\p_1 d\p_2\, d\q
\frac{1}{(\p_1-\p_2)^2} \frac{D_0(q,\nu)}{q^2(q^2-4\pi e^2 D_0(q,\nu))} \times 
\nonumber \\
&& \hspace*{10pt}\left[ -\frac{n_{\p_2+\q}(n_{\p_1+\q}-n_{\p_1})}{(-i\nu + \ve_{\p_1+\q}-\ve_{\p_1})^2 } 
 - \frac{\beta n_{\p_1} (1-n_{\p_1}) n_{\p_2} }{-i\nu + \ve_{\p_1}-\ve_{\p_1+\q}}\right] \nonumber \\
 &=& \frac{V}{\beta} \frac{(4\pi e^2)^3}{(2\pi)^9}\sum_\nu \int d\p_1 d\p_2 \,d\q
\frac{1}{(\p_1-\p_2)^2} \frac{D_0(q,\nu)}{q^2(q^2-4\pi e^2 D_0(q,\nu))} \times 
\nonumber \\
&& \hspace*{10pt}\left[ -\frac{(n_{\p_2+\q} - n_{\p_2})(n_{\p_1+\q}-n_{\p_1})}
{(-i\nu + \ve_{\p_1+\q}-\ve_{\p_1})^2 } 
 + \frac{2\beta n_{\p_1} (1-n_{\p_1}) n_{\p_2} }{-i\nu + \ve_{\p_1+\q}-\ve_{\p_1}}\right] 
 \nonumber \\
 &\xrightarrow[T \to 0]{|\q| \to 0}& (- V m) \, \frac{(4\pi e^2)^3}{(2\pi)^9}\int_{-\infty}^\infty \frac{du}{2\pi}
\int_{-1}^1 \frac{\cos \vt_1\,\cos \vt_2\, d\cos \vt_1\, d\cos \vt_2}{(-iu+p_F\cos \vt_1)^2} 
  \times \nonumber \\ 
 &&\hspace*{30pt} \frac{(2\pi)^2 p_F^2/2}{|\cos \vt_1 - \cos \vt_2|} 
 \int 4\pi q dq \frac{D_0(q,uq/m)}{q^2-4\pi e^2 D_0(q,uq/m)} \nonumber \\
 && + (2V) \frac{(4\pi e^2)^3}{(2\pi)^3}\int_{-\infty}^\infty 
  \frac{du}{2\pi} \int_{-1}^1 \frac{ 4\pi m p_F \, d\!\cos \vt }{-iu + p_F \cos \vt}
  \int \frac{dq }{(2\pi)^2} 
  \frac{D_0(q,uq/m)}{q^2-4\pi e^2 D_0(q,uq/m)} \times \nonumber \\
&& \hspace*{15pt}  \left[\int \frac{ q'^2 dq' d\!\cos \vt'  }{(2\pi)^2} 
   \frac{1}{(\q')^2} \theta(-\cos \vt' - \frac{q'}{2p_F}) \right] \; 
\label{Gamma.2.2.b.1}   \\
&\equiv & (-Vm)L_1 +(2V)L_2 \; , \nonumber
\eea
where the first term inside the square brackets after the 
 second equal sign is obtained by
 adding to it an equivalent expression with $\p_1+\q \to -\p_1$, $\p_2+\q \to -\p_2$,
  $\nu \to -\nu$, and then taking the average, while  
 the second term inside the same square brackets   
  results from changing $\nu \to -\nu$.  The $L_2$ part, where the dummy variable of 
  the integral is switched from $\p_2$ to $\q' \equiv \p_2 - \p_1$, can be
   cancelled by one of the terms contributing to the (b) diagram of (\ref{Gamma.2.3.decomp}).
  The $L_1$ part, however, upon taking the $T \to 0$ and 
   $q \to 0$ limits, may be combined with (\ref{Gamma.2.1.b.1}) to yield a finite 
  expression 
\bea 
&&\hspace*{-10pt} \frac{V m p_F^3}{2} \, \frac{(4\pi e^2)^3}{(2\pi)^7}\int_{-\infty}^\infty \frac{du}{2\pi}
\int_{-1}^1 \frac{\cos \vt_1\, d\cos \vt_1}{(-iu+p_F\cos \vt_1)^2} \, 
\frac{\cos \vt_2\, d\cos \vt_2}{-iu+p_F\cos \vt_2}
  \times \nonumber \\ 
 &&\hspace*{10pt} \frac{\cos \vt_1 - \cos \vt_2}{|\cos \vt_1 - \cos \vt_2|} 
 \int 2\pi \; d(q^2) \frac{D_0(q,uq/m)}{q^2-4\pi e^2 D_0(q,uq/m)} \; .
\eea 
Note that this expression, in agreement with Carr and Maradudin,\cite{CM_64} 
 carries a different sign when compared to the original result of DuBois.\cite{DuBois_59a}  

It was conjectured before~\cite{VF_97_arxiv} 
that cancellation of diagrams of the following form always hold true for HEG\vspace*{-6pt}
\[
\begin{picture}(30,15)(0,-4)
\Arc[arrow,arrowpos=0.50,arrowlength=3.0,arrowwidth=0.8,arrowinset=0.2](15,0)(15,0,180)
\Arc[arrow,arrowpos=0.50,arrowlength=3.0,arrowwidth=0.8,arrowinset=0.2](15,0)(15,180,360)
\Vertex(0,0){8}
\Vertex(30,0){5} 
\end{picture} \;\;\; + \;\;\;\;\; 
\begin{picture}(80,15)(0,-4)
\Arc[arrow,arrowpos=0.50,arrowlength=3.0,arrowwidth=0.8,arrowinset=0.2](15,0)(15,0,180)
\Arc[arrow,arrowpos=0.50,arrowlength=3.0,arrowwidth=0.8,arrowinset=0.2](15,0)(15,180,360)
\Vertex(0,0){8}
\Vertex(30,0){1.8}
\Vertex(45,0){1.8} 
\Line[double](30,0)(45,0)
\Arc[arrow,arrowpos=0.50,arrowlength=3.0,arrowwidth=0.8,arrowinset=0.2](60,0)(15,0,180)
\Arc[arrow,arrowpos=0.50,arrowlength=3.0,arrowwidth=0.8,arrowinset=0.2](60,0)(15,180,360)
\Vertex(75,0){5} 
\end{picture} \; = \; 0 \; . \vspace*{4pt}
\]
Here, black circles denote parts of the diagram that are connected to each
other via two propagators. If this is the case, then the contributions of the 
(b) diagrams of (\ref{Gamma.2.2.decomp}) and (\ref{Gamma.2.3.decomp}) will cancel each other.
 We don't expect this to happen since the (b) diagram of (\ref{Gamma.2.2.decomp}) 
  is not anomalous. To illustrate that the (b) diagram of (\ref{Gamma.2.3.decomp}) does not
 eliminate the (b) diagram of (\ref{Gamma.2.2.decomp}), we now proceed to 
  evaluate the (b) diagram of (\ref{Gamma.2.3.decomp}).

Here, $B_2(\q',\nu') = 4\pi e^2 /(\q')^2$ 
 while $B_1(\q,\nu) = (4\pi e^2)^2 D_0(\q,\nu)/\left\{\q^2[\q^2- 4\pi e^2 D_0(\q,\nu)] \right\}$. 
Summing over $\nu'$ via methods described in 
section~\ref{sec:KS_systematics}, one obtains 
\bea
-\frac{1}{\beta}\; \begin{picture}(90,40)(0,-4)
\Arc[arrow,arrowpos=0.50,arrowlength=3.0,arrowwidth=0.8,arrowinset=0.2,clock](20,0)(20,-90,90)
\Arc[arrow,arrowpos=0.50,arrowlength=3.0,arrowwidth=0.8,arrowinset=0.2,clock](20,0)(20,90,0)
\Arc[arrow,arrowpos=0.50,arrowlength=3.0,arrowwidth=0.8,arrowinset=0.2,clock](20,0)(20,0,-90)
\Line[dash,dashsize=1.5](20,-20)(20,20)
\Vertex(20,-20){1.8} \Vertex(20,20){1.8}
\Line[double](40,0)(50,0)
\Arc[arrow,arrowpos=0.50,arrowlength=3.0,arrowwidth=0.8,arrowinset=0.2](70,0)(20,-90,90)
\Arc[arrow,arrowpos=0.50,arrowlength=3.0,arrowwidth=0.8,arrowinset=0.2](70,0)(20,90,0)
\Arc[arrow,arrowpos=0.50,arrowlength=3.0,arrowwidth=0.8,arrowinset=0.2](70,0)(20,0,-90)
\Vertex(40,0){1.8}
\Vertex(50,0){1.8}
\Line[dash,dashsize=1.5](70,20)(70,10)
\Arc[](70,4)(6,0,360)
\Photon[](70,-2)(70,-20){2}{4}
\Vertex(70,-20){1.8} \Vertex(70,20){1.8}
\end{picture} &=& -(2V) \frac{(4\pi e^2)^3 /(2\pi)^{12}} 
{2\beta \int \frac{d\p}{(2\pi)^3} n_\p(1-n_\p)} 
\frac{2}{\beta} \sum_\nu \int d\p_1 d\p_2\, d\q\, d\q' \frac{n_{\p_2+\q'}}{(\q')^2}  \times 
\nonumber \\
&& \frac{D_0(q,\nu) \beta n_{\p_2}(1-n_{\p_2})} {q^2(q^2-4\pi e^2 D_0(q,\nu))} 
\left[\frac{\beta n_{\p_1}(1- n_{\p_1})}{-i\nu + \ve_{\p_1+\q}-\ve_{\p_1} } 
 + \frac{n_{\p_1 + \q} -n_{\p_1} }{(-i\nu + \ve_{\p_1+\q}-\ve_{\p_1})^2}\right] \nonumber \\
 &\xrightarrow[T \to 0]{|\q| \to 0}& 
  \frac{-(2V)(4\pi e^2)^3 /(2\pi)^{3}} 
{2 \int \frac{d\p}{(2\pi)^3} \delta(\mu-\ve_\p) } \int \frac{du}{2\pi}
  \int d\p_1 \frac{d\q}{(2\pi)^3} \frac{q}{m}
 \frac{D_0(q,uq/m)}{q^2(q^2-4\pi e^2 D_0(q,uq/m))} \times 
\nonumber \\
&& \hspace*{5pt}\left[ 2 \int \frac{d\p_2\, d\q'}{(2\pi)^6} \delta (\mu - \ve_{\p_2}) \frac{n_{\p_2+\q'}}{(\q')^2} \right] \times \nonumber \\
&& \hspace*{5pt}\left[ \frac{\delta(\mu-\ve_{\p_1}) m/q}{-iu + p_F \cos \vt_1} + 
\frac{\delta (\mu-\ve_{\p_1}) m^2/q^2}{(-iu + p_F \cos \vt_1)^2}\frac{qp_F \cos \vt_1}{m} \right] \nonumber \\
&=&  \frac{-(2V)(4\pi e^2)^3 /(2\pi)^{3}} 
{2 \int \frac{d\p}{(2\pi)^3} \delta(\mu-\ve_\p) } \int \frac{du}{2\pi}
  \int d\p_1 \frac{d\q}{(2\pi)^3}
 \frac{D_0(q,uq/m)}{q^2(q^2-4\pi e^2 D_0(q,uq/m))} \times 
\nonumber \\
&& \hspace*{5pt}\left[ 2 \int \frac{d\p_2}{(2\pi)^3} \delta (\mu - \ve_{\p_2})  \right]  
\left[ \int \frac{d\q'}{(2\pi)^3} \frac{\theta(-\cos \vt' - \frac{q'}{2p_F})}{(\q')^2} \right]\times \nonumber \\
&& \hspace*{5pt}\left[ \frac{\delta (p_1-p_F) m/p_F}{-iu + p_F \cos \vt_1} + 
\frac{\delta (p_1-p_F) m \cos \vt_1}{(-iu + p_F \cos \vt_1)^2} \right] \nonumber \\
&=& -(2V) \frac{(4\pi e^2)^3}{(2\pi)^{3}} \int_{-\infty}^\infty \frac{du}{2\pi} \int_{-1}^1 
 \frac{4\pi m p_F d\cos \vt_1}{-iu+p_F \cos \vt_1} \left( 1 + \frac{p_F\cos \vt_1}{-iu+p_F \cos \vt_1}\right)
 \times \nonumber \\
&& \hspace*{5pt} \left[\int \frac{dq}{(2\pi)^2} \frac{D_0(q,uq/m)}{q^2-4\pi e^2 D_0(q,uq/m)}
\right] \left[ \int \frac{d\q'}{(2\pi)^3} \frac{\theta(-\cos \vt' - \frac{q'}{2p_F})}{(\q')^2}\right] \; .
\label{Gamma.2.3.b.1}
\eea
The contribution of (\ref{Gamma.2.3.b.1}) can be easily divided in two by explicitly 
 expanding the two parts inside the round parentheses. 
 It is obvious that the contribution associated with  
$1$ cancels exactly the $(2V)L_2$ part of
 (\ref{Gamma.2.2.b.1}) while the contribution associated with
 $\frac{p_F\cos \vt_1}{-iu+p_F\cos \vt_1}$ cannot cancel 
 the $(-Vm)L_1$ part of (\ref{Gamma.2.2.b.1}).

For the (c) diagrams of (\ref{Gamma.2.1.decomp}-\ref{Gamma.2.3.decomp}), both $B_1$ and $B_2$ are of order 
$(4\pi e^2)^2$, leading to contributions of order $(e^2)^4$ and higher. The (c) diagrams thus already lead us 
  beyond what was studied  by DuBois~\cite{DuBois_59a} and by Carr and Maradudin.\cite{CM_64} 
 The last two diagrams of $\Gamma_2[n]$ (shown in Figure~\ref{fig:Gamma.2}) gives rise to
  the $E'_3$ term of Carr and Maradudin~\cite{CM_64} when the $\tilde {\mathcal D}_0$ lines
   are each replaced by $U$, the first term in the decomposition of $\tilde {\mathcal D}_0$.    
  In principle, one may go on to study terms of order $(e^2)^4$; we will not, however, 
   delve into this endeavor since this is not the primary aim here.  
 We would like to emphasize the following points. First, unlike conventional 
  $e^2$ based perturbation theory, the formalism presented here naturally avoids 
 divergence. This is shown by the fact that each diagram in our formalism contains no
  singularity while attempts to perform $e^2$ based purturbation necessarily require 
   further re-grouping, such as combining the (b) diagrams of (\ref{Gamma.2.1.decomp}) 
   and (\ref{Gamma.2.2.decomp}), to tame the divergence.  
  Second, even if one were to pursue $e^2$ expansion, using Dyson's equation (\ref{tD.Dyson})
   within our formalism still makes the task straightforward. In fact, as shown in this section,
    the $\tilde \Gamma_0[n] + \Gamma_1[n]+\Gamma_2[n]$ part when applied to the HEG already contain the 
 celebrated results of Carr and Maradudin.\cite{CM_64} 
   Third, the removal of anomalous contributions that required special attention within the formalism of
    Luttinger {\it et al.}~\cite{KL_60,LW_60} becomes automatic under this formalism.

\section{Excitations} \label{sec:excitation}
  To obtain information regarding the excitations, one needs a time-dependent probe, although
   one with infinitesimal amplitude is sufficient.  
 Runge and Gross~\cite{RG_84} extended the correspondence between the 
   external probe potential and the ground state charge density of the DFT to the time-dependent case. 
    This relationship provides the foundation for studying excitation energies under DFT.
Fukuda {\it et al.}~\cite{FKSY_94} expressed the excitation 
   energy condition using effective action formalism, without explicit connection to the Kohn-Sham formalism. 
  By introducing time-dependent Kohn-Sham orbitals while assuming time idependence of 
  the orbital occupation numbers, Casida~\cite{Casida_95a} derived via linear response theory 
  a self-consistent condition on the density matrix response that leads to determination of excitation energies. 
 Along a similar line, Petersilka {\it et al.}~\cite{PGG_96} proposed the so-called optimized effective potential
  (expanded in time-dependent Kohn-Sham orbitals) to tackle the problem of excitation energies.  
 Since there exist formalisms to extract the excitation energies of the system provided
 that the UDF is known, our result for excitation energies should not be considered novel.
 The reasons for this section are twofold. First, we  would like to explicitly 
 show that the excitation energies can be obtained using the formalism of
  section~\ref{sec:formulation} without introducing time-dependent orbitals. 
 Second, although it is possible to find derivations of formulas~\cite{VF_97_arxiv} similar 
 to those that will be shown here, they do appear slightly different and thus a self-contained 
 exposition may be helpful. 
    
 Intuitively speaking, by varying the frequency of the probe, 
  one seeks the frequency/energy where the amplitude of the response function diverges. 
Indeed, it is known that the spectral representation of the correlation  function
 has poles at the excitation energies of the system~\cite{Kobe_62}.  Having obtained the effective action
  $\Gamma[n]$, we also note that the second (functional) derivative of $\Gamma[n]$ 
 with respect to the local electron density is the inverse of the 
 density-density correlation function. Therefore, any pole associated with 
  the correlation function becomes a root of the effective action. It can be shown that 
  upon analytic continuation the correlation function, obtained using the imaginary time 
  (finite temperature) formalism, can be turned into the response function of the real time. 
 Below we briefly illustrate this point. Readers interested in more details can find an
  extensive exposition in reference~\onlinecite{FW_71}.
      
From Eqs.~(\ref{W2n}) and (\ref{Gamma2J}), we know that 
\bean
n(x) &=& {\delta (\beta W[J])\over \delta J(x)} \; , \\
{\rm and~~~~}J(x) &=& - \frac{\delta \Gamma[n]}{\delta n(x)}  \; .
\eean     
One then considers
\be \label{GnW.id.0}
\delta (x-y) = \frac{\delta J(x)}{\delta J(y)}
 = -\int dz \frac{\delta^2 \Gamma[n]}{\delta n(x)\delta n(z)} \frac{\delta n(z)}{\delta J(y)}
 = -\int dz \frac{\delta^2 \Gamma[n]}{\delta n(x)\delta n(z)} \frac{\delta^2 (\beta W[J])}{\delta J(z)\delta J(y)} \; .
\ee
Although a time-dependence of $J$ is introduced to probe the excitations, in the end we
 will return to a time-independent source ($J(x) \to J(\x)$) while computing the excitation energies. 
As we will show below, it is most convenient to go to the zero temperature limit to compute the
 excitation energies.

Note that 
\bea
\frac{\delta^2 (\beta W[J])}{\delta J(x)\delta J(y)} &=& -\left[ \hat n(x) \hat n(y) \right]_T + 
\left[ \hat n(x) \right]_T  \left[ \hat n(y) \right]_T  = -\left[ \hat n(x) \hat n(y) \right]_T 
 + n(x) n(y) \nonumber \\
 & = & -\left[ (\hat n(x) -n(x)) (\hat n(y)-n(y)) \right]_T \equiv -\left[ \tilde n(x)\,\tilde n(y) \right]_T
\eea 
 where $ \hat n(x) = \psid(x) \psi(x)$ is the electron density operator
 and  the square  bracket $[ ]_T$ indicates an (imaginary) time-ordered thermal average 
  such that  
\[ 
 \left[\hat {\mathcal O}_1(t_1) \hat {\mathcal O}_2(t_2) \right]_T = \frac
 {\Tr \left[ e^{-\beta H[J]} T\left( \hat {\mathcal O}_1(t_1) \hat {\mathcal O}_2(t_2) \right) \right]}
 {\Tr \left[ e^{-\beta H[J]} \right]} 
 = \frac{\Tr \left[ e^{-\beta H[J]} T \left( \hat {\mathcal O}_1(t_1)\hat {\mathcal O}_2(t_2) \right) \right]}{Z[J]} \; .
 \]

 Let us denote by $\{\vert \ell \ra_{ex} \}_{\ell=0}^\infty $  the eigenstates of the Hamiltonian $H[J]$ with the corresponding 
  eigenenergies $\{{\mathcal E}_{\ell} \}_{\ell =0}^\infty$. 
The spectral representation is obtained by first writing the time-ordered product (assuming that 
 operators $\hat {\mathcal O}_1(t_1)$ and $\hat {\mathcal O}_2(t_2)$ are bosonic) as
\[
\phantom{}_{ex}\!\la \ell \vert T\left( \hat {\mathcal O}_1(t_1) \hat {\mathcal O}_2(t_2) \right) \vert \ell \ra_{ex} 
 = \theta(t_1-t_2)\, \phantom{}_{ex}\!\la \ell \vert \hat {\mathcal O}_1(t_1) \hat {\mathcal O}_2(t_2) \vert \ell \ra_{ex}  
  + \theta(t_2-t_1) \, \phantom{}_{ex}\!\la \ell \vert \hat {\mathcal O}_2(t_2) \hat {\mathcal O}_1(t_1) \vert \ell \ra_{ex} \; ,  
\]
 then by inserting the identity operator $\sum_{\ell\, '} \vert \ell\, ' \ra_{ex} \;\phantom{}_{ex}\!\la \ell\, ' \vert$ between the two 
  operators, and finally by multiplying by $e^{i\omega (t_1-t_2)}$ and then integrating over the time variable $t_1-t_2$.
 Proceeding in this way, one will then obtain information on ${\mathcal E}_{\ell\, '} - {\mathcal E}_\ell$. Since knowing  
 $\Omega_{\ell\,'} \equiv {\mathcal E}_{\ell\, '} - {\mathcal E}_0$ also provides complete information 
 on ${\mathcal E}_{\ell\, '}-{\mathcal E}_\ell$, one may also focus on $\ell = 0$  
  by taking the limit $\beta \to \infty$.  Let
\be  \label{W2.def}
-W^{(2)}(x,y) \equiv  -\frac{\delta^2 (\beta W[J])}{\delta J(x)\delta J(y)} \; ,
\ee   
and 
\be 
-W^{(2)}(\x,\y,i\nu_n) \equiv  \int_0^\beta d(\tau_x-\tau_y) e^{i\nu_n (\tau_x -\tau_y)}\; W^{(2)}(x,y) 
\equiv   \int_{-\infty}^\infty \frac{d\omega'}{2\pi} \frac{A(\omega')}{i\nu_n -\omega'} \; ,
\label{W2.freq}
\ee   
with $\nu_n = 2\pi n /\beta$ and 
\be \label{spectral.weight}
A(\omega) = e^{\beta W[J]} \sum_{\ell,m} e^{-\beta {\mathcal E}_\ell } \left(e^{-\beta \omega} -1 \right)
 2\pi \delta (\omega + {\mathcal E}_\ell - {\mathcal E}_m )\,  \phantom{}_{ex}\!\la \ell \vert \tilde n(\x) \vert m\ra_{ex}
 \, \phantom{}_{ex}\!\la m \vert \tilde n(\y) \vert \ell \ra_{ex}  \; .
\ee
Note that $\tilde n(\x)$ measures the deviation from the thermally averaged electronic density $n_T(\x)$.
Since the expression (\ref{W2.def}) is evaluated at static $J(\x)$, the Hamiltonian contains no time dependence.
 We may thus write $\tilde n(x) = e^{H \tau_x} \tilde n(\x) e^{- H\tau_x}$.

Let's now express using real time the retarded correlation function (${\mathcal R}$), also called the response function, 
 and the advanced correlation function (${\mathcal A}$) as follows (with $n(x) = e^{it_xH} n(\x) e^{-it_xH}$) 
\bea
\left( \begin{array}{c}{\cal R}(x,y)\\ {\mathcal A} (x,y)\end{array} \right) 
&=& \left(\begin{array}{r}-i\theta(t_x-t_y)\\
i\theta (t_y-t_x) \end{array} \right) e^{\beta W[J]} \Tr \left(e^{-\beta H} \left[\tilde n(x), \tilde n(y) \right] \right) \nonumber \\
 &=& \left(\begin{array}{r}-i\theta(t_x-t_y)\\
i\theta (t_y-t_x) \end{array} \right)  \sum_{\ell, m} \, e^{\beta( W[J]- {\mathcal E}_\ell)} \left[ e^{i(t_x-t_y)({\mathcal E}_\ell-{\mathcal E}_m)}\, \phantom{}_{ex}\!\la \ell \vert \tilde n(\x)
  \vert m \ra_{ex}\,   \phantom{}_{ex}\!\la m \vert \tilde n(\y) \vert \ell \ra_{ex} \right. \nonumber \\
  && \left. \hspace*{90pt} - e^{-i(t_x-t_y)({\mathcal E}_\ell-{\mathcal E}_m)} 
   \phantom{}_{ex}\!\la \ell \vert \tilde n(\y)
  \vert m \ra_{ex}\,   \phantom{}_{ex}\!\la m \vert \tilde n(\x) \vert \ell \ra_{ex} \right] \; . \label{response.def}
\eea
Taking the Fourier transform of the response function, we consider
\bea
\left(\!\! \begin{array}{c}{\cal R}(\x,\y, \omega)\\ {\mathcal A} (\x,\y, \omega)\end{array} \!\! \right) 
&=& \int_{-\infty}^\infty d(t_x-t_y)\;  e^{i\omega(t_x-t_y)} \
\left( \begin{array}{c}{\cal R}(x,y)\\ {\mathcal A} (x,y)\end{array} \right)  \nonumber \\
&=& e^{\beta W[J]} \sum_{\ell,m} e^{-\beta {\mathcal E}_\ell }  \frac{e^{\beta ({\mathcal E}_\ell - {\mathcal E}_m )}-1}
{\omega+ {\mathcal E}_\ell - {\mathcal E}_m \pm i\eta}\,  \phantom{}_{ex}\!\la \ell \vert \tilde n(\x) \vert m\ra_{ex}\, 
 \phantom{}_{ex}\!\la m \vert \tilde n(\y) \vert \ell \ra_{ex} \nonumber \\
&=& \int_{-\infty}^\infty \frac{d\omega'}{2\pi} \frac{A(\omega')}{\omega -\omega' \pm i\eta} \; . \label{response.freq}
\eea  
Comparing Eqs.~(\ref{W2.freq}) and (\ref{response.freq}), one finds that substituting $i\nu_n \to \omega + i\eta$ ($\omega-i\eta$) 
in the imaginary time-ordered  correlation function leads to the retarded (advanced) correlation function. The validity of this
 analytic continuation was discussed by Baym and Mermin~\cite{BM_61}. 

As $T\to 0$, $e^{-\beta W[J]} = e^{-\beta {\mathcal E}_0} \left[ 1+ {\mathcal O}(e^{-\beta ({\mathcal E}_1 - {\mathcal E}_0)}) \right]$, and $e^{\beta W[J]} = e^{\beta {\mathcal E}_0} \left[ 1 - {\mathcal O}(e^{-\beta ({\mathcal E}_1 - {\mathcal E}_0)}) \right]$. Under this limit, we may rewrite the spectral weight $A(\omega)$ as 
\bea 
\lim_{\beta \to \infty} \frac{A(\omega)}{2\pi} &=&  \sum_{\ell,m} \left( e^{-\beta ({\mathcal E}_m - {\mathcal E}_0)} - 
 e^{-\beta ({\mathcal E}_\ell - {\mathcal E}_0)} \right)
\delta (\omega + {\mathcal E}_\ell - {\mathcal E}_m ) 
\phantom{}_{ex}\!\la \ell \vert \tilde n(\x) \vert m\ra_{ex}\,  \phantom{}_{ex}\!\la m \vert \tilde n(\y) \vert \ell \ra_{ex} \nonumber \\
&&  + {\mathcal O}(e^{-\beta ({\mathcal E}_1 - {\mathcal E}_0 )} )\nonumber \\
&=& \sum_{\ell} \, \left[ \, \delta (\omega + {\mathcal E}_\ell - {\mathcal E}_0 ) 
 \phantom{}_{ex}\!\la 0 \vert \tilde n(\y) \vert \ell \ra_{ex}
 \, \phantom{}_{ex}\!\la \ell \vert \tilde n(\x) \vert 0\ra_{ex} \right. \nonumber \\
 && \hspace*{10pt} \left.  - \delta (\omega - ({\mathcal E}_\ell - {\mathcal E}_0 ))
   \phantom{}_{ex}\!\la 0 \vert \tilde n(\x) \vert \ell \ra_{ex}\,  \phantom{}_{ex}\!\la \ell \vert \tilde n(\y) \vert 0 \ra_{ex}
 \right] \nonumber \\
 & \equiv & \sum_{\ell} \left[ \delta (\omega + {\mathcal E}_\ell - {\mathcal E}_0 ) n_{\ell,e}^*(\y) n_{\ell,e}(\x)
  - \delta (\omega - ({\mathcal E}_\ell - {\mathcal E}_0 ) )n_{\ell,e}^*(\x) n_{\ell,e}(\y)   \right]
\label{spectral.weight.ZT}
\eea
where $n_{\ell,e}(\y) \equiv \phantom{}_{ex}\!\la \ell \vert \tilde n(\y) \vert 0 \ra_{ex} $ .

When continued to the retarded correlation function (response function), the density correlation function
 $W^{(2)}$ reads 
\be
\lim_{\beta \to \infty} W^{(2)}(\x,\y,\omega ) =
 \sum_\ell \left[ \frac{n^*_{\ell,e}(\x) \, n_{\ell,e} (\y)}{\omega-\Omega_\ell + i\eta} 
       - \frac{n^*_{\ell,e}(\y)\, n_{\ell,e}(\x)}{\omega + \Omega_\ell + i\eta}\right] \; , \label{W2.pole}
\ee 
 with
\be \label{excitation.energy}
\Omega_\ell \equiv {\mathcal E}_\ell -{\mathcal E}_0  \; . 
\ee
 Provided that the amplitudes $\phantom{}_{ex}\!\la 0 \vert \tilde n\vert \ell \ra_{ex} $ are nonzero, we see from Eq.~(\ref{W2.pole}) that  
 $\omega +i\eta = \pm \left( {\mathcal E}_l - {\mathcal E}_0 \right)$ 
 are simple poles of $W^{(2)}(\x,\y,\omega )$. Furthermore, we also see that  
\be \label{W2*.def}
\left(W^{(2)}(\x,\y,\omega )\right)^* = W^{(2)}(\x,\y,-\omega^* )  \; ,
\ee 
and when $\omega$ is real  
\[
\left(W^{(2)}(\x,\y,\omega )\right)^* = W^{(2)}(\x,\y,-\omega ) \; .
\]

Let us also define 
\[
\Gamma^{(2)}(x,y) \equiv \frac{\delta^2 \Gamma[n]}{\delta n(x) \delta n(y)} \; .
\]
Eq.~(\ref{GnW.id.0}) may thus be rewritten as 
\be \label{GnW.id.1}
-\delta(x-y) = \int dz \; \Gamma^{(2)}(x,z)\, W^{(2)}(z,y) \; .
\ee
Since eventually, $\tau_x $ must agree with $\tau_y$ in the equation above,
 if $\tau_z > \tau_y$, we must have $\tau_z > \tau_x$ as well. 
Similarly to Eq.~(\ref{W2.freq}), the Fourier transform for $\Gamma^{(2)}$can be written as
\be\label{Gn2.freq}
\Gamma^{(2)}(\x,\z,i\nu_n) \equiv \int_0^\beta \!\! d(\tau_z-\tau_x)\, e^{i\nu_n (\tau_z -\tau_x)}\, \Gamma^{(2)}(x,z) \; .
\ee
The inverse transform of (\ref{W2.freq}) and (\ref{Gn2.freq}) can be written as 
\bea
W^{(2)}(z,y) & = & \frac{1}{\beta} \sum_n e^{-i\nu_n(\tau_z-\tau_y)} W^{(2)}(\z,\y,i\nu_n) \nonumber \\
\Gamma^{(2)}(x,z) &=& \frac{1}{\beta} \sum_n e^{-i\nu_n(\tau_z-\tau_x)} \Gamma^{(2)}(\x,\z,i\nu_n) \nonumber
\eea  
and 
\[
\int_0^\beta\!\!\!\! d\tau_z \!\! \int\!\! d\z \;\Gamma^{(2)}(x,z) W^{(2)}(z,y) = \int d\z \, \frac{1}{\beta}\sum_n \, e^{-i\nu_n(\tau_x-\tau_y)} 
 \Gamma^{(2)}(\x,\z,-i\nu_n)\, W^{(2)}(\z,\y,i\nu_n) \; .
\]  
Since $\delta (\tau_x-\tau_y) = \frac{1}{\beta} \sum_n e^{-i\nu_n(\tau_x-\tau_y)} $, one obtains 
\be
-\delta(\x-\y) = \int \!\! d\z \; \Gamma^{(2)}(\x,\z,-i\nu_n) W^{(2)}(\z,\y,i\nu_n) 
= \int \!\! d\z \; W^{(2)}(\x,\z,-i\nu_n) \Gamma^{(2)}(\z,\y,i\nu_n)\;,
\ee
which under analytic continuation becomes 
\be \label{GnW.id.space}
-\delta(\x-\y) = \int \!\! d\z \; \Gamma^{(2)}(\x,\z,-\omega ) W^{(2)}(\z,\y,\omega ) 
 = \int \!\! d\z \; W^{(2)}(\x,\z,-\omega  ) \Gamma^{(2)}(\z,\y,\omega )\; . 
\ee 
 From Eqs.~(\ref{W2*.def}) and (\ref{GnW.id.space}), one has 
\be \label{Gn2*.def}
\left( \Gamma^{(2)}(\x,\z,\omega) \right)^* = \Gamma^{(2)}(\x,\z,-\omega^*) \; .
\ee 
Multiplying the LHS and the middle expression 
of Eq.~(\ref{GnW.id.space}) by $(\omega \mp \Omega_\ell + i\eta)$ and then setting  
 $\omega \to \pm \, \Omega_\ell - i\eta $, we see that 
 \bea
 \int\!\! d\z  \, \Gamma^{(2)}(\x,\z, -\omega \to -\Omega_{\ell} -i\eta) \, n^*_{\ell,e} (\z) &=& 0  \; , \label{eigen.G.1}\\
{\rm and~~~~} \int\!\! d\z  \, \Gamma^{(2)}(\x,\z, -\omega \to +\Omega_{\ell} -i\eta ) \, n_{\ell,e} (\z) &=& 0  \; . 
\label{eigen.G.2}\;, 
 \eea
 That is, $n_{\ell,e}^{(*)}(\y)$ become eigenvectors of $\Gamma^{(2)}(\x,\y, -\omega)$ with zero eigenvalues. 
 The matter of finding excitation energies and the corresponding electronic densities 
  thus reduces to finding for $\Gamma^{(2)}$ the eigenvectors with zero eigenvalue.\cite{FKSY_94}  
  Since we are interested in obtaining the excitation energy  
 under the physical condition, $J(\x) \to 0$, this also means that 
  the derivatives of $\Gamma$ above are evaluated at the ground state  electronic density
   in the zero temperature limit.

From the expressions (\ref{Gamma.loop}) and (\ref{Gamma.0.def}), one sees that 
the effective action is split into  the free particle part $\Gamma_0$, the Hartree functional and the exchange-correlational functional
\be 
\Gamma[n] = \Gamma_0[n] + \frac{1}{2} n \4di \, U \4di n + \Gamma_{xc}[n] \equiv 
 \Gamma_0[n] + \Gamma_{\rm int}[n],  
\ee
where $\Gamma_{xc}[n] = \sum_{l=1}^\infty \Gamma_l [n]$. 
Letting $\Delta (x)$ be the eigenvector, the excitation condition becomes 
\be \label{eigen.G.3} 
\int d\z \Gamma^{(2)}(\x,\z) \Delta (\z) = 0 \;. 
\ee
 After splitting the effective action into $\Gamma_0$ and
 $\Gamma_{\rm int}$ the eigenvalue equations (\ref{eigen.G.1}-\ref{eigen.G.2})
 may be expressed as 
\be \label{Gn2.eq.0}
-\int\!\! d\z  \, \Gamma_0^{(2)}(\y,\z, -\omega) \Delta (\z) = \int \!\! d\z  \, 
\Gamma_{\rm int}^{(2)}(\y,\z, -\omega) \Delta (\z)  \; .
\ee
Multiplying both sides of (\ref{Gn2.eq.0}) by $W_0^{(2)}(\x,\y,\omega)$ 
and then integrating over $d\y$, one obtains
\be \label{Gn2.eq.1}
\Delta (\x) = \int d\y d\z \; W_0^{(2)}(\x,\y,\omega) \Gamma_{\rm int}^{(2)}(\y,\z, -\omega) \Delta (\z) 
\;.
\ee
Note that $W_0[J_0]$ describes our Kohn-Sham system, a constructed non-interacting system that produces
 the same ground state electron density as that of the physical system considered. 
 From Eq.~(\ref{W2.pole}), one can write down $W^{(2)}_0(\x,\y,\omega)$ in terms of the 
  excitation energies associated with the Kohn-Sham non-interacting system:
\be \label{W20.pole}
W^{(2)}_0(\x,\y,\omega) = \sum_\ell \left[ \frac{n^*_{\ell}(\x) \, n_{\ell} (\y)}{\omega-\omega_{\ell } + i\eta} 
       - \frac{n^*_{\ell}(\y)\, n_{\ell}(\x)}{\omega + \omega_{\ell} + i\eta}\right] \; ,
\ee   
where 
\be\label{excitation_energy.KS}
\omega_{\ell}\equiv E_\ell - E_0  \; ,
\ee 
and $E_\ell$ is the energy of $\vert \ell \ra_{ks}$, the $\ell$th  state of the many-particle Kohn-Sham system with $J_0(\x)$ 
chosen to generate the correct ground state electron density. Therefore, for any $\ell$, $E_\ell$ is simply
the sum of single-particle energies $\ve_m$.   Note that in Eq.~(\ref{W20.pole}), $n_{\ell}(\x)$ is  defined
 by $\phantom{}_{ks}\!\la \ell \vert \tilde n(\x) \vert 0 \ra_{ks}$ except that $\vert \ell \ra_{ks}$ now describes the 
 $\ell$th state of the Kohn-Sham system, not the physical system considered.

 We seek a general solution for $\Delta (\x)$ of the form
\be \label{eigenvec.ansatz}
\Delta (\x) = \sum_\ell\, \left[ a_\ell \, n_{\ell }(\x) + b_\ell  \,n^*_{\ell } (\x) \right] \; .
\ee       
Evidently, if a frequency $\hat \omega$ leads to a solution $\{ b_\ell \}$, 
 then $-\hat \omega^* $ should lead to a solution $\{ a_\ell \}$, which plays the role of $\{ b_\ell^* \}$.  
Substituting Eqs.~(\ref{W20.pole}) and (\ref{eigenvec.ansatz}) into Eq.~(\ref{Gn2.eq.1}), we find that 
\bea
&& \hspace*{-30pt}\sum_{\ell} \left[ a_\ell n_{\ell} (\x) + b_\ell n^*_{\ell}(\x) \right] \nonumber \\
&& \hspace*{-20pt}=  
 \sum_{\ell,\ell\,'} \int \!\! d\y d\z \left[ \frac{n^*_{\ell}(\x)n_{\ell}(\y)}{\omega-\omega_\ell + i\eta}
  - \frac{n_{\ell}(\x)n^*_{\ell}(\y)}{\omega+\omega_\ell+i\eta}
 \right] \Gamma^{(2)}_{\rm int}(\y,\z,-\omega) \left[ a_{\ell\,'} n_{\ell\,'} (\z) + b_{\ell\,'} n^*_{\ell\,'}(\z) \right] .
\label{excitation.0}
\eea
Equating the coefficients associated with $n_\ell(\x)$ and $n_\ell^*(\x)$, we find that 
\bea
a_\ell &=& - \sum_{\ell\,'} \int d\y d\z \frac{n_{\ell}^*(\y)}{\omega+\omega_\ell+i\eta} 
 \Gamma^{(2)}_{\rm int} (\y,\z,-\omega) 
 \left[ a_{\ell\, '} n_{\ell\,'}(\z) + b_{\ell\, '}n^*_{\ell\, '}(\z) \right] \; , \nonumber \\
 b_\ell &=&  \sum_{\ell\,'} \int d\y d\z \frac{n_{\ell}(\y)}{\omega - \omega_\ell+i\eta} 
 \Gamma^{(2)}_{\rm int} (\y,\z,-\omega) 
 \left[ a_{\ell\, '} n_{\ell\,'}(\z) + b_{\ell\, '}n^*_{\ell\, '}(\z) \right] \; .\nonumber
\eea   
Let us define 
\bea
Y_{\ell,\ell\,'}(\omega) &=& \int d\y d\z \; n_{\ell}^*(\y)
 \Gamma^{(2)}_{\rm int} (\y,\z,-\omega) n_{\ell\, '}(\z) \; + \omega_\ell \,\delta_{\ell,\ell\,'} \; , \label{Y.def}\\
 K_{\ell,\ell\,'}(\omega) &=& \int d\y d\z \; n_{\ell}^*(\y)
 \Gamma^{(2)}_{\rm int} (\y,\z,-\omega) n^*_{\ell\, '}(\z) \; , \; \label{K.def}
\eea 
and we obtain the following matrix equation
\be \label{excitation.key}
 \left( \begin{array}{cc} {\mathbf Y}(\omega) & {\mathbf K}(\omega) \\
 {\mathbf K}^*(-\omega^*) & {\mathbf Y}^*(-\omega^*) \end{array} \right)
 \left( \begin{array}{c} A\\B \end{array} \right) = \left(\omega + i\eta \right) 
 \left( \begin{array}{rr} -1 & 0 \\ 0 & 1\end{array} \right) \left( \begin{array}{c} A\\B \end{array} \; , \right)
\ee
where $(A)_\ell = a_\ell$ and $(B)_\ell = b_\ell$. Evidently, one seeks  $\hat \omega$ such that
\be \label{excitation.key.det}
\det \left[ \left( \begin{array}{cc} {\mathbf Y}(\hat \omega) & {\mathbf K}(\hat \omega) \\
 {\mathbf K}^*(-\hat \omega^*) & {\mathbf Y}^*(-\hat \omega^*) \end{array} \right) - 
 \left(\hat \omega + i\eta \right) 
 \left( \begin{array}{rr} -1 & 0 \\ 0 & 1\end{array} \right) \right] = 0 \; .
\ee

As mentioned earlier, one anticipates $\left( A(\hat \omega) \right) = \left( B^*(-\hat \omega^*) \right)$.
To see this, we perform the change $ \omega \Leftrightarrow -\omega^* $ 
in (\ref{excitation.key}) and find that one can then rearrange the resulting equation into
\be 
 \left( \begin{array}{cc} {\mathbf Y}(\omega) & {\mathbf K}(\omega) \\
 {\mathbf K}^*(-\omega^*) & {\mathbf Y}^*(-\omega^*) \end{array} \right)
 \left( \begin{array}{c} B^*(-\omega^*)\\A^*(-\omega^*) \end{array} \right) 
  = \left(\omega + i\eta \right) 
 \left( \begin{array}{rr} -1 & 0 \\ 0 & 1\end{array} \right) \left( \begin{array}{c} B^*(-\omega^*)\\A^*(-\omega^*) \end{array} \right) \;  ,
\ee 
which is identical to Eq.~(\ref{excitation.key}) except with $\left( B^*(-\hat \omega^*) \right)$ playing the role of 
 $\left( A(\hat \omega) \right)$ and  $\left( A^*(-\hat \omega^*) \right)$ playing the role of 
 $\left( B(\hat \omega) \right)$ .

We now compare Eq.~(\ref{excitation.key}) with similar existing results. In references~\onlinecite{BA_96} and~\onlinecite{VF_97_arxiv}, 
 equations similar to (\ref{excitation.key}) were obtained, and those will be identical to Eq.~(\ref{excitation.key}) 
  provided that ${\mathbf K}^*(-\omega^*) = {\mathbf K}^*(\omega)$. This will happen if $\Gamma(\x,\y,-\omega)
   = \Gamma(\x,\y,\omega)$ for real $\omega$.

\section{Saddle-Point as an Alternative Formalism} \label{sec:saddle}
Below, we will obtain the effective action using a classical variable $i\varphi_c$ that corresponds to the 
 saddle-point of the auxiliary field path integral. At the physical condition $J=0$, $i\varphi_c$ 
 is interpreted as the electron density of a self-consistent Hartree solution. 
 Our Hartee problem is not of the conventional type, but rather similar to what Kohn described in his 
  Nobel lecture.~\cite{Kohn_99} In the conventional Hartree calculation, the wave functions obtained 
   may not be orthogonal to each other due to the fact that
 each particle's wave function is solved with a different potential.~\cite{Slater_30} 
  In the method below and mentioned by Kohn~\cite{Kohn_99}, the electric potential 
   experienced by every electron is the same. Another difference between the method below
    and the aforementioned Hartree methods~\cite{Slater_30,Kohn_99} is that the integral of the Hartree density
     in our method is not necessarily an integer due to the possibility that
 the density correction term may have a nonzero integral.

The saddle-point method below is quite different from what was described in the previous sections.
 First, although the diagrams in the saddle-point method are all connected diagrams,
 they are not one-particle irreducible (1PI). Second, unlike the method presented in previous sections,
  the computation of the effective action now requires no further functional derivatives 
of $\beta W [J]$   with respect to $J$ evaluated at $J_c$, while in the formalism mentioned in previous sections,
 one needs to compute higher order derivatives of $\beta W[J_0]$ with respect to $J_0$ 
 (see Eqs.~(\ref{Gamma_order_l.new}-\ref{J2}) ).

\subsection{Evaluation of $e^{-\beta  W_\phi [J]}$ via expansion around the saddle-point}
\label{sec:saddle_point}
The path integral (\ref{z1})
\[
e^{-\beta W_\phi[J]} \equiv \int D\phi  \; \exp \left\{ -I\left[ \phi \right] - iJ\4di \phi  \right\} \; .
\]
may be evaluated by the saddle point method. The extrema condition gives
\[
\left. \frac{\delta I\left[ \phi \right]}{\delta \phi(x)}\right|_{\varphi_c} = -iJ(x) \; .
\]
Since 
\bea
\left. \frac{\delta I\left[ \phi \right]}{\delta \phi(x)}\right|_{\varphi_c} &=&  \left( U\4di \varphi_c \right)_x -
 \Tr  \left( G_\phi \frac{\delta G_\phi^{-1}}{\delta \phi(x)} \right)_{\phi \to \varphi_c}  \nonumber \\
&=& \left( U\4di \varphi_c \right)_x - \int dy \; \PH_c(y,y) \left( i U(y,x) \right) \; , 
\eea
where $\PH_c(y,y) \equiv G_{\phi \to \varphi_c} (y,y)$, we obtain with $U(x,y) = U(y,x)$
\bea \label{J2phic}
J(x) = \left( U\4di (i\varphi_c) \right)_x + \int dy \; U(x,y) \PH_c(y,y) 
 = \int dy \; U(x,y) \left[ i\varphi_c(y) + \PH_c(y,y) \right] \; .
\eea
As $J \to 0$ (the physical condition), when $U$ is invertible such as Coulomb interaction, one must have
 $i\varphi_c(x) = -\PH_c(x,x)$. Note that the negative of the diagonal element
 of the Green's function $-\PH_c(x,x)$ is the particle density corresponding to the following Hamiltonian
\[
 \int dx \; \hpsid (x) \left[ 
 -\frac{\nabla ^2}{2m} + \upsilon _{\rm ion}({\x})-\mu + i (U \4di \varphi_c)_x \right]
 \hpsi(x) \; .
\]
The saddle-point equation therefore produces a Hartree-like equation: the Green's function depends on the input particle density $i\varphi_c(x)$
 and is required to produce the same particle density $i\varphi_c(x)$ in the end.    

When $J \ne 0$, one can still view (\ref{J2phic}) as a generalized Hartree equation in the following sense.
Remember that the inverse Green's function $\PH^{-1}_c(x,y)$ is given by (with $\delta(x-x') = 
 \delta(\tau - \tau') \delta(\x-\x')$) 
\be \label{Green_H}
\PH^{-1}_c(x,x') = \left[ \partial_\tau -\frac{\nabla^2}{2m} + \upsilon _{\rm ion}({\x}) -\mu 
 + U \4di (i\varphi_c) \right] \delta (x-x') \; ,
\ee
and may be rewritten as 
\be \label{Hartree_J_finite}
\PH^{-1}_c(x,x') =  \left[ \partial_\tau -\frac{\nabla^2}{2m} + \upsilon _{\rm ion}({\x}) + J(x) -\mu 
 + U \4di n_H \right] \delta (x-x') \; .
\ee
That is, we now view the potential as given by $\upsilon_{\rm ion}(\x) + J(x)$ and
 the Hartree particle density $n_H(x) = i\varphi_c(x) - (U^{-1}\4di J)_x$. This interpretation indeed 
 agrees with our equation (\ref{J2phic}) which can be written as 
\be\label{J2phic_interpret}
0 = \int dy \; U(x,y) \left[ \left( i\varphi_c(y) - \int dz \; U^{-1}(y,z) J(z) \right) + \PH_c(y,y) \right] \; .
\ee
That is, for a given $J(x) \ne 0$, one will solve as before the Hartree equation but with $\upsilon _{\rm ion}({\x})
 \to \upsilon _{\rm ion}({\x}) + J(x)$. Once the Hartree particle 
 density $n_H$ is obtained, one obtains $i\varphi_c = n_H +  U^{-1} \4di J$. 
Since $U \4di (i\varphi_c)$ always appears as a unit inside the Green's function $\PH_c(x,x')$,
  for convenience, we define 
\[  
J_c \equiv \, U\4di (i\varphi_c) = J +  U \4di n_H \; . 
\]

Once $\varphi_c(x)$ is obtained, we shift $\phi$ by $\varphi_c$ and 
 re-express the exponent in the integrand as
\[
-I[\phi] - iJ\4di \phi \Rightarrow -I[\varphi_c+\phi] - iJ\4di (\varphi_c+\phi) \; , 
\]
and then expand around $\varphi_c$. 
To do the expansion, we first rewrite (\ref{Green.general}) as
\be \label{Green.general.1}
G_\phi^{-1}(x,x') = \PH_c^{-1}(x,x') + i\, b(x)\, \delta(x-x')
 \equiv \PH^{-1}_c(x,x') +  V(x,x') \; ,
\ee
where $b = U \4di \phi$. 
We may then write down
\[
G_\phi^{-1} = \PH^{-1}_c \left[ {\mathbf I} + \PH_c \4di {\mathbf V} \right] \; ,  
\]
and 
\be \label{Green.general.expand_c}
\ln \left( G_\phi^{-1} \right) = 
\ln \left( \PH^{-1}_c \right) + \sum_{k=1}^\infty \frac{(-1)^{k-1}}{k}
 \left[ \PH_c \4di  {\mathbf V} \right]^k \; .
\ee
Note that 
\[
\left[ \PH_c \4di {\mathbf V} \right]_{x,z} 
 = \int\!\! dy \, \PH_c(x,y) V(y,z)
 =  \int dy \PH_c(x,y) \delta(y-z) \left( i\, b(y)\, \right)
 = \PH_c(x,z) \left( i b(z) \right) \; .
\]

Consequently, 
\bea 
\Tr \ln \left( G_\phi^{-1} \right) & = &  \Tr \ln \left( \PH^{-1}_c \right)
 + \int \! dx_1 \,\PH_c(x_1,x_1) (i b(x_1))  \nonumber \\
&& \hspace*{-50pt}
 - \frac{1}{2} \int \! dx_1 dx_2 \, \PH_c(x_1,x_2) \PH_c(x_2,x_1)
 (ib(x_1))(i b(x_2))  \nonumber \\
&& \hspace*{-50pt} + \sum_{k=3}^\infty \frac{(-1)^{k-1}}{k} \int \! dx_1 \ldots dx_k  \; 
\PH_c(x_k,x_1) \ldots \PH_c(x_{k-1},x_k)
  (ib(x_1))\ldots(ib(x_k)) \; .
\label{Green.general.trace.1}
\eea
Note that in the final expression of the exponent of the integrand in
 (\ref{z1}) the terms linear in $\phi$ (or $b$) cancel out as one may verify
 and we arrive at  
\bea
-I[\varphi_c+\phi] - iJ\4di (\varphi_c+\phi) & =& 
 - \frac{1}{2}\varphi_c \4di \, U \4di \varphi_c - \varphi_c \4di b 
 - \frac{1}{2}  b \4di \, U^{-1} \4di b + \Tr \ln (G_\phi^{-1} ) \nonumber \\
&& +\frac{1}{2}\Tr \ln (U) - iJ \4di \varphi_c - i J \4di \phi  \nonumber \\
&=& \frac{1}{2}\Tr \ln (U)- \frac{1}{2}\varphi_c \4di \, U \4di \varphi_c + \Tr \ln (\PH_c^{-1} ) 
 - iJ \4di \varphi_c \nonumber \\ 
&& - \frac{1}{2} b \4di \tilde {\cal D}_c^{-1} \4di b + \sum_{k=3}^\infty
  I^{(k)}[\varphi_c] \4di \, b_1 \ldots \4di\, b_k  \; ,
\eea
where 
\[
\tilde {\cal D}_c^{-1} = U^{-1} -  D_c  \; ,
\]
\[
D_c(x,y) = \PH_c(x,y)\, \PH_c(y,x) \; ,
\]
and 
\be 
 I^{(k)}[\varphi_c]  \4di\, b_1 \ldots \4di\, b_k
\equiv   \frac{(-1)^{k-1}}{k} \int \! dx_1 \ldots dx_k  
\PH_c(x_k,x_1) \ldots 
 \PH_c(x_{k-1},x_k)\left[
  (ib(x_1))\ldots(ib(x_k)) \right] \; .
\ee
We therefore have (based on the linked-cluster theorem)
\bea
\beta W_\phi [J] &=&  \frac{1}{2} \varphi_c \4di \, U \4di \varphi_c
- \Tr \ln \left( \PH_c^{-1} \right) + i J \4di \varphi_c \nonumber \\
&& + \frac{1}{2} \Tr \ln \left( \tilde {\cal D}_c^{-1} \4di \, U \right) 
 - \sum_{n=1}^\infty \frac{1}{n!} \la \left[ \sum_{k=3}^\infty
 I^{(k)}[\varphi_c] \4di \,b_1 \ldots \4di\, b_k  \right]^n \ra_{\rm conn.} \; ,
\eea
where the subscript ``${\rm conn.}$" is used to indicate connected Feynman diagrams.

Consequently, the effective action, defined  by
\[
\Gamma[n] = \beta W_\phi[J] - \frac{1}{2} J \4di \, U^{-1}  \4di J  - J \4di n 
 =  \beta W_\phi[J] + \frac{1}{2} J \4di \, U^{-1}  \4di J - i J \4di \varphi \;  
\]
and with $\varphi = \varphi_c + \la \phi \ra \equiv \varphi_c + \tilde \varphi$
  as well as with $J = J_c - U\4di n_H$, can now be written as
\bea
\Gamma[n] &=&  
- \Tr \ln \left( \PH_c^{-1} \right) - J_c \4di n_H 
 + \frac{1}{2} n_H \4di \, U \4di n_H  + \frac{1}{2} \Tr \ln \left( \tilde {\cal D}_c^{-1} \4di \, U \right) 
  \nonumber \\
&& - i J \4di {\tilde \varphi} 
 - \sum_{n=1}^\infty \frac{1}{n!} \la \left[ \sum_{k=3}^\infty
 I^{(k)}[\varphi_c] \4di \, b_1 \ldots \4di\, b_k  \right]^n \ra_{\rm conn.} \;.
 \label{G1.varphic}
\eea
This expression should be contrasted with Eq.~(\ref{G1.1}) where the true particle density is introduced as
 the natural variable. 

From the definition of $\varphi$ (see Eq.~(\ref{phiJ4.def}) we have  
\be
i \varphi(x) = \frac{\int D\phi  \left(i\phi(x)\right) e^{-I[\phi] -iJ\4di \phi}}{\int D\phi  
 \, e^{-I[\phi] -iJ\4di \phi}} = i\varphi_c(x) + \int dy \, U^{-1}(x,y)
 \frac{\int D b  \left(i b(y)\right) e^{-I[\varphi_c+ \phi] -iJ\4di (\varphi_c +\phi)}}
 {\int D b \,  e^{-I[\varphi_c+ \phi] -iJ\4di (\varphi_c +\phi)}} \;,
\ee
which means that
\be
{i\tilde \varphi} = 
\frac{\int D\, b   \left(i U^{-1} \4di  b \right) e^{- \frac{1}{2} b \4di  \tilde {\cal D}_c^{-1} \4di b
 + \sum_{k=3}^\infty \frac{1}{k!}I^{(k)}[\varphi_c] \4di \, b_1 \ldots \4di\, b_k }}
{\int D b \,  e^{- \frac{1}{2} b \4di  \tilde{\cal D}_c^{-1} \4di b
 + \sum_{k=3}^\infty \frac{1}{k!}I^{(k)}[\varphi_c] \4di \, b_1 \ldots \4di\, b_k }}\; .
\ee
On the basis of a simple replica argument~\cite{Negele_Orland_book}, one sees that the above expression may
 be written as 
\bea
{i\tilde \varphi}(x) &=& \sum_{n=0}^\infty \frac{1}{n!} \int dy\, U^{-1}(x,y)\la \left[ \sum_{k=3}^\infty
 I^{(k)}[\varphi_c] \4di \,b_1 \ldots \4di\, b_k  \right]^n i b(y) \ra_{\rm conn.} \nonumber \\
& = & \sum_{n=1}^\infty \frac{1}{n!} \int dy\, U^{-1}(x,y) \la \left[ \sum_{k=3}^\infty
 I^{(k)}[\varphi_c] \4di \,b_1 \ldots \4di\, b_k  \right]^n i b(y) \ra_{\rm conn.} \; ,
\label{tvarphi}
\eea
where the $n=0$ term vanishes because 
$\int D b \, e^{- \frac{1}{2} b \4di  \tilde {\cal D}_c^{-1} \4di b} i b(x) = 0$.
Eq.~(\ref{tvarphi}) implies that $i\tilde \varphi(x)$ at each point $x$ 
 is a functional of $\varphi_c$. Returning to (\ref{G1.varphic}), this implies that 
 one may express the effective action $\Gamma$ in terms of 
 $i\varphi_c$ although its canonical argument is supposed to be $n$. 
 The relation between $n$ and $i\varphi_c$ is obtained through 
\be \label{phic2n}
n = i\varphi_c + i\tilde \varphi [i\varphi_c] - U^{-1} \4di J [ i \varphi_c ] = n_H + i\tilde \varphi [i\varphi_c] 
\ee
with $i\tilde \varphi(x)$ expressed as a functional of $i\varphi_c$.
 This implies that given an $i\varphi_c$, we may obtain $n$ through (i) the 
 diagrammatic expansion of (\ref{tvarphi}) to produce the corresponding $i\varphi$,  and (ii) equation 
 (\ref{J2phic}) to find $-U^{-1} \4di J$.  Since in general, we have 
 \[
 \int d\x\, \left(i\tilde \varphi (\x) \right) \ne 0 \;, 
 \]
one cannot expect the integral of the Hartree electron density
to be $N_e$, the total number of electrons. Instead, we have
\be \label{ptcle_const}
N_e = \int d\x\, n(\x) =\int d\x \, \left[  n_H(\x) + i\tilde \varphi(\x) \right]
\ee 
and $\int d\x \, n_H(\x)$ is not necessarily an integer. In particular, when 
$\int d\x \, n_H(\x)$ deviates significantly from $N_e$ or $\int d\x \, |i\tilde \varphi (\x)| \gg 1$,  
forcing $\int d\x \, n_H(\x) = N_e$ 
 may lead to the occurrence of a self-consistent solution that 
 differs significantly from the true solution.  

Note that $n_H(\x) = -\PH_c(x,x)$ and in the absence of the source term $n_H(\x)=i\varphi_c(\x)$.
 Furthermore, since $(U^{-1}\4di J)_x \propto \nabla_{\x}^2 J(x)$, $\int n_H(\x) \, d\x = \int (i\varphi_c(\x))\,  d\x$. 
The constraint (\ref{ptcle_const}) for the total number of electrons can also be written as
\be \label{ptcle_const.1} 
N_e = \int d\x\, n(\x) =\int d\x \, \left[  n_H(\x) + i\tilde \varphi(\x) \right]
 =\int d\x \, \left[ i\varphi_c(\x) + i\tilde \varphi(\x) \right] \; .
\ee 

 The Hartree-like Green's function $\PH_c(x,x')$ shown in (\ref{Green_H}) 
 may be viewed as a functional of $i\varphi_c(x)$. 
 When expressing $\PH_c(x,x')$ using only single particle orbitals,  
  we define in Eq.~(\ref{single_ptcle_h}) $v(\x) \equiv \upsilon_{\rm ion}(\x) -\mu + U\3di (i\varphi_c)_{\x}$. 
   For the evaluation of $\PH_c(x,y)$, we solve first the eigensystem (\ref{Eigen_1ptle}). 
The single-particle wave functions (\ref{1ptle_basis}) associated with $\hat h$ are to be obtained 
 self-consistently. 
 Basically, one starts with a {\it guess} for the electronic density $i\varphi_c(\x)$ satisfying 
 $\int d\x (i\varphi_c(\x)) \approx N_e$, where $N_e$ is the number of electrons. One then obtains the 
 single-particle wave functions, and then computes the corresponding Green's function $\PH_c$, 
  obtains $\int d\x (i\tilde \varphi (\x)) $ and
 tunes the chemical potential to ensure that $- \int d\x \; \PH_c(\x,\x) = N_e - \int d\x (i\tilde \varphi(\x))$.
  One then takes $-\PH_c(\x, \x)$ in place of $i\varphi_c(\x)$ in the next round of iteration until convergence is reached.

The procedure of the saddle-point method is now obvious. One starts with an external potential, and then 
 determines the Hartree electron density via $n_H(\x) = -\PH_c(x,x)$, $n_H = i\varphi_c - U^{-1}\4di J$
  and Eq.~(\ref{ptcle_const.1}). This self-consistent procedure will also provide the physical electron density $n(\x)$. The ground state energy is obtained by using (\ref{G1.varphic}) 
  to calculate the effective action, which is the ground state energy times $\beta$ 
  in the limit $T \to 0$. When one wishes to obtain
  the density functional $\Gamma [n]$ at an electron density other than $n_T$, one adds the source term into
   the potential $\upsilon_{\rm ion}(\x)$ and then solves for the Hartree density, shown
   in  (\ref{Hartree_J_finite}), as outlined above.

\subsection{Remark on the single-particle limit}
The single electron limit for the Hartree-method is more complicated than for the 
 method presented in section~\ref{one_ptcle_limit}. 
Because $\int n_H(\x) \, d\x$ is not necessarily $1$ in the single electron limit, due to
 Eq.~(\ref{ptcle_const.1}), in general one needs to obtain $n_1$ self-consistently. 
 The other issue is that the diagrammatic expansion contained in (\ref{G1.varphic}) does not 
 cover all the diagrams for a given order of $U$. This means that one can't use the Hugenholtz diagram
  argument to eliminate vertex matrix element even when $n_1 \le 1$.
   To see this point explicitly, we rewrite (\ref{G1.varphic})  as
\bea  
\Gamma[n] &=&  
- \Tr \ln \left( \PH_c^{-1} \right) + (J - J_c )\4di n_H 
 + \frac{1}{2} n_H \4di \, U \4di n_H  + \frac{1}{2} \Tr \ln \left( \tilde {\cal D}_c^{-1} \4di \, U \right)  \nonumber \\
&& 
 - \sum_{n=1}^\infty \frac{1}{n!} \la \left[ \sum_{k=3}^\infty
 I^{(k)}[\varphi_c] \4di \,b_1 \ldots \4di\, b_k  \right]^n \ra_{\rm conn.} 
- J \4di n 
 \;. \label{G1.varphic.new}
\eea  
Evidently, except for the last term above, the rest of the terms must constitute $\beta W[J]$. 
Since $J-J_c = - U\4di n_H$, one can also write $\beta W[J]$ as
\bea
\beta W[J] &=& - \Tr \ln \left( \PH_c^{-1} \right) - n_H\4di \, U \4di n_H
+ \frac{1}{2} n_H \4di u \4di n_H  + \frac{1}{2} \Tr \ln \left( \tilde{\cal D}_c^{-1} \4di \, U \right) \nonumber \\
&&  - \sum_{n=1}^\infty \frac{1}{n!} \la \left[ \sum_{k=3}^\infty
 I^{(k)}[\varphi_c] \4di \, b_1 \ldots \4di\, b_k  \right]^n \ra_{\rm conn.}\;\; . \nonumber 
\eea
Diagrammatic expansion of the last two terms shows that the following diagrams 
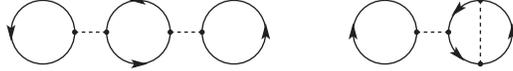
\[
\begin{picture}(96,20)(0,-3)
\Arc[arrow,arrowpos=0.50,arrowlength=3.0,arrowwidth=0.8,arrowinset=0.2](12,0)(12,0,360)
\Arc[arrow,arrowpos=0.50,arrowlength=3.0,arrowwidth=0.8,arrowinset=0.2](48,0)(12,0,180)
\Arc[arrow,arrowpos=0.50,arrowlength=3.0,arrowwidth=0.8,arrowinset=0.2](48,0)(12,180,360)
\Arc[arrow,arrowpos=0.50,arrowlength=3.0,arrowwidth=0.8,arrowinset=0.2](84,0)(12,-180,180)
\Vertex(24,0){1.0}
\Vertex(36,0){1.0}
\Vertex(60,0){1.0}
\Vertex(72,0){1.0} 
\Line[dash,dashsize=1.5](24,0)(36,0)
\Line[dash,dashsize=1.5](60,0)(72,0) 
\end{picture} \; \; \hspace*{10pt} \; \;\hspace*{10pt}
\begin{picture}(60,20)(0,-3)
\Arc[arrow,arrowpos=0.50,arrowlength=3.0,arrowwidth=0.8,arrowinset=0.2,clock](12,0)(12,0,360)
\Arc[arrow,arrowpos=0.50,arrowlength=3.0,arrowwidth=0.8,arrowinset=0.2](48,0)(12,-90,90)
\Arc[arrow,arrowpos=0.50,arrowlength=3.0,arrowwidth=0.8,arrowinset=0.2](48,0)(12,90,180)
\Arc[arrow,arrowpos=0.50,arrowlength=3.0,arrowwidth=0.8,arrowinset=0.2](48,0)(12,180,270)
\Line[dash,dashsize=1.5](48,-12)(48,12)
\Vertex(48,-12){1.0}
\Vertex(48,12){1.0}
\Vertex(24,0){1.0}
\Vertex(36,0){1.0}
\Line[dash,dashsize=1.5](24,0)(36,0)
\end{picture} \; 
\] 
of order $U^2$ are absent, 
when compared to the regular field theoretic perturbation calculation. This is not a disadvantage of the 
 method. Instead, what our derivation shows is that the missing diagrams eventually will be compensated by the
 $-n_H \4di \, U \4di n_H$ term. However, it is obvious that the saddle-point formalism  
  makes the single-electron limit hard to analyze.    
 
 When $n_1 > 0$ but $n_1 \ll 1$ at the low temperature limit, 
    we know that $D_c(x,y) = \PH_c(x,y) \PH_c(y,x)$ will be of order $n_1$.  This is because 
\[
\PH_c(x,y) = \sum_\alpha \phi_\alpha(\x) \phi_\alpha^*(\y) e^{-(\ve_\alpha-\mu)(\tau_x - \tau_y)} \times \left\{ \begin{array}{l r}
(-n_\alpha) & {\rm if~} \tau_x \le \tau_y \\
(1-n_\alpha) & {\rm if~} \tau_x > \tau_y \end{array} \right. \; ,
\]
and whenever $\tau_x \le \tau_y$, the propagator is of order $n_1$. Since
 $D_c(x,y) = \PH_c(x,y) \PH_c(y,x)$, one of the propagators in the product must be of order $n_1$. 
 In principle,  
 one needs to solve for the occupation number $n_1$ of the 
 lowest energy state using  $i\varphi_c(\x) - \left(U^{-1} \4di J\right)_{x} = - \PH_c(x,x)$ and
  Eq.~(\ref{ptcle_const.1}). Nevertheless, the correct one particle limit 
  can be seen if one starts with a chemical potential $\mu$ such that $n_1 \approx 0$.
  In this case, we have at the $J=0$ limit $i\varphi_c (x) = n_H(\x) \approx 0$ 
   as well as $D_c(x,y) \propto n_1 \approx 0$. This way, the  higher order exchange-correlation terms 
 may be viewed as the having smaller contributions and one may control the accuracy by
  controlling the number of higher order terms included.    
Of course, one then has $n(\x) \approx i\tilde \varphi(\x)$ and the condition 
$\int d\x \left( i\tilde \varphi(\x) \right) = 1 - n_1$ must be satisfied.   
 
\subsection{Obtaining excitations using the Hartree method} 
In general the excitations are determined by Eqs.~(\ref{eigen.G.1}) and (\ref{eigen.G.2}). 
Under the Hartree formalism described in this section, the natural variable used is the 
 Hartree density $n_H$ rather than the true particle density $n_T$. 
  One thus must transform the variable used in Eqs.~(\ref{eigen.G.1}) and (\ref{eigen.G.2}) 
   from $n_T$ to $n_H$. We describe below how this can be achieved. 
 
 At the physical condition ($J=0$), one has 
 \[
 0 =  \left. \frac{\delta \Gamma}{\delta n(x)} \right|_{n=n_T} = \left.
 \int dx_1 dx_2 \frac{\delta \Gamma}{\delta n_H(x_2)}\right|_{n=n_T} \left. 
 \frac{\delta n_H(x_2)}{\delta J_c(x_1)}\right|_{n=n_T}
  \left. \frac{\delta J_c(x_1)}{\delta n(x)}\right|_{n=n_T} \; .
 \] 
 Note that $\delta n_H/\delta J_c$ contains no zero mode because $n_H = \delta W_H/\delta J_c$,
   where $W_H[J_c] \equiv -\Tr \ln \left( {\PH_c}^{-1} \right)$, 
  and $\delta^2 W_H/\delta J_c \delta J_c$ is known to be strictly negative from a theorem proved by 
   Valiev and Fernando.~\cite{VF_97_arxiv} The strict negative-definiteness of 
  $\delta^2 W/\delta J \delta J$ can also be interpreted as the stability condition
\[  
 \int dx \, \delta n(x) \,  \delta J (x) < 0  \; , 
\]  
which means raising the local one-particle potential leads to an average  decrease 
 of the local particle concentration and vice versa.     
  The strict negative-definiteness means that $\delta n_H/\delta J_c$ contains
    no zero modes and is invertible. Also, because  $n = n_H + i\tilde \varphi_c$, 
 $\delta n /\delta J_c = \delta n_H /\delta J_c + \delta (i\tilde \varphi)/\delta J_c$ 
  exists via diagrammatic expansion of $i\tilde \varphi$ in terms of 
  $J_c$. The existence of $\delta n/\delta J_c$ 
 implies that $\delta J_c/\delta n$ is invertible (i.e., has 
 no zero eigenvalues). Therefore, after multiplying the inverse of $\delta J_c/\delta n$ and 
  $\delta n_H/\delta J_c$ on both sides of the above equation, one has 
  \[
  \left. \frac{\delta \Gamma}{\delta n_H(x)}\right|_{n=n_T} = 0 \; ,
  \]
which then leads to    
 \bea
 \Gamma^{(2)}(x,y) &=& 
\left. \frac{\delta^2 \Gamma}{\delta n(x) \delta n(y)} \right|_{n=n_T} \nonumber \\
 & = & 
  \left[ \int dx_1 dx_2 dy_1 dy_2 \frac{\delta J_c(x_1)}{\delta n(x)} \frac{\delta n_H(x_2)}{\delta J_c(x_1)}
 \frac{\delta^2 \Gamma}{\delta n_H(x_2) \delta n_H(y_2)} \frac{\delta n_H(y_2)}{\delta J_c(y_1)} 
 \frac{\delta J_c(y_1)}{\delta n(y)}\right]_{n=n_T} \; .\nonumber
 \eea
Using Eq.~(\ref{Gn2.freq}), one may write $\Gamma^{(2)}(\x,\y,\omega)$ as the analytic continuation of
 the integral  of $\Gamma^{(2)}(x,y)$ over time. To achieve this goal, one first writes down  
\bea
\Gamma^{(2)}(\x,\y,i\nu_n) &=& \int_0^\beta d(\tau_y-\tau_x) \,e^{i\nu_n (\tau_y-\tau_x)} \, \Gamma^{(2)}(x,y) 
\nonumber \\
& = & \int_0^\beta d(\tau_y-\tau_x) \,e^{i\nu_n (\tau_y -\tau_{y_1} + \tau_{y_1} -\tau_{y_2} +\tau_{y_2}
  -\tau_{x_2} + \tau_{x_2}-\tau_{x_1} + \tau_{x_1} -\tau_x)} \, \Gamma^{(2)}(x,y) \nonumber \\
&\equiv& \int d\x_1 d\x_2 d\y_1 d\y_2   f^{\x_1}_{\x}\!(-i\nu_n) g^{\x_2}_{\x_1}\!(-i\nu_n) \tilde \Gamma^{(2)}(\x_2,\y_2,i\nu_n)
g^{\y_2}_{\y_1}\!(i\nu_n) f^{\y_1}_{\y}\!(i\nu_n) \nonumber \\
&=& \int d\x_2 d\y_2 h^{\x_2}_{\x}(-i\nu_n) \tilde \Gamma^{(2)}(\x_2,\y_2,i\nu_n) h^{\y_2}_{\y}(i\nu_n) \; , 
\label{Hartree.excitation.0}
\eea 
where 
\bea
\tilde \Gamma^{(2)}(\x,\y,i\nu_n) &=& \int_0^\beta d(\tau_y-\tau_x) 
 \frac{\delta^2 \Gamma}{\delta n_H(x) \delta n_H(y)} \nonumber \\
g^{\y_2}_{\y_1}\!(i\nu_n) &\equiv & \left. \int_0^\beta d(\tau_{y_1}-\tau_{y_2}) \, e^{i\nu_n(\tau_{y_1}-\tau_{y_2})} \, 
 \frac{\delta n_H(y_2)}{\delta J_c(y_1)} \right|_{n=n_T} \nonumber \\
f^{\y_1}_{\y}\!(i\nu_n) &\equiv & \left. \int_0^\beta d(\tau_{y}-\tau_{y_1}) \, e^{i\nu_n(\tau_{y}-\tau_{y_1})} \,
   \frac{\delta J_c(y_1)}{\delta n(y)} \right|_{n=n_T}  \nonumber \\
h^{\y_2}_{\y}(i\nu_n) & \equiv & \int d\y_1 \;  g^{\y_2}_{\y_1}\!(i\nu_n)  f^{\y_1}_{\y}\!(i\nu_n)  \nonumber \; .
\eea
Therefore, Eq.~(\ref{eigen.G.3}) for finding the excitations becomes
\be \label{eigen.G.4}
\int\!\! d\z  \, \tilde \Gamma^{(2)}(\x,\z, -\omega ) \, \tilde \Delta (\z) = 0  \;, 
\ee
with $\tilde \Delta(\z)$ given by 
\[
\tilde \Delta (\z) = \int d\y \, h^{\z}_{\y}(-\omega) \Delta (\y) \; .
\]  
One therefore obtains the excitation energy via solving Eq.~(\ref{eigen.G.4}). Since $h^{\z}_{\y}(-\omega)$
 is invertible, one may also obtain $\Delta (\z)$ via $\tilde \Delta (\z)$ if desired.  
 Within this framework, the first two terms of (\ref{G1.varphic}) correspond to the 
 non-interacting part $\Gamma_0$ of the effective action
  in section~\ref{sec:excitation}. The protocol for obtaining the excitation energy is then the same as 
   described in section~\ref{sec:excitation} with the one exception that the ground state of the 
   Hartree system is not the same as the ground state of the real system.

\section{Discussion and Future Directions} \label{sec:final} 
In this paper we focus on the auxiliary field method applied to the development of the density
 functional. It is natural to inquire into the physical meaning of the auxiliary, 
 bosonic field $\phi$ introduced in eqs~(\ref{genfunc.1}-\ref{Green.general}). 
  In the path integral treatment of relativistic quantum electrodynamics, 
  if one were to integrate out the photon field, one generates the current-current interaction which 
  is quartic in fermionic field. When viewing this process backwards, one sees that the quartic 
   fermionic interaction is disentangled by introducing the photon field. The $\phi$ field here thus
    plays a similar role to the photon field as it disentangles the quartic fermionic interaction term.
 As we will argue below, the $\phi$ variable is closely related to the ``time" component of the photon field,
  i.e., the electric potential.

  To make the connection between the photon field and $\phi$, let us first seek the nonrelativistic limit    
  of the Lagrangian density, ${\mathcal L}_{EM} = - F^{\mu \nu}F_{\mu \nu}/ (16 \pi)$ , associated with 
  the relativistic photon field. Note that in the limit
    when the speed of light approaches infinity (removal of terms involving time derivative of the 
  three-vector potential), the Lagrangian density turns into $(\nabla A_0)^2/ 8\pi $, which contains  
   no quadratic derivative with respect to time. Therefore,  at finite temperature (with $it \to \tau$) 
  this implies that the exponent associated with the photon field path integral will behave as
 \be \label{photon.field}
i \int dt\, d\x \;{\mathcal L}_{EM} \to 
 \int\! d\tau \! \int \! d\x  \frac{(\nabla A_0)^2} {8\pi} = - \int \! d\tau \! \int \!d\x  \frac{1} {8\pi} \,A_0 \nabla^2 A_0 \; .
 \ee  
 Setting $U(\x-\y) = e^2/|\x-\y|$ [thus $U^{-1}(x,y) = -\frac{1}{4\pi e^2} \nabla^2_{\x} \delta(\x-\y) $]
  and comparing (\ref{photon.field}) with Eqs.~(\ref{genfunc.1}-\ref{action.1}), 
 we make the identification $(i U\4di \phi)/e = A_0$,  because now
\[
-\frac{1}{2} \phi \4di \, U \4di \phi \to \frac{1}{2} A_0 \,\4di (e^2 u^{-1} ) \,\4di A_0 = 
- \int d\tau \int d\x  \frac{1} {8\pi} \,A_0 \nabla^2 A_0 \; .
\]   

In the non-relativistic limit, the electric part and the magnetic part of electromagnetism are decoupled. 
  Starting with a non-relativistic many--electron system, one may ask what is the
  quantum mechanical analog of the Poisson equation that forms the basis of electrostatics. 
 It turns out that this is 
  easily obtained via computing $\left. \delta Z[J]/\delta J \right|_{J=0}$ 
in two different ways.
First, upon taking the derivative of (\ref{genfunc.1}) and using Eqs.~(\ref{action.1}) and (\ref{GJ}), one obtains
\[
\left. \frac{\delta Z[J]}{\delta J(\x )} \right|_{J=0} = -\beta Z_{J=0} \la \psid(\x) \psi(\x) \ra =
 -\beta Z_{J=0} \frac{\la e \psid(\x) \psi(\x)  \ra}{e}  \; .
\]
Second, while taking the derivative of (\ref{genfunc.1}), if using Eqs.~(\ref{action.2}) 
and (\ref{Green.general}) one arrives at
\[
\left. \frac{\delta Z[J]}{\delta J(\x )} \right|_{J=0} = -\beta Z_{J=0} \la i\phi \ra =
-\beta Z_{J=0}\,  e\; U^{-1} \4di \la A_0 \ra =  \beta Z_{J=0} \frac{1}{4\pi e} \nabla^2_{\x} \la A_0(\x) \ra \;.
\]  
We therefore obtain the thermal quantum-mechanical analog of the Poisson equation
\[
\nabla^2_{\x} \la A_0(\x) \ra = -4\pi \la e\psid (\x) \psi(\x) \ra \; ,
\]  
a result also obtained in reference~\onlinecite{VF_96}. 
 This connection to classical electrostatics is essential
 since it provides the quantum-mechanical correspondence of an important 
 ingredient in (bio)molecular interactions that have been extensively studied in the presence of 
     dielectrics.~\cite{Yu_03,DY_04,DY_06,ODRY_09}

The UDF described in this paper is systematically constructed, uniquely determined, and in principle exact. 
 However, in terms of real computations, one can only keep $\Gamma_i$ terms up to some order in $\lambda$. 
 A natural question thus arises. How well will the truncated version work? In general, this question can only be answered with
   numerical results. However, by providing theoretical arguments and comparisons to other approaches, 
 we wish to convey that this method is likely to produce good results and thus to attract 
 computational efforts towards using the proposed approach.  

It is worth pointing out the relation between the expression (\ref{W.varphi.3}) and the scheme~\cite{PS_02}
motivated by the renormalization-group. The  vertex functions $I^{(j\le l)}$ in (\ref{W.varphi.3}) 
 will contribute to the so-called $l$-local approximation of reference~\onlinecite{PS_02}.
 The absence of $I^{(2)}$ manifests the absence of the correction term due to the bi-local contribution, 
 as shown in reference~\onlinecite{PS_02}. Furthermore, with the bilocal approximation~\cite{PS_02} 
 included, Polonyi and Sailer obtained an approximate energy functional which corresponds exactly to our 
 $\Gamma_0 + \Gamma_1$. To reach an equivalent form of the proposed $l$-local approximation of  reference~\onlinecite{PS_02},  
  we simply keep terms  up to $\Gamma_{l/2}$. Therefore, our formulation provides an explicit means for   
  achieving an $l$-local approximation without resorting to the Hellmann-Feynman theorem.

As shown in (\ref{tD.exp}), the propagator $\tilde {\mathcal D}_0$ can be expanded as 
\[
\tilde {\mathcal D}_0 = U + U \4di D_0 \4di \, U +  U\4di D_0 \4di \, U \4di D_0 \4di \, U + \ldots \;,
\]
when $U$ can be viewed as a small quantity and treated perturbatively. When one is not
 allowed to treat $U$ as a small parameter (say in the strong coupling regime), or when one needs to treat
 $U^{-1}$ as  a small parameter instead, the conventional perturbative expansion in $e^2$ breaks down completely
 while our approach is still applicable. In the case when $U^{-1}$ must be treated as small,
  we expand $\tilde {\mathcal D}_0$ as 
\be \label{tD.exp.sc} 
\tilde {\mathcal D}_0 = -D_0^{-1} - D_0^{-1}\4di \, U^{-1} \4di D_0^{-1} -  D_0^{-1}\4di \, U^{-1} 
\4di D_0^{-1} \4di \, U^{-1} \4di D_0^{-1}
 - \ldots \; .
\ee    
And in this case, $U \gg 1$, our effective action expansion does have the Hartree term $\frac{1}{2} n \4di \, U \4di n$ 
as the leading order, followed by terms of order $U^0$ and then the expansion of $\tilde {\mathcal D}_0$ 
 provides series in powers of $U^{-1}$. Note that in this case, the exchange-correlation functional is not led
  by order $U$ at all, but is led by order $U^0$. This feature is not present in the conventional perturbative
   approach using $U$ (or $e^2$) as the expansion parameter.

As mentioned earlier, there also exist different functional methods for many electron systems. 
For example, the exchange correlation functional outlined by Sham~\cite{Sham_85} 
is founded on the perturbative  functional approach developed by Luttinger and Ward~\cite{LW_60}
 or equivalently by Klein.\cite{Klein_61}   A succinct review of the 
   Luttinger-Ward/Klein functional and its applications can be found in reference \onlinecite{KSHetal_06}. 
   The Luttinger-Ward/Klein functional
   yields the grand-potential/ground-state-energy only when the functional argument 
   is equal to the {\it fully-interacting}, physical, one-particle Green's function. 
 Instead of allowing the physical, full, one-particle Green's function,
  Dahlen {\it et al.}~\cite{DvLvB_06} proposed to find that stationary point of   
  the Luttinger-Ward/Klein functional while restricting 
  the argument to the Hartree-Fock Green's functions or
  the Green's functions of non-interacting systems. 
 However, even if
 the Luttinger-Ward/Klein functional is computed to all orders, the 
 error in the value of the grand potential (or the ground state energy) due to restriction 
 on the Green's functions remains unknown.
 
Furthermore, it should be noted that when the functional argument is equal to the 
physical one-particle Green's function, the Luttinger-Ward/Klein functional reaches 
a stationary point, {\it not} the minimum.\cite{Klein_61} This means that it is possible
 for the Klein/Luttinger-Ward functional to assume an even lower value than 
 the ground state energy (or the grand potential) when the functional argument deviates from
  the physical Green's function. In other words,  
 the Klein/Luttinger-Ward functional only retains its meaning as the
   ground state energy (or grand potential) when the Green's function takes the value of  
   the true (physical) Green's function. 
  Our effective action expression of the energy functional, on the other hand, truly represents 
   energy of the system. 
   Our effective action energy functional,  when no truncation on the series is made, 
   reaches its {\it minimum} when the electron density assumes the 
  true (physical) density, and for any other $\upsilon$-representable density profile prescribed, 
  it represents the lowest energy possible associated with that prescribed density profile.

The method of Hedin~\cite{Hedin_65} is largely identical to that of Luttinger and Ward.\cite{LW_60} 
 This includes the fact that the energy functional reaches a stationary point, rather than the minimum,
  when the functional argument is the fully-interacting one-particle Green's function.  
 However, Hedin aims to replace the $e^2$-based (bare Coulomb) perturbative 
expansion of the electron self energy by another expansion using a screened 
interaction ${\mathcal W}$. Hedin expresses the electron self energy and the 
screened interaction as a functional of 
  the electron Green's function of the interacting system. Interestingly, the first order result,
   also termed the GW approximation, of Hedin~\cite{Hedin_65} has been shown~\cite{DvLvB_06}  
 to produce good results when compared to other density functionals. This suggests that not treating 
  $e^2$ as small might have some advantage. It is worth pointing out that the first order term,  
 $\Gamma_1 = -\frac{1}{2} \Tr \ln (\tilde {\mathcal D}_0^{-1} \4di \, U )$, 
  of the UDF described here is equivalent to the celebrated GW approximation.
  
Like the ${\mathcal W}$ propagator of Hedin,\cite{Hedin_65} 
  the $\tilde {\mathcal D}_0$ propagator introduced here also corresponds to that of a screened 
  interaction (see section~\ref{sec:screening}), thereby avoiding any  possible infrared 
  divergence associated with perturbative expansion based on bare Coulomb interactions. 
    However, the screening associated with $\tilde {\mathcal D}_0$ is
   from the KS particles and thus keeps the same form no matter how many orders one wishes to include.
 This is different from that of Hedin's where the expression of ${\mathcal W}$ in terms of 
  the electron Green's function changes with the order included. The other difference between the 
  proposed approach and reference~\onlinecite{Hedin_65} is that the UDF proposed here depends on $J_0$, a function of
   three spatial variables (and possibly with one additional time variable),  
   while the method of reference~\onlinecite{Hedin_65} expresses  via ${\mathcal W}$ the electron self energy 
   as a functional of the Green's function, a function of six spatial variables 
   (and possibly with two additional time variables). 
   
It is well known that a loopwise expansion may also be viewed as an $\hbar$ expansion~\cite{IZ_80}, that is,
an expansion of quantum-mechanical effects. By first integrating out the fermionic degrees of freedom completely,
  the proposed method is an expansion of bosonic loops formed by $\tilde {\mathcal D}_0$ propagators 
 associated with the auxiliary field $b$. The $b$ field describes the potential produced by 
  electron density fluctuations around $n_g$. Since the ground state charge 
  density $n_g$ captures the full quantum information
   of the ground state thanks to the HK theorem, one anticipates a weaker quantum effect
   associated with the auxiliary $b$ field than with the fermionic field. 
   This makes the auxiliary $b$ field a suitable candidate  
   for loop (or quantum effect) expansion, the approach pursued in this paper.

Finally, let us remark on the issue of convexity. 
The full $\Gamma[n]$ is supposed to be convex,~\cite{VF_97_arxiv} thus guaranteeing a unique solution 
 without any local minima when  searching for the minimum of $\Gamma[n]$. However, 
 in real computations only a finite number of terms of the effective action can be kept. This approximate/truncated 
 expression may not warrant convexity and thus it is not guaranteed to be free of local minima
 while numerically searching for the ground state density $n_g$ (or thermal averaged density $n_T$ 
 at finite temperature). In the near future, we plan to implement numerically the methods presented 
 in this paper, and will describe in a separate publication 
 the results obtained as well as the investigation on the issue of convexity.

\section*{Acknowledgement} 
This research was supported by the Intramural Research Program of the National Library of Medicine of
 the National Institutes of Health. 
 The author thanks Dr. Oleg Obolensky and Professor John Neumeier for useful comments. He is particularly indebted to 
  Professor Richard Friedberg, who has provided numerous useful suggestions and correspondence during the 
   writing of the paper.   
 The Feynman diagrams are made using the style file contained in the package developed in reference~\onlinecite{BT_04}
 

\end{document}